\documentclass{iopart}       % journal version
\usepackage{iopams}
\usepackage{setstack}
\usepackage[dvips]{graphics}

\def\Reals{{\mathbb R}}
\def\Complexes{{\mathbb C}}

\input diagrams

\def\Ahat{{\hat A}}
\def\Ad{{\mathop{\rm Ad}\nolimits}}
\def\ahat{{\hat a}}
\def\avec{{\bf a}}
\def\Bhat{{\hat B}}

\def\CS{{\cal C}}

\def\Ihat{{\hat I}}
\def\Im{{\mathop{\rm Im}\nolimits}}
\def\Id{{\rm Id}}
\def\jvec{{\bf j}}
\def\Jhat{{\hat J}}
\def\Jvec{{\bf J}}
\def\Jvechat{{\hat{\bf J}}}
\def\Hhat{{\hat H}}
\def\HS{{\cal H}}

\def\Liealgebra#1{{\mathfrak #1}}
\def\Liederiv{{\cal L}}
\def\max{{\mathop{\rm max}\nolimits}}
\def\min{{\mathop{\rm min}\nolimits}}	
\def\mvec{{\bf m}}
\def\nvec{{\bf n}}
\def\phat{{\hat p}}
\def\rank{{\mathop{\rm rank}\nolimits}}
\def\Thetahat{{\hat \Theta}}
\def\xhat{{\hat x}}
\def\SS{{\cal S}}
\def\ZS{{\cal Z}}  % subspace of rotational invariants
\def\zbar{{\bar z}}
\def\zerovec{{\bf 0}}

\def\ket#1{\vert#1\rangle}
\def\bra#1{\langle#1\vert}
\def\braket#1#2{\langle#1\vert#2\rangle}

\def\Wket#1#2#3#4#5#6#7#8{\left| 
    \begin{array}{@{\;}c@{\,}c@{\,}c@{\;}c@{}}
           #1 & #2 & #3 & #4 \\
           #5 & #6 & #7 & #8
           \end{array}\right>}

\def\Kket#1#2#3#4#5#6{\left|
    \begin{array}{@{\;}c@{\,}c@{\;}c@{}}
      #1 & #2 & #3 \\
      #4 & #5 & #6
    \end{array}\right>}

\def\ISchwingermodel{3}
\def\Iabmatrixelement{3}
\def\Iabmatrixelementone{5}
\def\Iudef{18}
\def\IRuproj{21}
\def\IIflow{27}
\def\IndotJeqns{28}
\def\IndotJflow{29}
\def\Izaction{64}
\def\Irotaction{74}
\def\Itwozquantone{85}

\def\IOmegasoln{92}
\def\Icomputingactions{9.2}
\def\Iampdet{11}

\begin{document}
\title{Semiclassical Mechanics of the Wigner $6j$-Symbol}
\author{Vincenzo Aquilanti}
\address{Dipartimento di Chimica, Universit\`a di Perugia, Perugia, 
Italy 06100}
\author{Hal M. Haggard, Austin Hedeman, Nadir Jeevanjee, 
Robert G. Littlejohn, and Liang Yu}
\address{Department of Physics, University of
California, Berkeley, California 94720 USA}

\ead{robert@wigner.berkeley.edu}

\begin{abstract}
The semiclassical mechanics of the Wigner $6j$-symbol is examined from
the standpoint of WKB theory for multidimensional, integrable systems,
to explore the geometrical issues surrounding the Ponzano-Regge
formula.  The relations among the methods of Roberts and others for
deriving the Ponzano-Regge formula are discussed, and a new approach,
based on the recoupling of four angular momenta, is presented.  A
generalization of the Yutsis-type of spin network is developed for
this purpose.  Special attention is devoted to symplectic reduction,
the reduced phase space of the $6j$-symbol (the 2-sphere of Kapovich
and Millson), and the reduction of Poisson bracket expressions for
semiclassical amplitudes.  General principles for the semiclassical
study of arbitrary spin networks are laid down; some of these were
used in our recent derivation of the asymptotic formula for the Wigner
$9j$-symbol.
\end{abstract}

\pacs{03.65.Sq, 02.20.Qs, 02.30.Ik, 02.40.Yy}

%\maketitle %use this if you want to have title on separate page

\section{Introduction}

The Wigner $6j$-symbol (or Racah $W$-coefficient) is a central object
in angular momentum theory, with many applications in atomic,
molecular and nuclear physics.  These usually involve the recoupling
of three angular momenta, that is, the $6j$-symbol contains the
unitary matrix elements of the transformation connecting the two bases
that arise when three angular momenta are added in two different ways.
Such applications and the definition of the $6j$-symbol based on them
are described by Edmonds (1960).  More recently the $6j$- and other
$3nj$-symbols have found applications in quantum computing (Marzuoli
and Rasetti 2005) and in algorithms for molecular scattering
calculations (De~Fazio \etal\ 2003, Anderson and Aquilanti 2006),
which make use of their connection with discrete orthogonal
polynomials (Aquilanti \etal\ 1995, 2001a,b).

The $6j$-symbol is an example of a spin network, a graphical
representation for contractions between tensors that occur in angular
momentum theory.  The graphical notation has been developed by Yutsis
\etal (1962), El Baz and Castel (1972), Lindgren and Morrison (1986),
Varshalovich \etal (1981), Stedman (1990), Danos and Fano (1998),
Wormer and Paldus (2006), Balcar and Lovesey (2009) and others.  The
$6j$-symbol is the simplest, nontrivial, closed spin network (one that
represents a rotational invariant). Spin networks are important in
lattice QCD and in loop quantum gravity where they provide a
gauge-invariant basis for the field.  Applications in quantum gravity
are described by Rovelli and Smolin (1995), Baez (1996), Carlip
(1998), Barrett and Crane (1998), Regge and Williams (2000), Rovelli
(2004) and Thiemann (2007), among others.

Alongside the Yutsis school of graphical notation and the
Clebsch-Gordan school of algebraic manipulation there is a third
approach to the evaluation of rotational ($SU(2)$) invariants. The
third method, sometimes called chromatic evaluation, grew out of
Penrose's doctoral work on the graphical representation of tensors and
is closely related to knot theory. We will not have further occasion
to mention this school, see Penrose (1971) for its introduction,
Rovelli (2004) for an overview and Kauffman and Lins (1994) for its
full development. 

The asymptotics of spin networks and especially the $6j$-symbol has
played an important role in many areas.  By ``asymptotics'' we refer
to the asymptotic expansion for the spin network when all $j$'s are
large, equivalent to a semiclassical approximation since large $j$ is
equivalent to small $\hbar$.  The asymptotic expression for the
$6j$-symbol (the leading term in the asymptotic series) was first
obtained by Ponzano and Regge (1968), or, more precisely, they
obtained several formulas, valid inside and outside the classically
allowed region and in the neighborhood of the caustics.  In the same
paper those authors gave the first spin foam model (a discretized path
integral) for quantum gravity.  The formula of Ponzano and Regge is
notable for its high symmetry and the manner in which it is related to
the geometry of a tetrahedron in three-dimensional space.  It is also
remarkable because the phase of the asymptotic expression is identical
to the Einstein-Hilbert action for three-dimensional gravity
integrated over a tetrahedron, in Regge's (1961) simplicial
approximation to general relativity.  The semiclassical limit of the
$6j$-symbol thus plays a crucial role in simplicial approaches to the
quantization of the gravitational field.

For all these reasons, the asymptotic formula of Ponzano and Regge for the
$6j$-symbol has attracted a great deal of attention.  Ponzano and Regge
obtained their formula by inspired guesswork, supporting their
conclusion both with numerical evidence and arguments of consistency
and plausibility.  The formula itself is of the one-dimensional
WKB-type, a reflection of the fact that the $6j$-symbol fundamentally
represents a dynamical system of one degree of freedom.

The Ponzano-Regge formula was first derived by Neville (1971), using
the recursion relations satisfied by the $6j$-symbol and a discrete
version of WKB theory.  Similar techniques were later used by Schulten
and Gordon (1975a,b), who also presented stable algorithms for
evaluating the $6j$-symbol numerically.  A proof of a different sort
was later given by Biedenharn and Louck (1981), based on showing that
the Ponzano-Regge formula satisfies a set of defining properties of
the $6j$-symbol.

More recently there have appeared more geometrical treatments of the
asymptotics of the $6j$-symbol, that is, those based on geometric
quantization (Kirillov 1976, Guillemin and Sternberg 1977, Woodhouse
1991), symplectic geometry and symplectic and Poisson reduction
(Abraham and Marsden 1978, Arnold 1989, Marsden and Ratiu 1999) and
other techniques.  Among these are the works by Roberts (1999) and by
Charles (2008).  In addition, the $6j$-symbol has been taken as a test
case for asymptotic studies of amplitudes that occur in quantum
gravity (Barrett and Steele 2003, Freidel and Louapre 2003), in which
the authors developed integral representations for the $6j$-symbol as
integrals over products of the group manifold.  There have also been
quite a few other studies of asymptotics of particular spin networks,
including Barrett and Williams (1999), Baez \etal (2002), Rovelli and
Speziale (2006), Hackett and Speziale (2007), Conrady and Freidel
(2008), Alesci \etal (2008), Barrett \etal (2009), among others.  We
also mention the works of Gurau (2008), which applies standard
asymptotic techniques (Stirling's approximation, etc) directly to
Racah's sum for the $6j$-symbol; of Ragni \etal (2010) on the
computation of $6j$-symbols and on the asymptotics of the $6j$-symbol
when some quantum numbers are large and others small; and of
Littlejohn and Yu (2009) on uniform approximations for the
$6j$-symbol.

In addition there has been some work on the $q$-deformed $6j$-symbol,
important for the regularization of the Ponzano-Regge spin-foam model
(Turaev and Viro 1992, Ooguri 1992a,b) and for its possible connection
to quantum gravity with cosmological constant.  In particular, Taylor
and Woodward (2004) applied the recursion and WKB method of Schulten
and Gordon to the $q$-deformed $6j$-symbol.  The results are
geometrically interesting (the tetrahedron of Ponzano and Regge is
moved from $\Reals^3$ to $S^3$ when the $q$-deformation is turned on),
but it seems that at present there is no geometrical treatment of the
asymptotics of the $q$-deformed $6j$-symbol, analogous to what is
available for the ordinary $6j$-symbol.  There is also the recent work
of Van der Veen (2010) on the asymptotics of general $q$-deformed spin
networks, which treats the problem from the standpoint of knot theory
and representation theory.  Among other things, this work creates a
broad generalization of the Schwinger-Bargmann generating function of
the $6j$-symbol.

In Aquilanti \etal (2007) we applied multidimensional WKB theory for
integrable systems to the asymptotics of the $3j$-symbol, and in this
paper we apply similar techniques to the $6j$-symbol.  These methods
bear the closest relationship to the works of Roberts (1999) and of
Charles (2008).  The point of this paper is not another derivation of
the Ponzano-Regge formula, although one is provided, but rather to
clarify the relationship among some of the methods used in the past,
to reveal useful calculational techniques, and to lay the basis for
the development of new results.  Among the latter we mention our own
work on uniform approximations for the $6j$-symbol (Littlejohn and Yu,
2009), our recent derivation of the asymptotic form for the
$9j$-symbol (Haggard and Littlejohn, 2010), both of which relied on
techniques explained in this paper, and our work on the
Bohr-Sommerfeld quantization of the volume operator in loop quantum
gravity (Bianchi and Haggard, 2011).  Previous and current work on the
volume operator includes Chakrabarti (1964), L\'evy-Leblond and
L\'evy-Nahas (1965), Lewandowski (1996), Major and Seifert (2001),
Carbone \etal (2002), Neville (2006), Brunnemann and Rideout (2008,
2010) and Ding and Rovelli (2010).

In addition, this paper is distinguished by its use of what we call
the ``$4j$-model'' for the $6j$-symbol, in contrast to the
``$12j$-model'' used by Roberts (1999).  The $4j$-model is less
symmetrical than the $12$-model, but it is closer to the manner in
which the $6j$-symbol is commonly used in recoupling theory.  In
addition, in loop quantum gravity (Rovelli 2004) angular momenta
represent area vectors, which in the case of four-valent nodes
correspond by Minkowski's (1897) theorem to a tetrahedron.  In this
context the $4j$-model is closer to the applications than the
$12j$-model, indeed, it played an important role in the work of Bianchi
and Haggard (2011).

In this paper we refer to Aquilanti \etal (2007) as I, for example
writing eqn.~(I.13) for an equation from that paper.  We note two errata
in I, namely, $\sigma(x)$ in (I.89) should read $\sigma(u)$, and $j_3$
and $j_4$ should be swapped in the $6j$-symbol in (I.112).

\section{Spin network notation}
\label{spinnetworknotation}

We begin by explaining our notation for spin networks, which is based
on that of Yutsis \etal (1962) with modifications due to Stedman
(1990).  At the end of this section we compare our conventions for
spin networks with others in the Yutsis tradition.

\subsection{The $3j$-symbol and Wigner intertwiner}
\label{3jwigner}

\begin{figure}[htb]
\begin{center}
\scalebox{0.43}{\includegraphics{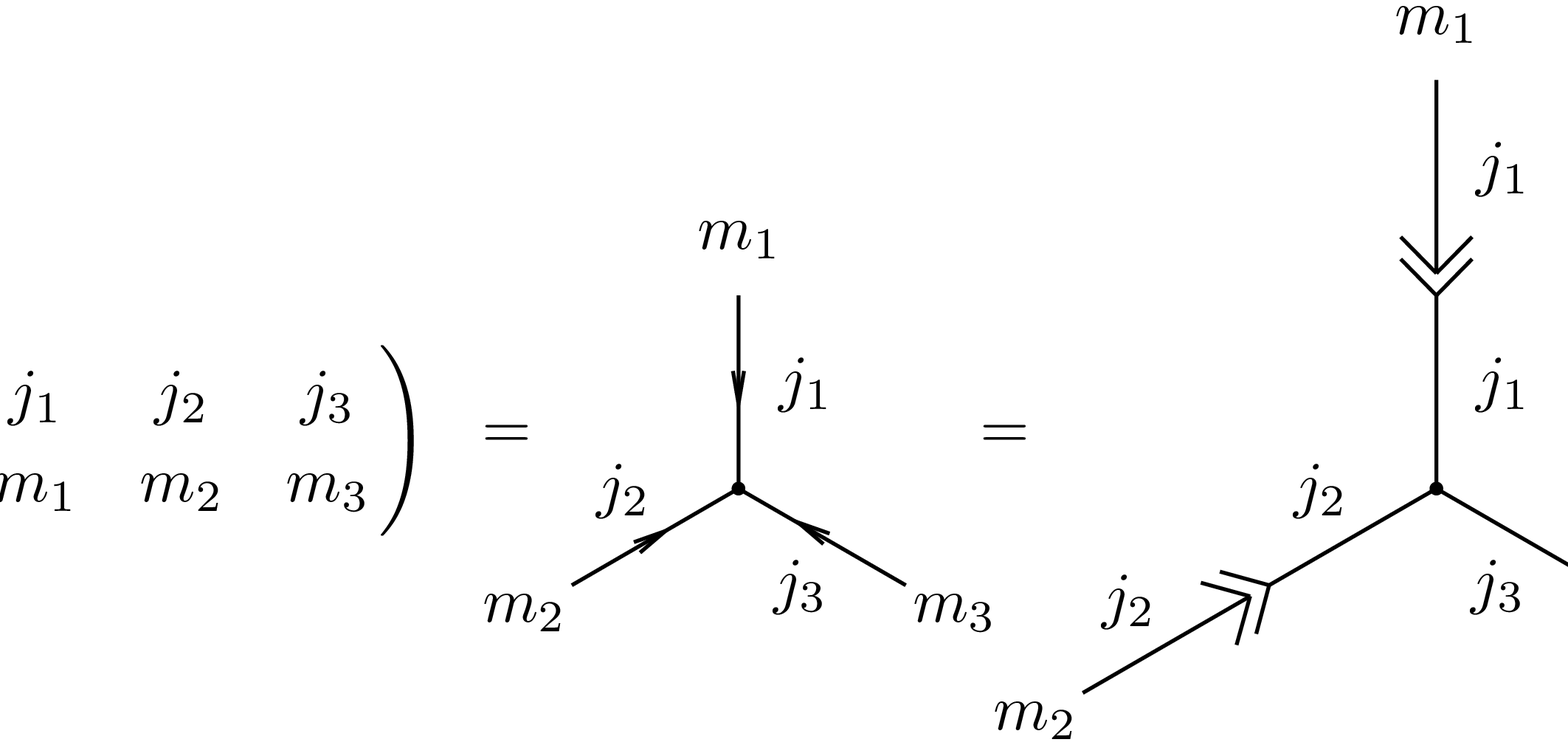}}
\end{center}
\caption[rps]{\label{3jnetwork} The $3j$-symbol contains the
components of the standard three-valent intertwiner.}
\end{figure}

The $3j$-symbol is a number that can be regarded as the components of
an intertwiner $W:\CS_{j_1}\otimes\CS_{j_2}\otimes\CS_{j_3
}\to\Complexes$ with respect to the standard basis $\ket{j_1m_1
}\otimes\ket{j_2m_2}\otimes\ket{j_3m_3}$, as indicated in
Fig.~\ref{3jnetwork}.  In this paper $\CS_j$ denotes a carrier
space for unitary irrep $j$ of $SU(2)$, so that $\dim\CS_j = 2j+1$. We
call $W$ the ``Wigner'' intertwiner.  We will gradually explain the
features of Fig.~\ref{3jnetwork} as we proceed.

The standard notation for the $3j$-symbol is on the left of
Fig.~\ref{3jnetwork}, while the central diagram is the standard Yutsis
spin network for the $3j$-symbol, with small arrows presented as in
the Yutsis notation.  The indices $(m_1,m_2,m_3)$ in the central
diagram are covariant, that is, they transform under rotations as the
components of a dual vector (in contrast to an ordinary vector).  In a
Hilbert space we regard ordinary wave functions or ket vectors as
``vectors,'' while bra vectors are regarded as ``dual vectors.''
Thus, contravariant indices are those that transform as the components
of a vector.  In the Yutsis notation the arrows indicate the
transformation properties of the corresponding $m$ index, and there
are rules for ``raising and lowering'' indices, that is, reversing the
direction of the arrow.  The rules do not, however, make use of the
metric as in ordinary tensor analysis.  Our definition of the arrow
(explained below) is different from that of Yutsis, but designed so
that the two notations agree as much as possible.  In particular, our
notation for the $3j$-symbol is the same as the Yutsis diagram in
Fig.~\ref{3jnetwork}.

A trivalent node of a spin network such as those illustrated in
Fig.~\ref{3jnetwork} is assumed to have a positive or counterclockwise
orientation, unless otherwise indicated (thus we dispense with the $+$
sign used by Yutsis).

\subsection{Bras, kets and scalar products}
\label{brasketssps}

\begin{figure}[htb]
\begin{center}
\scalebox{0.43}{\includegraphics{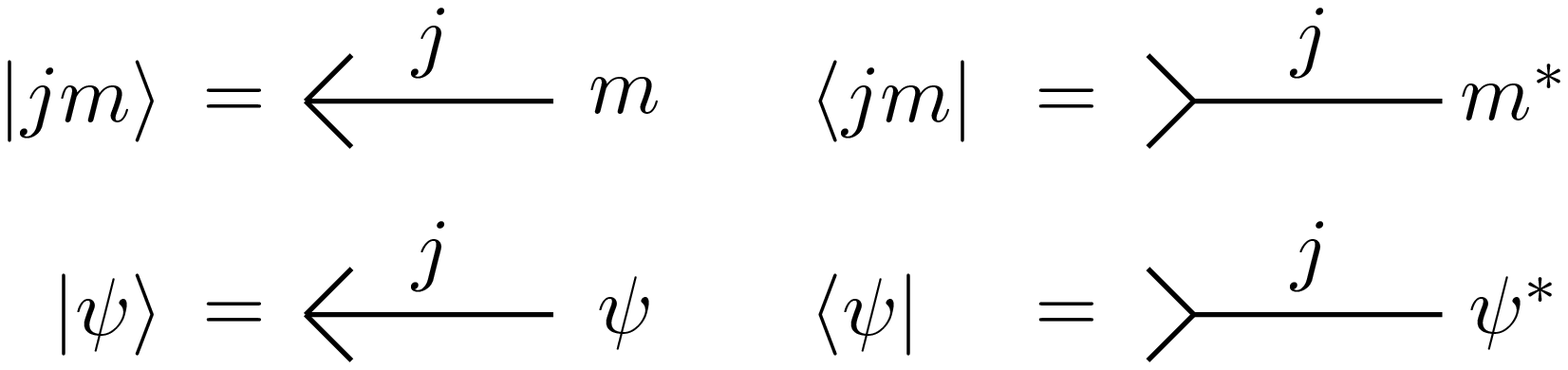}}
\end{center}
\caption[rps]{\label{basisvectors} Spin networks for bras and 
kets.}
\end{figure}

The diagram on the right of Fig.~\ref{3jnetwork} makes use of the
standard basis vectors $\ket{j_im_i}$ in $\CS_{j_i}$, $i=1,2,3$.  The
spin networks for these basis vectors and their duals are shown in
Fig.~\ref{basisvectors}.  The large, broadly open arrow is a
``chevron'' (Stedman 1990).  When pointing outward (inward), the
chevron indicates a ket (bra) vector.  Also shown in
Fig.~\ref{basisvectors} is the spin network for an arbitrary vector
$\ket{\psi}$ in $\CS_j$, and the dual bra vector $\bra{\psi}$ obtained
by Hermitian conjugation.  In the Dirac notation it is customary to
label bras by the same symbol as kets, it being understood that the
two are related by Hermitian conjugation.  This convention is so
deeply ingrained that we dare not change it.  But in spin networks
there are two different ways of converting kets into bras and vice
versa, and this presents some notational challenges.  One can see in
Fig.~\ref{basisvectors} that Hermitian conjugation applied to bras and
kets is notationally the changing of bra chevrons to ket chevrons and
vice versa, and the starring of identifying symbols, with a double
star being removed.  Full rules for Hermitian conjugation of any spin
network are given in Sec.~\ref{Hermitianconjugation}.

The lines of a spin network will be referred to as ``edges,''
including cases like those shown in Fig.~\ref{basisvectors}.

An edge of a spin network ending in an unstarred $m$ index represents
a contraction with the basis ket $\ket{jm}$, so the $m$ index
transforms under rotations as a covariant index.  The explicit
insertion of basis kets may be seen in Fig.~\ref{3jnetwork}.  An edge
ending in a starred $m$ index represents the insertion of a basis bra
$\bra{jm}$, so starred indices are contravariant.  Thus, the star
on an $m$ index indicates its transformation property, and invariant
contractions can only take place between a pair of starred and
unstarred $m$ indices.

\begin{figure}[htb]
\begin{center}
\scalebox{0.43}{\includegraphics{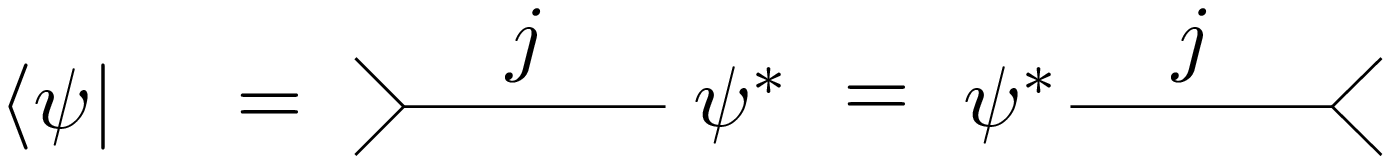}}
\end{center}
\caption[rps]{\label{orientation} Orientation of a spin network does not
matter.}
\end{figure}

As illustrated in Fig.~\ref{orientation}, the orientation of a spin
network on the page does not affect its value.  The spin network in
the figure has been rotated by $180^\circ$. 

\begin{figure}[htb]
\begin{center}
\scalebox{0.43}{\includegraphics{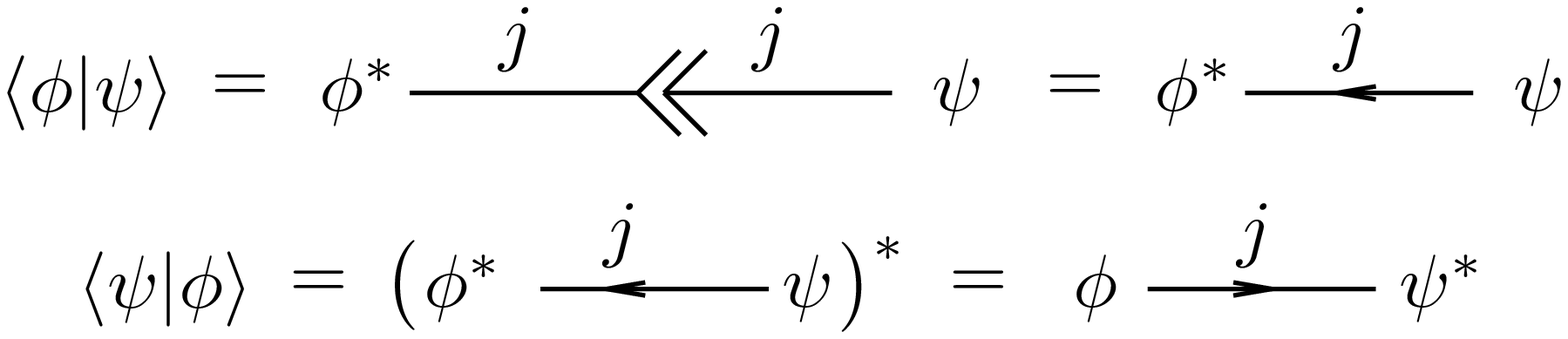}}
\end{center}
\caption[rps]{\label{scalarprod} Spin networks for scalar products or
contractions.  Illustration of rules for complex conjugation.}
\end{figure}

As illustrated in the final diagram in Fig.~\ref{3jnetwork} and in
Fig.~\ref{scalarprod}, when a bra chevron and a ket chevron are
juxtaposed, it represents the scalar product or contraction.  In
effect, the bra chevron acts as a receptacle for the ket chevron, and
vice versa.  After the contraction the two edges may be joined, with a
small arrow remaining to indicate which was the bra and which the ket
in the contraction.  One might suppose that the star would carry the
same information, but, as shown below, it is possible to change the
direction of the arrow without changing the stars.
Figure~\ref{scalarprod} also presents another example of Hermitian
conjugation (complex conjugation, in this case).  Since the small
arrow represents the contraction of a bra and a ket chevron, when the
chevrons are reversed, the direction of the arrow changes.

\begin{figure}[htb]
\begin{center}
\scalebox{0.43}{\includegraphics{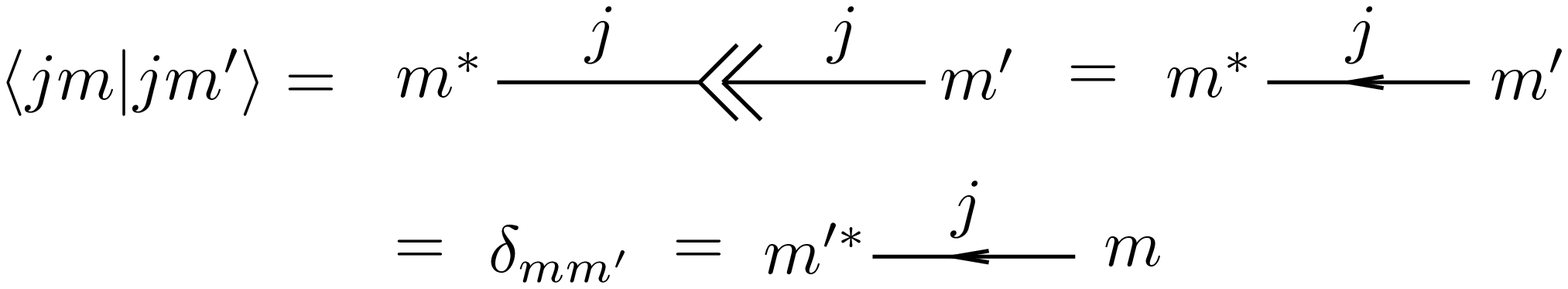}}
\end{center}
\caption[rps]{\label{basison} Orthonormality relations for basis
vectors.}
\end{figure}

As a special case of the scalar product, the orthonormality relations
of the basis vectors are illustrated in Fig.~\ref{basison}.  The final
spin network follows from the symmetry of the Kronecker delta, or,
alternatively, by the reality of $\delta_{mm'}$ and the rules for
complex conjugation.

The $j$ labels on the edges of the spin networks indicate which
carrier space $\CS_j$ the ket or bra lies in, or in which carrier
space a contraction has taken place.  If there are distinct carrier
spaces with the same $j$ label, then additional distinguishing
information must be supplied.

\subsection{Intertwiners}
\label{intertwiners} 

\begin{figure}[htb]
\begin{center}
\scalebox{0.43}{\includegraphics{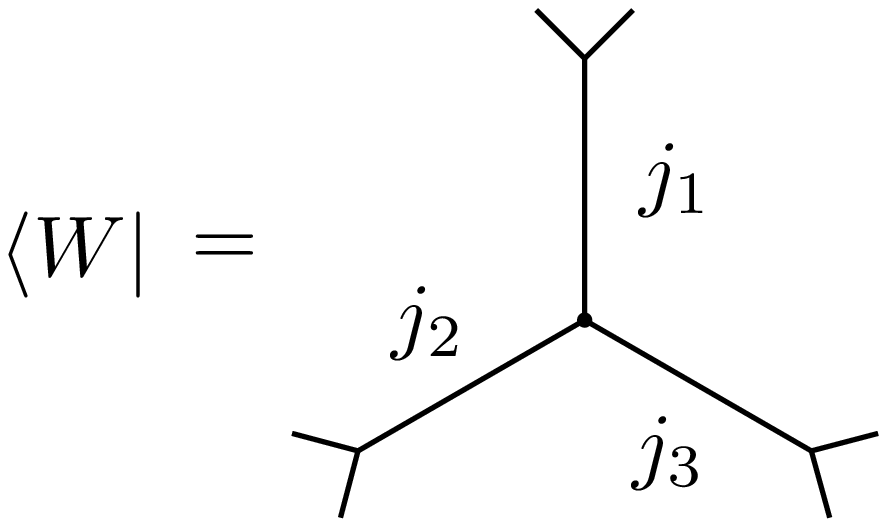}}
\end{center}
\caption[rps]{\label{Wintertwiner} Spin network for the Wigner or 
$3j$-intertwiner.}
\end{figure} 

An $SU(2)$ intertwiner (the only kind we are interested in) is a
linear map between two vector spaces that commutes with the action of
$SU(2)$ on the two spaces.  The only case of interest here is where
the target space is $\Complexes$, consisting of scalars, that is,
invariants under rotations.  The Wigner intertwiner
$W:\CS_{j_1}\otimes\CS_{j_2 }\otimes\CS_{j_3} \to \Complexes$ is of
this type.  But a linear map from a Hilbert space to $\Complexes$ can
be thought of as a dual or bra vector, for example, we can associate
the map $W$ with a bra vector $\bra{W}$ belonging to
$\CS^*_{j_1}\otimes\CS^*_{j_2}\otimes\CS^*_{ j_3}$.  The spin network
notation for $\bra{W}$ is shown in Fig.~\ref{Wintertwiner}.  The
components of the Wigner intertwiner, that is, the $3j$-symbol, are
obtained by inserting basis kets, the ket chevron first, into the bra
chevrons of the intertwiner, as in the final diagram of
Fig.~\ref{3jnetwork}.

More generally, a $\Complexes$-valued intertwiner on a Hilbert space
$\HS$ can be regarded as an $SU(2)$-invariant bra vector on this
space, that is, a member of $\HS^*$.  By Hermitian conjugation we
obtain an $SU(2)$-invariant ket vector in $\HS$.  Thus, there is a
one-to-one correspondence between the subspace of $\HS$ of rotationally
invariant vectors and the set of $\Complexes$-valued intertwiners on
$\HS$.  The subspace $\ZS$ introduced in Sec.~\ref{4jmodel} below is a
subspace of this type, consisting of rotationally invariant vectors.

\subsection{Tensor products and resolution of identity}
\label{tensorprodsroi}

\begin{figure}[htb]
\begin{center}
\scalebox{0.43}{\includegraphics{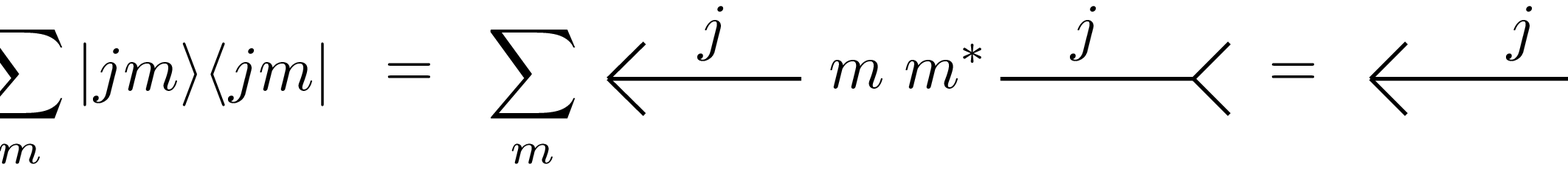}}
\end{center}
\caption[rps]{\label{roi} Resolution of identity, and an illustration
of the outer product.}
\end{figure}

The outer product of a ket with a bra is represented in spin network
language simply by placing the spin networks for the ket and the bra
on the same page, as illustrated in Fig.~\ref{roi}.  The orientations
of the bra and ket spin networks is immaterial, but in the figure they
have been placed with their $m$ indices adjacent in order to emphasize
the summation (a contraction over two indices, one covariant, one
contravariant).  The final diagram is the spin network for the
identity operator on space $\CS_j$.

\begin{figure}[htb]
\begin{center}
\scalebox{0.43}{\includegraphics{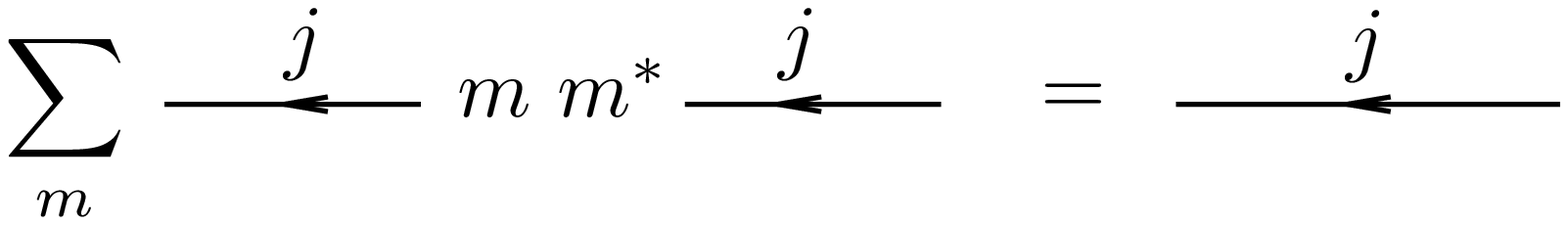}}
\end{center}
\caption[rps]{\label{msum} Summing on $m$ to join two edges.}
\end{figure}

Figure~\ref{roi} illustrates another technique, the replacement of a
sum on $m$ by a joining of edges.  The general usage is shown in
Fig.~\ref{msum}.  The directions of the arrows and the stars on one of
the $m$'s must be coordinated as shown for the identity to be used as
shown.  The small arrow in the final diagram in Fig.~\ref{msum} (a
fragment of a spin network) is a reminder of the directions of the
arrows before the sum.  This small arrow is omitted in the final
diagram of Fig.~\ref{roi} (the identity diagram) because the chevrons
already indicate the direction of the edge.  Recall that edges are
also joined on contracting a bra with a ket, as in
Fig.~\ref{scalarprod}. 

\begin{figure}[htb]
\begin{center}
\scalebox{0.43}{\includegraphics{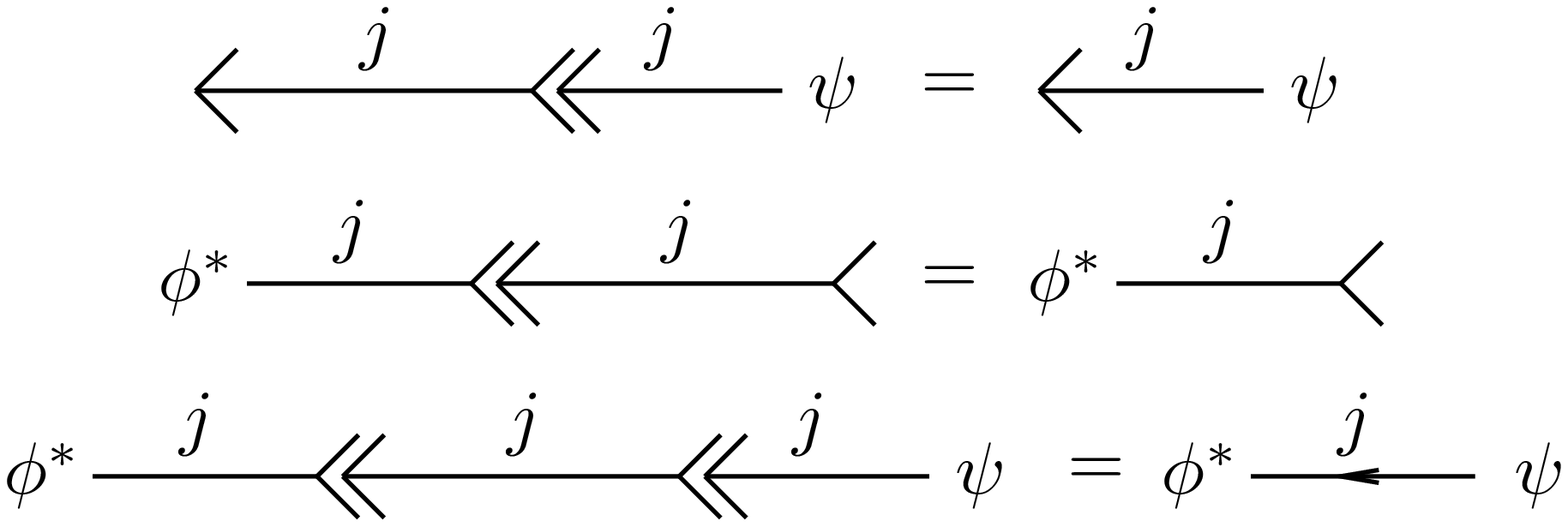}}
\end{center}
\caption[rps]{\label{idj} Summing on $m$ to join two edges.}
\end{figure}

The use of the identity diagram is illustrated in Fig.~\ref{idj}.  The
first row illustrates its action on kets (as a map $:\CS_j\to\CS_j$),
where bra and ket chevons are combined as in Fig.~\ref{scalarprod}.
The small arrow is omitted after the joining of the bra and ket
chevrons because the remaining chevron indicates the direction of the
edge.  The second row of the figure illustrates its action on bras
(the map $:\CS^*_j\to\CS_j^*$), and the third row illustrates its
action as a map $:\CS_j^*\otimes\CS_j\to\Complexes$, that is, the
scalar product map $:\bra{\phi}\otimes\ket{\psi}\mapsto\braket{\phi}
{\psi}$.  All of these usages are encompassed by the same spin
network.  The identity spin network can also be seen as an element of
$\CS_j\otimes\CS^*_j$, that is, the space dual to
$\CS_j^*\otimes\CS_j$ on which it, viewed as a $\Complexes$-valued
linear operator, acts (the third line of Fig.~\ref{idj}).

In general a spin network has some number of edges that terminate in
incoming or outgoing chevrons.  Examples are the identity diagram in
Fig.~\ref{roi} and the Wigner intertwiner in Fig.~\ref{Wintertwiner}.
In all cases there are multiple interpretations of the spin network as
a linear operator mapping one vector space to another, depending on
how many of the incoming and outgoing chevrons have ket and bra
chevrons plugged into them (specifying the domain), and how many are
left free (specifying the range).  The domain is the tensor product of
some number of $\CS_j$ times some number of $\CS_j^*$, and so is the
range.  In the extreme case that all incoming and outgoing chevrons on
the spin network have kets and bras plugged into them the result is
simply a number and the range is $\Complexes$.  In that case the spin
network, as a $\Complexes$-valued linear operator, can be seen as a
vector in the space dual to the domain.

This facile identification of closely associated operators, and their
reinterpretation as elements of vector spaces, is an important advantage
of spin networks.  It is difficult and awkward to do something
similar with the Dirac notation.

In general, a tensor is a multilinear operator acting on a tensor
product of some set of vector spaces and their duals.  Thus, a spin
network is a notation for a tensor on some product of the $\CS_j$ and
their duals.  The edges of the spin network terminating in ket or bra
chevrons indicate the nature of the space the tensor acts on.  In
general, the tensor product of two tensors in spin network notation is
indicated by the placing of the two spin networks together on the
page, in any orientation.  The outer product of a bra and a ket
illustrated in Fig.~\ref{roi} is a special case.  Partial or complete
contractions of tensors are indicated by joining some or all ket
chevrons with bra chevrons.

\subsection{The $2j$ symbol and intertwiner}

We now consider another intertwiner, which leads to an important
mapping between kets and bras, alternative to Hermitian conjugation.
The intertwiner acts on the Hilbert space $\CS_j\otimes\CS'_j$, the
tensor product of two carrier spaces of the same $j$.  In general we
wish to consider the second carrier space as distinct from the first,
which is the purpose of the prime on the second factor.  To within a
normalization and phase, there is a unique vector in this space that
is invariant under rotations; we call it $\ket{K}$, and express it in
terms of the Clebsch-Gordan coefficients by
	\begin{equation}
	\ket{K}= \sqrt{2j+1} \sum_{mm'} \ket{jm}\otimes\ket{jm'}\,
	 C^{00}_{jjmm'}.
	\label{ketKdef}
	\end{equation}
This vector can also be expressed in terms of the ``$2j$-symbol,''
which we define in terms of the usual $3j$-symbol by
	\begin{equation}
    \fl\left(\begin{array}{cc}
        j & j \\
        m & m'
    \end{array}\right)=
	\left(\begin{array}{ccc}
	j & j & 0 \\
	m & m' & 0
    \end{array}\right) = 
	C^{00}_{jjmm'} =
	\frac{(-1)^{j-m}}{\sqrt{2j+1}}
	\,\delta_{m,-m'}.
	\label{2jdef}
	\end{equation}
The terminology ``$2j$-symbol'' is not entirely standard, but it has
been used by Stedman (1990).  The invariant vector $\ket{K}$ can
also be written,
	\begin{equation}
	\fl\ket{K}=\sum_{mm'} \ket{jm}\otimes\ket{jm'}\,
	(-1)^{j-m} \,\delta_{m,-m'} = \sum_m
	\ket{jm}\otimes\ket{j,-m}\,(-1)^{j-m}.
	\label{K2jdef}
	\end{equation}

\begin{figure}[htb]
\begin{center}
\scalebox{0.43}{\includegraphics{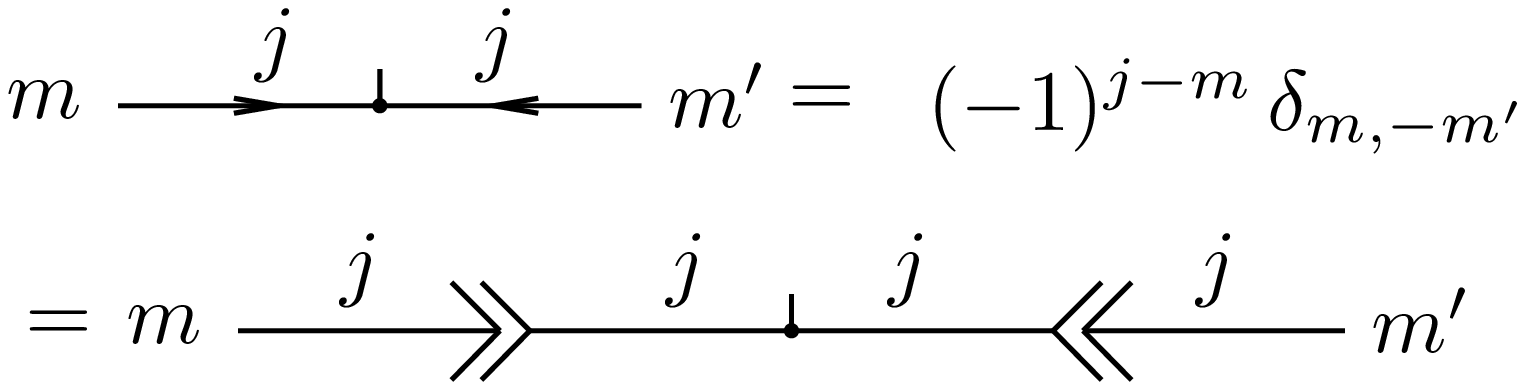}}
\end{center}
\caption[rps]{\label{2jnetwork} Components of the map $K$, a standard
bivalent intertwiner.}
\end{figure}

By Hermitian conjugation we convert $\ket{K}$ into the bra $\bra{K}$,
which is otherwise an intertwiner $K:\CS_j\otimes\CS'_j\to\Complexes$.
Just as the components of the intertwiner $W$ are the $3j$-symbol, the
components of the intertwiner $K$ are the $2j$-symbol, multiplied
however by $\sqrt{2j+1}$ because of a normalization convention.
Figure~\ref{2jnetwork} shows first the spin network for the components
of $K$, which is conceived of as a standard bivalent node or
intertwiner.  The small arrows indicate that the components on the
first line can be considered the result of plugging basis kets into
the intertwiner itself, as seen on the second line.  The short line
extending above the node is a ``stub'' (Stedman 1990), whose purpose
is to orient the node.  The convention is that if we start at the stub
and move in a positive (counterclockwise) direction, the first and
second edges we encounter are respectively the first and second
operands of $K$, conceived of as a map
$:\CS_j\otimes\CS'_j\to\Complexes$.

We note that if $V$ and $W$ are vector spaces, then $V\times W$ is not
the same as $V\otimes W$, but if we have a bilinear map on $V \times
W$ it can be extended to a linear map on $V\otimes W$ by linear
superposition.  For example, in the previous paragraph we have
regarded $K$ as a bilinear map $:\CS_j \times \CS'_j\to\Complexes$,
and computed its components as $K(\ket{jm},\ket{jm'})$.  The first and
second operands of this expression correspond to the first and second
edges as specified by the stub.

\begin{figure}[htb]
\begin{center}
\scalebox{0.43}{\includegraphics{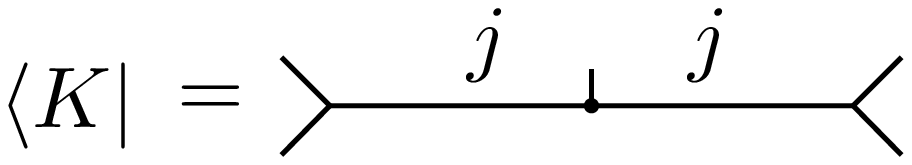}}
\end{center}
\caption[rps]{\label{2jKdef} Spin network for map
$K:\CS_j\otimes\CS'_j\to\Complexes$.}
\end{figure}

Figure~\ref{2jnetwork} also gives the numerical values of the
components of $K$, and in the final diagram, the network for $K$
itself, with two kets inserted.  The network for $K$ in isolation is
illustrated in Fig.~\ref{2jKdef}, regarded as a bra vector on $\CS_{
j}\otimes\CS'_j$, that is, as an element of
$\CS^*_j\otimes\CS^{\prime *}_j$.

\begin{figure}[htb]
\begin{center}
\scalebox{0.43}{\includegraphics{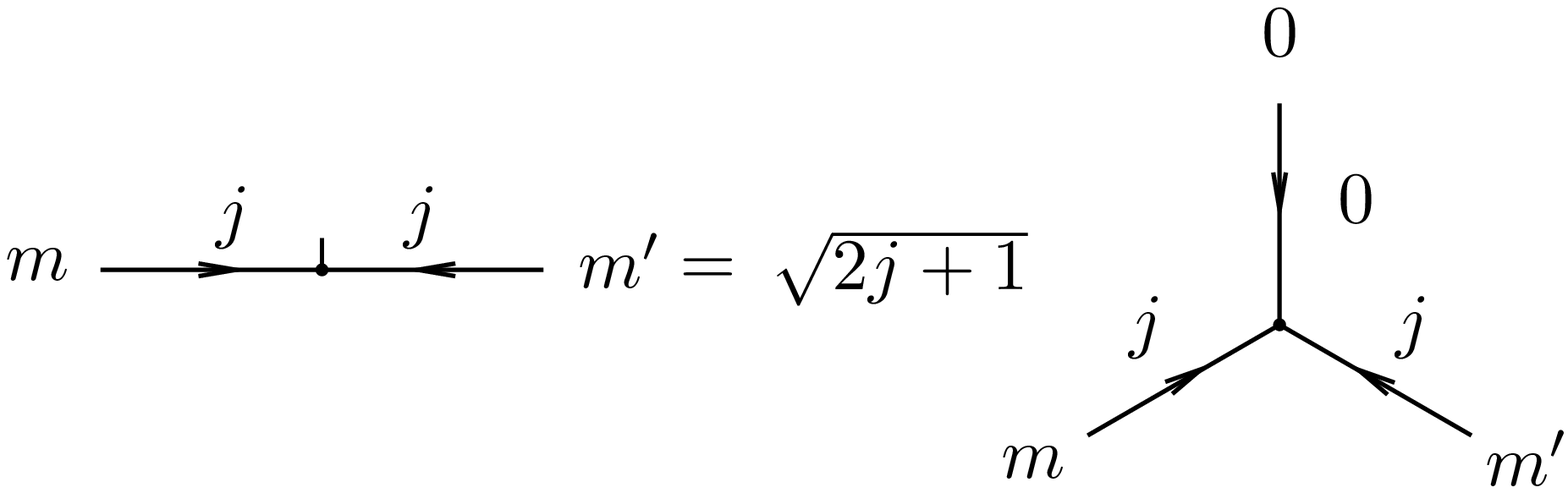}}
\end{center}
\caption[rps]{\label{vestigialstub} The stub can be regarded as a
vestigial edge of a $3j$-symbol with value $j=0$.}
\end{figure}  

The stub in Fig.~\ref{2jnetwork} can be regarded as a vestigial edge
of a $3j$-symbol or $W$-intertwiner with the value $j=0$, although one
must beware of the normalization convention.  This is illustrated in
Fig.~\ref{vestigialstub}, which is equivalent to (\ref{2jdef}).  The
value does not depend on the direction of the arrow on the zero edge
(see below for rules for reversing arrows).

\begin{figure}[htb]
\begin{center}
\scalebox{0.43}{\includegraphics{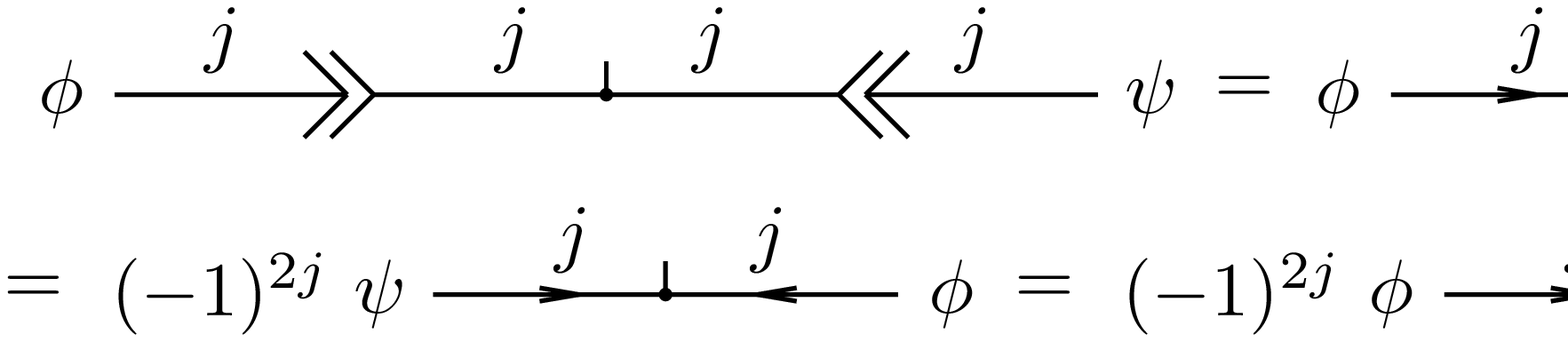}}
\end{center}
\caption[rps]{\label{Ksymm} The intertwiner $K$ acquires a phase
$(-1)^{2j}$ if the two operands are swapped.}
\end{figure} 

The components of $K:\CS_j\otimes\CS'_j\to\Complexes$, seen in
Fig.~\ref{2jnetwork}, acquire a phase of $(-1)^{2j}$ if $m$ and $m'$
are swapped.  This is equivalent to the statement
	\begin{equation}
	K(\ket{\phi},\ket{\psi}) = (-1)^{2j} K(\ket{\psi},\ket{\phi}),
	\label{Kswapphase}
	\end{equation}
for all $\ket{\psi}$, $\ket{\phi}$, which is illustrated in spin
network language in Fig.~\ref{Ksymm}.  The final diagram differs from
the preceding simply by a $180^\circ$ rotation, so the value is the
same.  But this leads to the rule, that a spin network acquires a
phase of $(-1)^{2j}$ when the stub at a bivalent node is inverted.  In
particular, the arrow on the null edge in Fig.~\ref{vestigialstub} can
be inverted without changing the value.

\subsection{Kets to bras}
\label{ketstobras}

\begin{figure}[htb]
\begin{center}
\scalebox{0.43}{\includegraphics{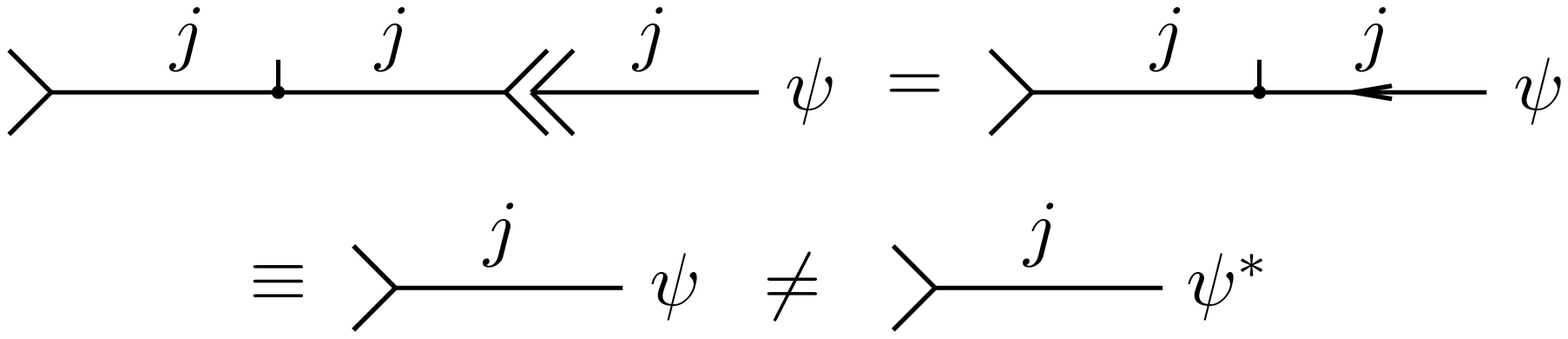}}
\end{center}
\caption[rps]{\label{ket2bra} The intertwiner $K$ can be used to
convert a ket to a bra.}
\end{figure}  

The intertwiner $K:\CS_j\otimes\CS_j\to\Complexes$ has an alternative
interpretation as a map $K_1:\CS_j\to\CS^*_j$, where for simplicity we
have dropped the prime on the second factor, and where the
$1$-subscript distinguishes the new map from the old. This alternative
interpretation is natural in spin network language, as seen in
Fig.~\ref{ket2bra}.  The spin network for $\ket{\psi} \in \CS_j$ is
plugged into the second operand of the spin network for $K$, resulting
in a spin network with one free bra chevron. The choice of the second
operand of $K$ for this purpose is conventional.  The result is an
element of $\CS_j^*$, that is, it is a bra. As indicated in the
figure, we abbreviate this spin network by drawing the same spin
network for $\ket{\psi}$ that we started with, except the ket chevron
is converted into a bra chevron. 

In other words, when we use $K_1$ to convert a ket into a bra, we just
flip the ket chevron, leaving everything else the same.  In
particular, we do not put a star on the label of the ket.  This
distinguishes the map $K_1:\CS_j \to \CS_j^*$ from the metric or
Hermitian conjugation, which is also a map $:\CS_j \to \CS_j^*$.
When the metric is used to convert a ket to a bra, not only is the ket
chevron flipped to a bra chevron, but a star is appended to the
label.  These two maps are quite distinct; in particular, $K_1$ is a
linear map, while the metric is an antilinear map.  As indicated in
Fig.~\ref{ket2bra}, the results are not the same.

When a ket is turned into a bra, the components with respect to some
basis change from contravariant to covariant.  But since there is more
than one way to do this, any notation based on the position (upper or
lower) of the indices is inadequate to represent the result. 

\begin{figure}[htb]
\begin{center}
\scalebox{0.43}{\includegraphics{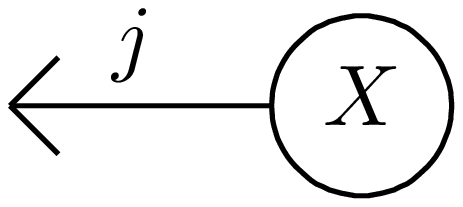}}
\end{center}
\caption[rps]{\label{Xket} An arbitrary spin network with one edge
terminating in a ket chevron.}
\end{figure}  

The map $K_1$ can be used to turn a ket chevron into a bra chevron on
any spin network, not only on kets themselves.  A notation for an
arbitrary spin network with an edge terminating in a ket chevron is
shown in Fig.~\ref{Xket}.  The circle around the $X$ indicates the
rest of the spin network, which may include other edges terminating in
ket or bra chevrons.  

\begin{figure}[htb]
\begin{center}
\scalebox{0.43}{\includegraphics{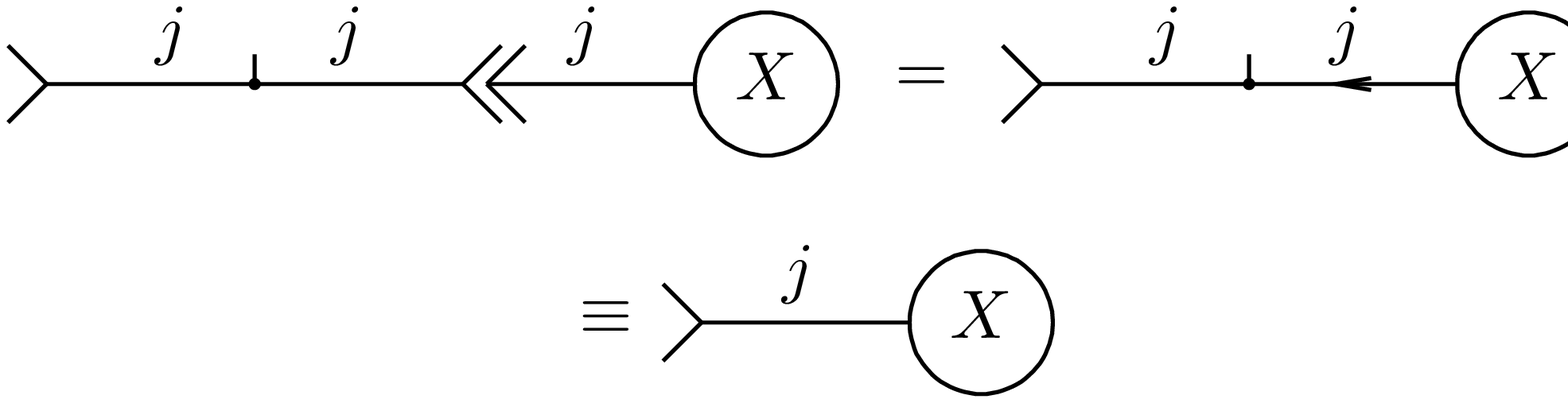}}
\end{center}
\caption[rps]{\label{lowerindex} Converting a ket chevron into a bra
chevron with the map $K_1:\CS_j\to\CS^*_j$.}
\end{figure}  

To convert the ket chevron in Fig.~\ref{Xket} into a bra
chevron we simply insert it into the second operand of
$K:\CS_j\otimes\CS_j\to\Complexes$, as shown in Fig.~\ref{lowerindex}.
The remainder of the spin network, indicated by the $X$, does not
change. 

\subsection{Bras to kets}
\label{brastokets}

\begin{figure}[htb]
\begin{center}
\scalebox{0.43}{\includegraphics{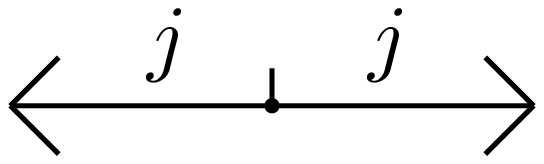}}
\end{center}
\caption[rps]{\label{Kinv} Spin network for the maps
$K^{-1}:\CS_j^*\otimes\CS_J^*\to\Complexes$ and 
$K_1^{-1}:\CS_j^*\to\CS_j$.}
\end{figure} 

\begin{figure}[htb]
\begin{center}
\scalebox{0.43}{\includegraphics{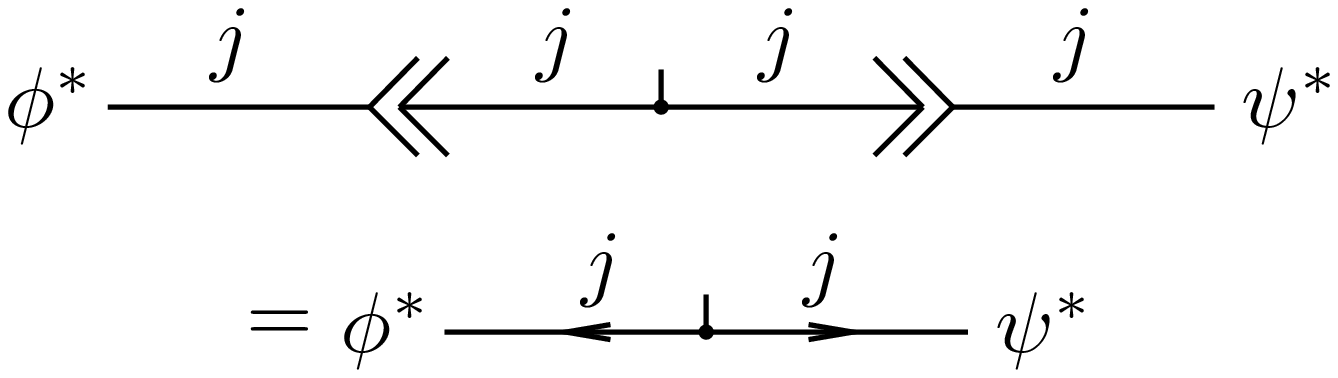}}
\end{center}
\caption[rps]{\label{Kinvsandwich} Spin network for the quantity 
$K^{-1}(\bra{\phi},\bra{\psi})$.}
\end{figure} 

The map $K_1:\CS_j\to\CS_j^*$ has an inverse, the map $K_1^{-1}
:\CS_j^*\to\CS_j$, that takes bras into kets.  We associate $K_1^{-1}$
with a closely related map $K^{-1}:\CS_j^*\otimes\CS_j^*\to\Complexes$
that is expressed by the spin network in Fig.~\ref{Kinv}.  This spin
network and the meaning of the $-1$ on $K^{-1}$ are to be defined, but
the spin network represents a linear operator $:\CS_j^
*\otimes\CS_j^*\to\Complexes$ with the ordering of the two operands
being specified by the stub.  The action of $K^{-1}$ on two bras is
illustrated in Fig.~\ref{Kinvsandwich}.

\begin{figure}[htb]
\begin{center}
\scalebox{0.43}{\includegraphics{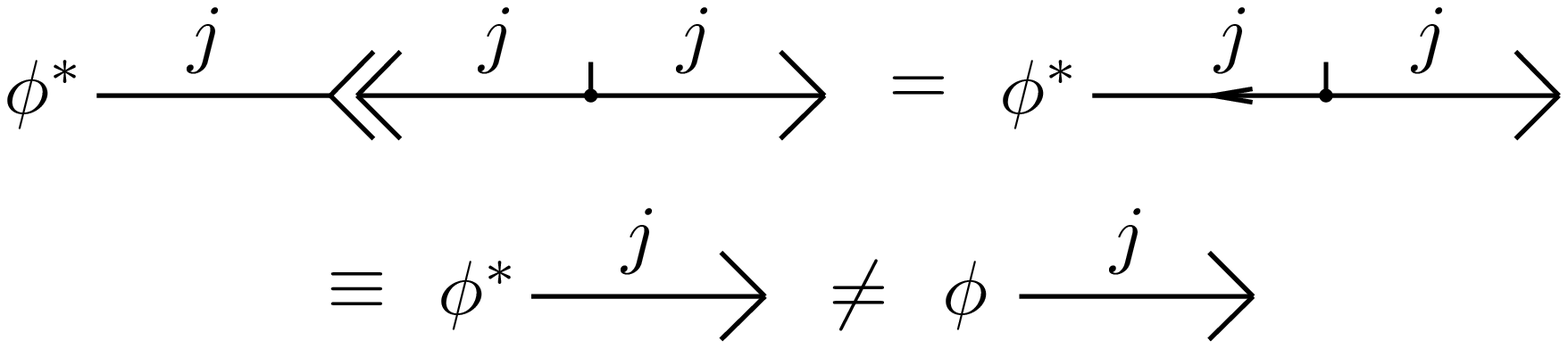}}
\end{center}
\caption[rps]{\label{K1invaction} Action of $K_1^{-1}$ on a bra.}
\end{figure}

We now define the network in Fig.~\ref{Kinv} by requiring that
$K_1^{-1}$ act on a bra by inserting it into the first operand of that
network, as illustrated in Fig.~\ref{K1invaction}, and by requiring
that $K_1^{-1}$ actually be the inverse of $K_1$.
Figure~\ref{K1invaction} shows the action of $K_1^{-1}$ on an
arbitrary bra $\bra{\phi}$.  The use of the first operand of $K^{-1}$
for this purpose is a convention, but one that makes our overall
notation for mapping kets to bras and vice versa consistent (see
Fig.~\ref{Kconsist} below).  As indicated, we abbreviate the result by
taking the original network for the bra $\bra{\phi}$ and simply
flipping the direction of the chevron.  We do not unstar the
identifying symbol.  As indicated, the result differs from Hermitian
conjugation applied to $\bra{\phi}$, which is the ket $\ket{\phi}$
(without the star).

\begin{figure}[htb]
\begin{center}
\scalebox{0.43}{\includegraphics{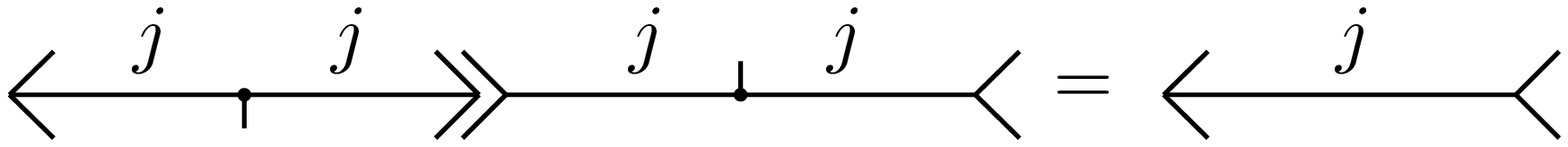}}
\end{center}
\caption[rps]{\label{K1invreq} The requirement that $K^{-1} \circ K 
=\Id$.}
\end{figure}

The requirement that $K_1^{-1}$ actually be the inverse of $K_1$ is
illustrated in Fig.~\ref{K1invreq}. The stub on the network for $K^{-1}$
is inverted so that the output of the first step is fed into the first
operand of $K^{-1}$.  

\begin{figure}[htb]
\begin{center}
\scalebox{0.43}{\includegraphics{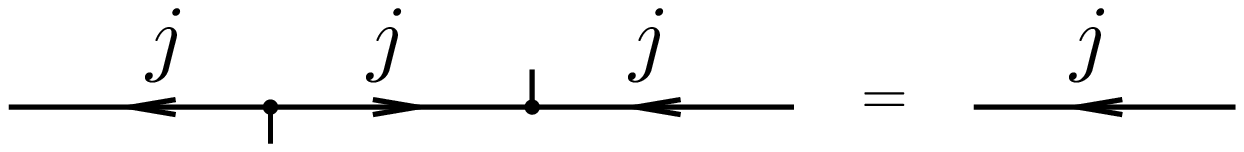}}
\end{center}
\caption[rps]{\label{2stubs} Two arrow flips, with stubs oppositely
oriented, annihilate each other.}
\end{figure}

The identity represented by Fig.~\ref{K1invreq} is usually encountered
in practice in the form shown in Fig.~\ref{2stubs} (a fragment of a
spin network).  The $2j$-node inverts the direction of the arrow.  Two
such inversions, with stubs pointing in opposite directions,
annihilate one another.

\begin{figure}[htb]
\begin{center}
\scalebox{0.43}{\includegraphics{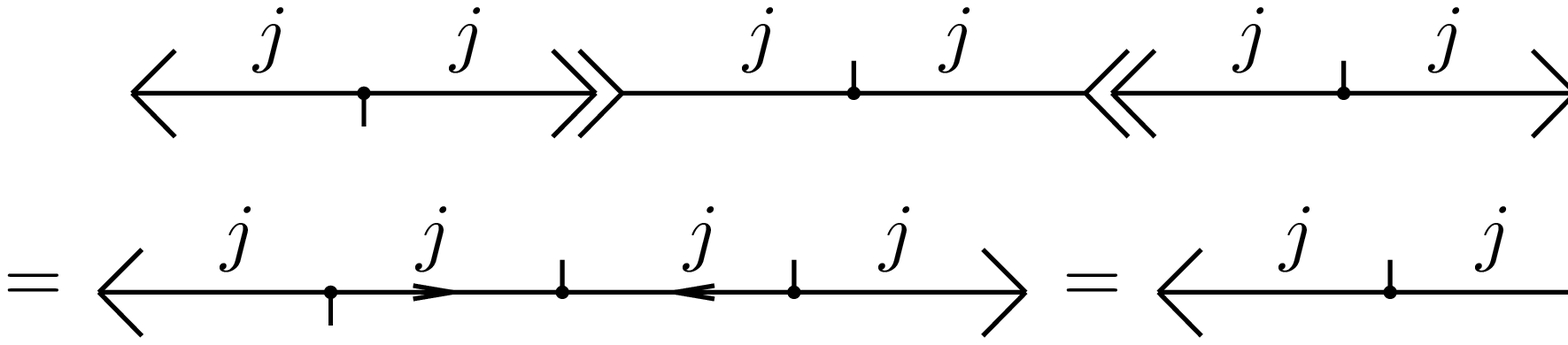}}
\end{center}
\caption[rps]{\label{Kconsist} Flipping the chevrons on $K$ gives 
$K^{-1}$.}
\end{figure}  

We have made an independent definition of the spin network in
Fig.~\ref{Kinv}, but it is the same as the spin network for $K$, shown
in Fig.~\ref{2jKdef}, with both bra chevrons flipped.  Since we now
have a convention for flipping bra chevrons (by applying $K_1^{-1}$),
for consistency we must show that the two results are the same.  This
is done in Fig.~\ref{Kconsist}, which uses the identity of
Fig.~\ref{2stubs}.

\begin{figure}[htb]
\begin{center}
\scalebox{0.43}{\includegraphics{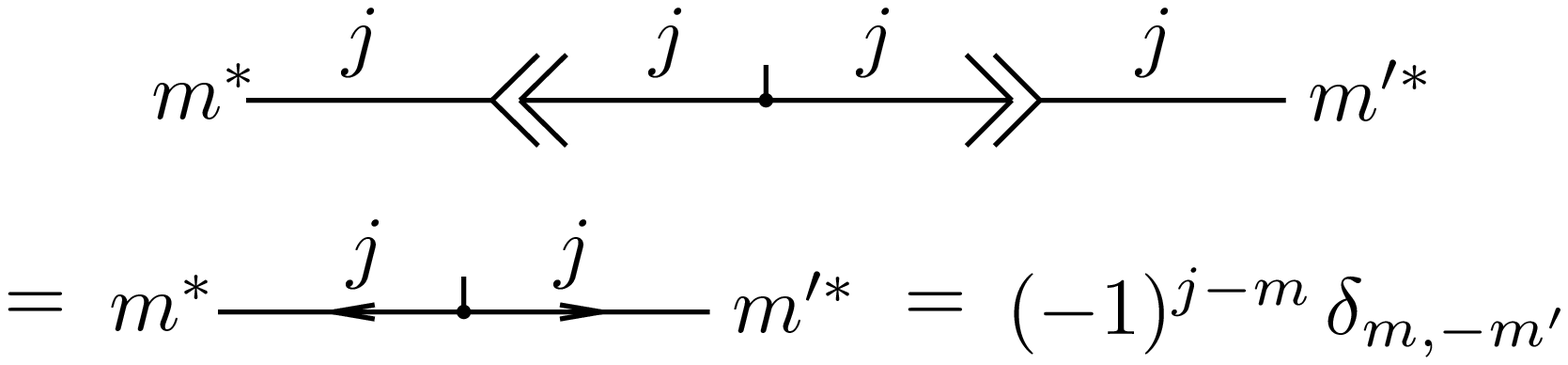}}
\end{center}
\caption[rps]{\label{Kinvcomponents} The components of $K^{-1}$ have
the same numerical values as the components of $K$.}
\end{figure}

By inserting resolutions of the identity into the diagram in
Fig.~\ref{K1invreq} it is easy to work out the components of
$K^{-1}$.  These are displayed in Fig.~\ref{Kinvcomponents}.  Notice
that they have the same numerical values as the components of $K$ (see
Fig.~\ref{2jnetwork}).

\begin{figure}[htb]
\begin{center}
\scalebox{0.43}{\includegraphics{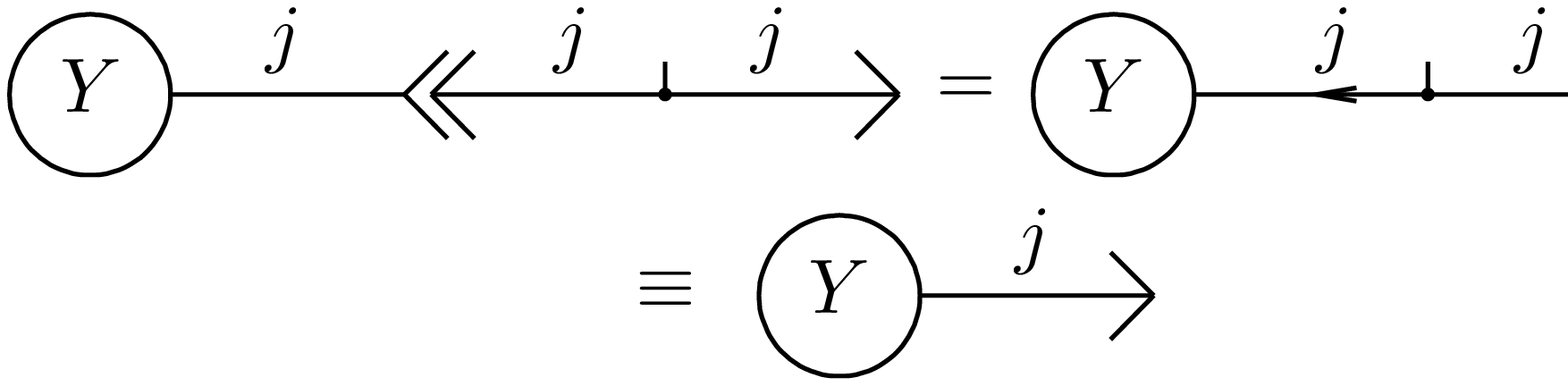}}
\end{center}
\caption[rps]{\label{raiseindex} Converting a bra chevron to a ket
chevron on an arbitrary spin network.}
\end{figure}

By using $K^{-1}$ we can convert a bra chevron into a ket chevron on
any spin network, not only on bras and kets themselves.  This is
illustrated in Fig.~\ref{raiseindex}, which may be compared to
Fig.~\ref{lowerindex}. 

\begin{figure}[htb]
\begin{center}
\scalebox{0.43}{\includegraphics{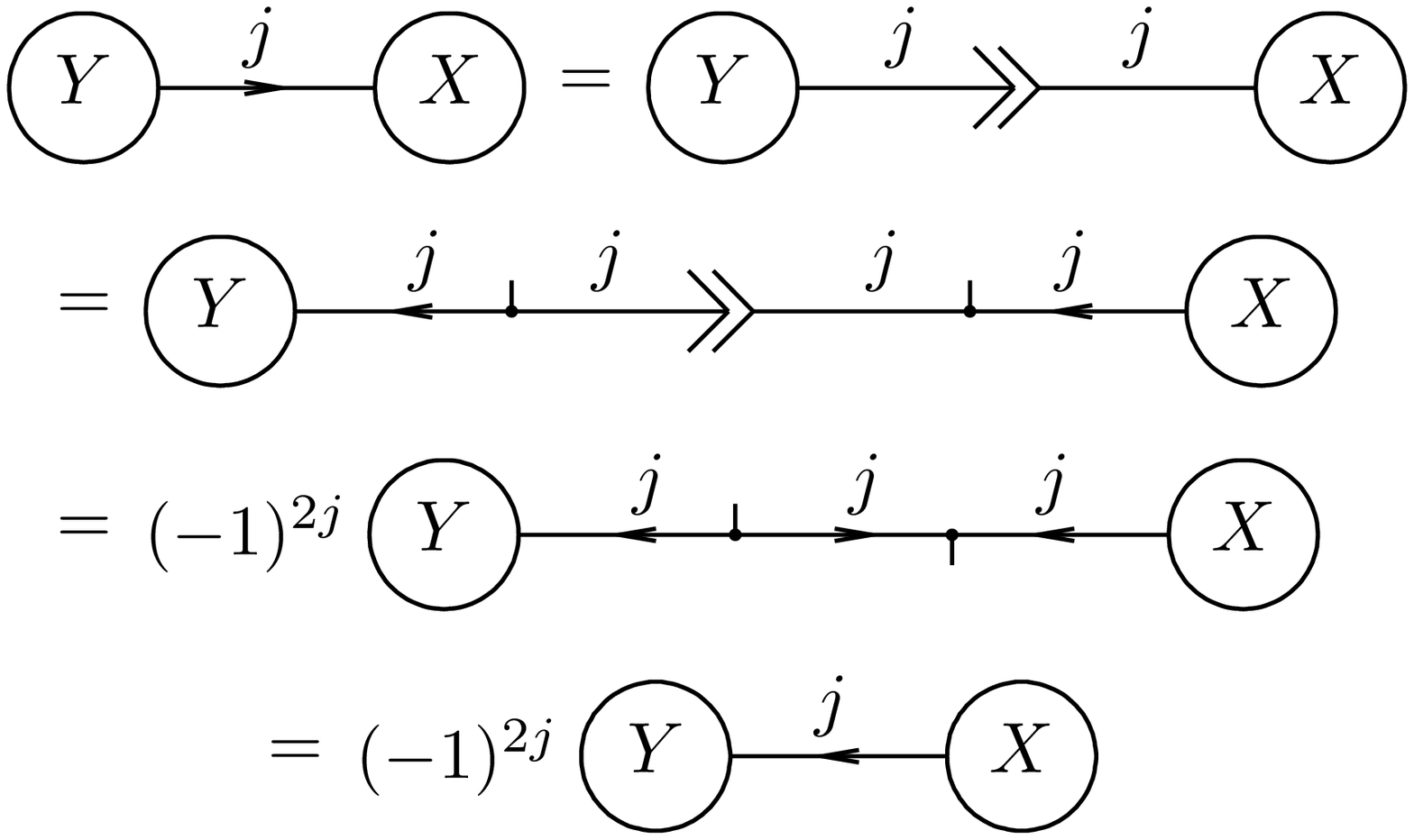}}
\end{center}
\caption[rps]{\label{reversearrow} Reversing the arrow on an edge of a
spin network incurs a phase of $(-1)^{2j}$.}
\end{figure}

Finally, by using Figs.~\ref{lowerindex}, \ref{raiseindex} and
\ref{2stubs}, it may be shown that when we reverse the arrow on an
edge of a spin network, we incur a phase of $(-1)^{2j}$.  This is done
in Fig.~\ref{reversearrow}.

\subsection{Raising and lowering indices}
\label{raisingandlowering}

When we convert a ket to a bra by the action of $K_1$, then the bra has
components with respect to the standard basis $\ket{jm}$ that are
simple functions of the components of the original ket with respect to
the standard basis $\bra{jm}$.  Mapping the one set of components to the
other is ``lowering the index.''  Using $K_1^{-1}$ to convert a bra to a
ket similarly amounts to ``raising the index.''  More generally the
procedure can be applied to an edge of any spin network terminating in
a starred (contravariant) or an unstarred (contravariant) index.  The
index can refer to any basis, not just the standard one.  

\begin{figure}[htb]
\begin{center}
\scalebox{0.43}{\includegraphics{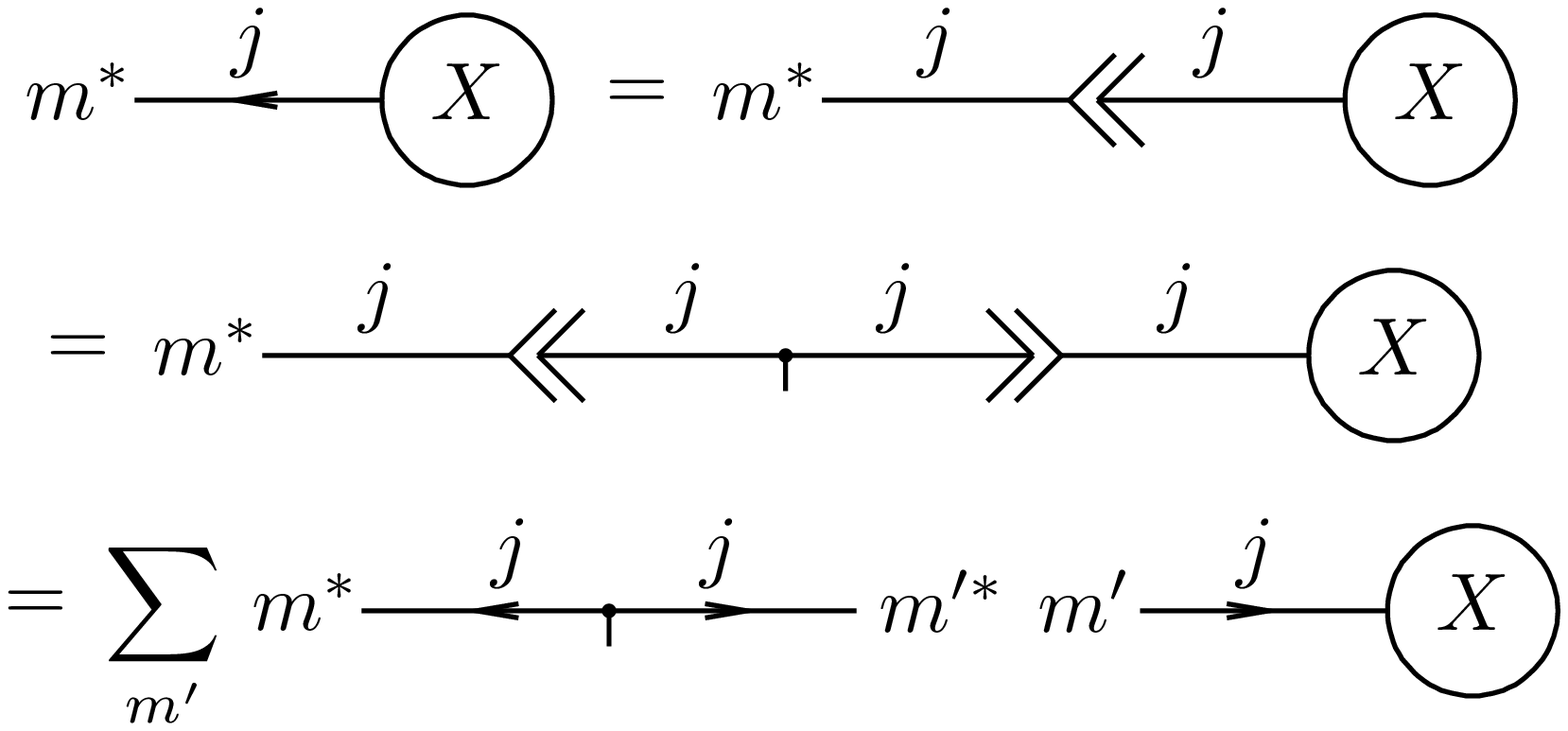}}
\end{center}
\caption[rps]{\label{raisem} Expressing contravariant components in
terms of covariant components.}
\end{figure}

\begin{figure}[htb]
\begin{center}
\scalebox{0.43}{\includegraphics{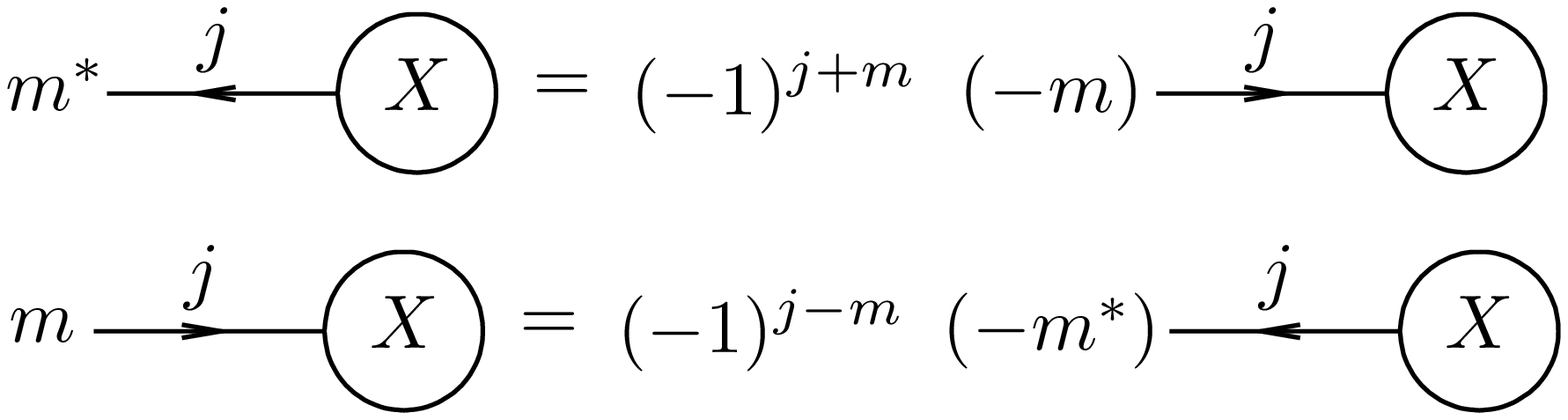}}
\end{center}
\caption[rps]{\label{raiseandlower} Rules for raising and lowering
indices.}
\end{figure}

Figure~\ref{raisem} shows how to the express contravariant components
in terms of the covariant components in the standard basis.  By
plugging in the numerical values of the components of $K$, we obtain
the first line of Fig.~\ref{raiseandlower}.  Similarly we derive the
second line of Fig.~\ref{raiseandlower} for expressing covariant
components in terms of contravariant components.

\begin{figure}[htb]
\begin{center}
\scalebox{0.43}{\includegraphics{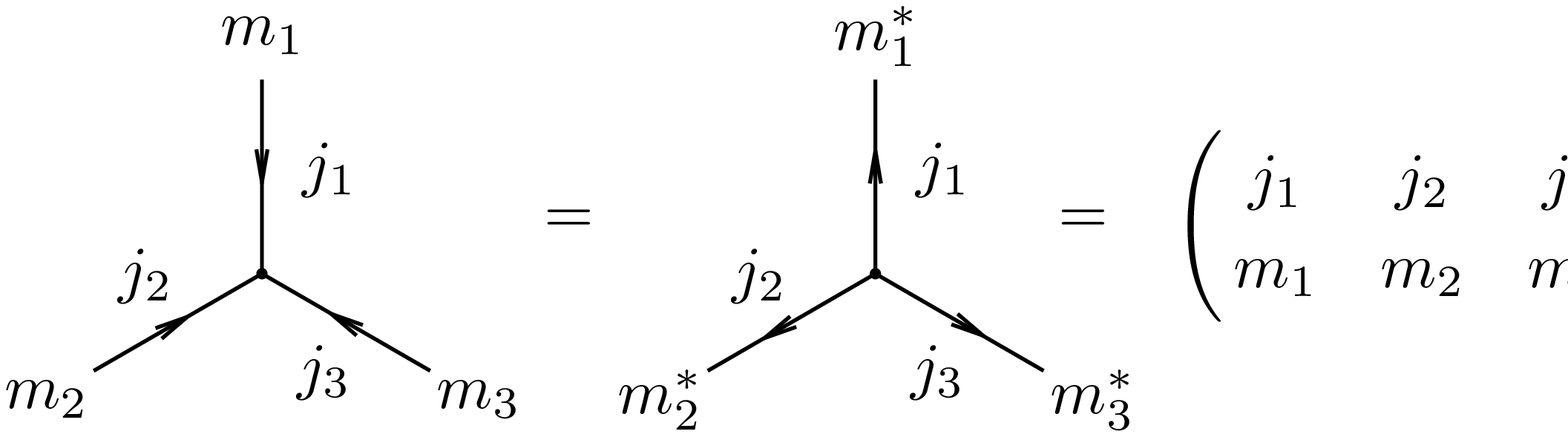}}
\end{center}
\caption[rps]{\label{3jversions} The completely covariant and
completely contravariant components of the $3j$-intertwiner are
numerically equal.}
\end{figure}

\begin{figure}[htb]
\begin{center}
\scalebox{0.43}{\includegraphics{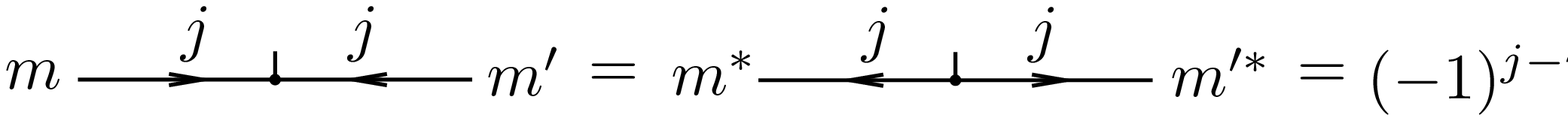}}
\end{center}
\caption[rps]{\label{2jversions} The completely covariant and
completely contravariant components of the $2j$-intertwiner are
numerically equal.}
\end{figure}

If these rules are used to raise all three covariant components of the
$3j$-intertwiner (Fig.~\ref{3jnetwork}), then we find that the
completely contravariant components have the same values, namely, the
$3j$-symbol.  This is illustrated in Fig.~\ref{3jversions}.  To show
this it is necessary to use the symmetry of the $3j$-symbol (see
Varshalovich \etal\ 1981, Eq.~(8.2.4.6)).  Then by setting one of the $j$'s
to zero and using Fig.~\ref{vestigialstub}, we find that the same is
true for the completely covariant and completely contravariant
components of the $2j$-intertwiner, as shown in Fig.~\ref{2jversions}.
The same result is obtained by comparing Figs.~\ref{2jnetwork} and
\ref{Kinvcomponents}.

\subsection{Hermitian conjugation of spin networks}
\label{Hermitianconjugation}

Consider a spin network of arbitrary complexity involving only $2j$-
and $3j$-nodes.  The network is allowed to have any number of edges
terminating in bra or ket chevrons, or in starred or unstarred labels
such as $m$ indices.  By using the identities above, possibly with the
insertion or removal of $2j$-nodes and the extraction of phases of the
form $(-1)^{2j}$, it is possible to bring the spin network into a
standard form, in which all edges joining $3j$-nodes have arrows
pointing toward the $3j$-node, all edges joining $2j$-nodes have
arrows pointing away from the $2j$-node, all edges terminating in a
starred symbol have arrows pointing toward that symbol, and all edges
terminating in an unstarred symbol have arrows pointing away from that
symbol.  Next, by inserting resolutions of the identity, which involve
$m$-sums, it is possible to express the spin network as sum over the
completely covariant components of $3j$-symbols and completely
contravariant components of $2j$-symbols, times a tensor product of
bras and kets.

In this form it is easy to take the Hermitian conjugate.  Under
Hermitian conjugation, bras go to kets and vice versa, while the
covariant components of $3j$-symbols and contravariant components of
$2j$-symbols do not change, since they are real.  By using
Figs.~\ref{3jversions} and \ref{2jversions}, however, these components
can be rewritten as the completely contravariant components of
$3j$-symbols and the completely covariant components of $2j$-symbols.
The $m$-sums can now be done, reversing the earlier insertions of
resolutions of the identity.  Then the other steps leading to the
standard form can be reversed.

The result is a simple rule for the Hermitian conjugation of any spin
network of the given form:  All ket chevrons are changed to bra
chevrons and vice versa, the directions of all arrows are reversed, and
all edges terminating in a symbol have a star added to the symbol,
with a double star being removed.  

\subsection{Discussion of spin network rules}
\label{spinnetworkdiscussion}

Our rules for spin networks differ from those of Yutsis \etal\ and
most of the literature in the Yutsis tradition primarily by our
ability to express abstract vectors (kets), dual vectors (bras) and
tensors in addition to the components of those objects.  Also, we
indicate the nature of an $m$ index (covariant or contravariant) by
the presence or absence of a star, rather than the direction of the
arrow.  One result is that our rules for reversing the direction of
the arrow are more uniform than in the Yutsis tradition, where such a
reversal picks up a phase $(-1)^{2j}$ only on internal edges.  In our
approach, the rule applies everywhere, including edges terminating in
an $m$ index.  In addition, our rules for Hermitian conjugation are
simpler than those in the Yutsis tradition, where phase factors must
be introduced.  The simplification is due to the explicit introduction
of $2j$-symbols, and the use of stubs. 

To translate a Yutsis spin network into one of ours, it is necessary
only to put stars on $m$ indices terminating edges with outward
pointing arrows.   

The standard form of a spin network discussed in
Sec.~\ref{Hermitianconjugation} has all $3j$-nodes with inward
pointing arrows, and all $2j$-nodes with outward pointing ones.  If we
assume the standard form, then the arrows become superfluous and can
be dropped.  Only the stubs and $2j$-nodes remain, in comparison to a
Yutsis-style spin network.  This is the procedure advocated by Stedman
(1990).  For the purposes of this paper we will keep the arrows, since
we wish to have finer control on the spin network than that offered by
the standard form.

\section{Models for the $6j$-symbol}
\label{models}

For given values of the six $j$'s, the $6j$-symbol is just a number,
but to study its semiclassical limit it is useful to write it as a
scalar product $\braket{B}{A}$ of wave functions in some Hilbert
space.  This can be done in many different ways, corresponding to what
we call different ``models'' of the $6j$-symbol.  In this section we
describe a class of such models that are related to one another.  We
begin by summarizing our notation for the Schwinger representation of
angular momentum operators.  Then we present what we call the
``$12j$-model,'' which was used by Roberts (1999) in his derivation of
the Ponzano-Regge formula.  Next we describe the ``$4j$-model'' which
we will use for the semiclassical analysis of this paper.  We also
mention an $8j$-model for the $6j$-symbol.

\subsection{The Schwinger Representation}
\label{schwingerrep}

Our notation for the Schwinger representation of angular momentum
operators is similar to that used in I.  We denote the Schwinger
Hilbert space by $\SS=L^2(\Reals^2)$; it is the space of wave
functions $\psi(x_1,x_2)$ for two harmonic oscillators of unit
frequency and mass.  The usual annihilation and creation operators are
$\ahat_\mu=(\xhat_\mu + i\phat_\mu) /\sqrt{2}$, $\ahat_\mu^\dagger=
(\xhat_\mu-i\phat_\mu) /\sqrt{2}$, for $\mu=1,2$; we use hats on
operators to distinguish them from their classical counterparts.  We
define operators
	\begin{equation}
	\Ihat = \frac{1}{2}\ahat^\dagger \ahat, 
	\qquad
	\Jhat_i = \frac{1}{2}\ahat^\dagger \sigma_i \ahat,
	\label{IhatJhatdef}
	\end{equation}
where $i=1,2,3$ and where $\ahat$ (without the $\mu$ index) is a
2-vector (or column spinor) of operators, with $\ahat^\dagger$ the
adjoint (or row spinor) and with obvious contractions against the
Pauli matrices $\sigma_i$.  These operators satisfy the commutation
relations $[\Ihat,\Jhat_i]=0$, $[\Jhat_i,\Jhat_j] = i\epsilon_{ijk} \,
\Jhat_k$.  We set $\hbar=1$.  Note that $\Ihat =
(\Hhat_1+\Hhat_2-1)/2$, where $\Hhat_\mu$, $\mu=1,2$, are the two
harmonic oscillators.  There is also the operator relation $\Jvechat^2
= \Ihat(\Ihat+1)$.  We denote the squares of 3-vectors in bold face.
The operators $\Jvechat$ generate an $SU(2)$ action on $\SS$, which
carries one copy of each irrep $j=0,1/2,1,\ldots$, that is,
	\begin{equation}
	\SS = \sum_j \oplus \,\CS_j.
	\label{SSdirectsum}
	\end{equation}
The irreducible subspace $\CS_j$ is an eigenspace of $\Ihat$ with
eigenvalue $j$.  In the semiclassical analysis of spin networks, the
spaces $\CS_j$ that the spin networks refer to are interpreted as one
of the irreducible subspaces of a Schwinger Hilbert space $\SS$.
Similarly, $\CS_j^*$ is interpreted as a subspace of a space $\SS^*$.
In this way the bra and ket vectors referred to by the spin network
are interpreted as wave functions on $\Reals^2$, and the spin network
itself can be interpreted as a wave function in $\Reals^{2N}$.

In the various $nj$-models we take tensor products of the Schwinger
Hilbert space, writing $\SS_r$ for the $r$-th copy.  Similarly, we put
an $r$ index on various operators, for example, $\ahat_r$, $\Ihat_r$,
$\Jvechat_r$, $r=1,\ldots,n$, or, with two indices, $\ahat_{r\mu}$,
$\mu=1,2$.

\subsection{The $12j$-model of the $6j$-symbol}
\label{12jmodel}

\begin{figure}[htb]
\begin{center}
\scalebox{0.5}{\includegraphics{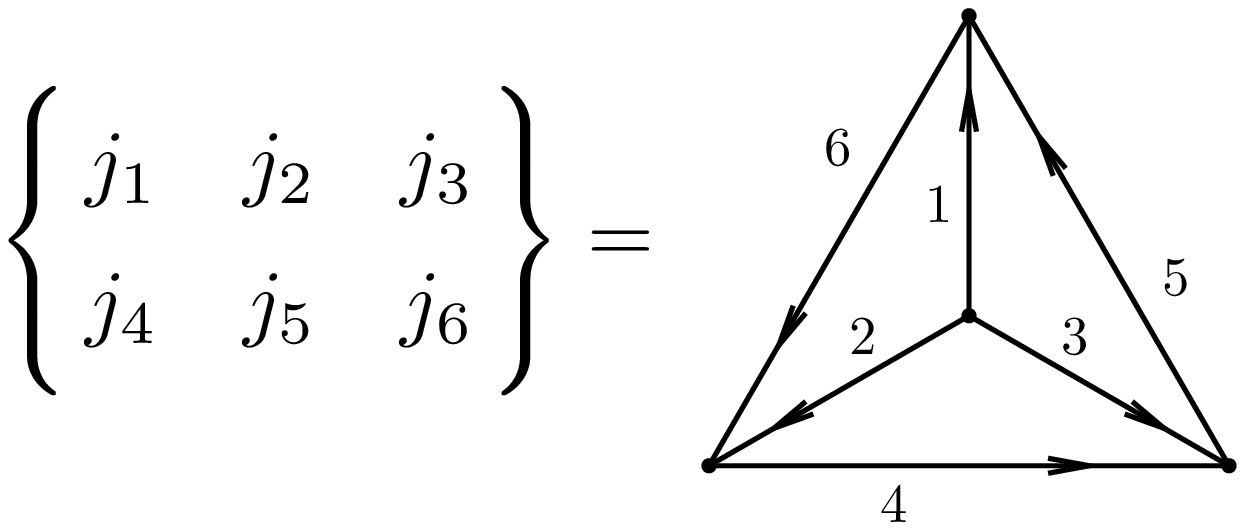}}
\end{center}
\caption[symm6j]{\label{symm6j} The Yutsis spin network for the $6j$-symbol,
  in a symmetrical labeling of the $j$'s.  The numbers $1$, $2$, etc
  on the spin network refer to $j_1$, $j_2$, etc.}
\end{figure}

We begin with the standard spin network (Yutsis \etal 1962) of the
$6j$-symbol, shown in Fig.~\ref{symm6j}.  According to the remarks in
Sec.~\ref{spinnetworkdiscussion}, this spin network can be
reinterpreted according to our conventions, presented in
Sec.~\ref{spinnetworknotation}, without modification.  We will refer to the
labeling of the six $j$'s in the $6j$-symbol shown in
Fig.~\ref{symm6j} as the ``symmetric'' labeling.

\begin{figure}[htb]
\begin{center}
\scalebox{0.5}{\includegraphics{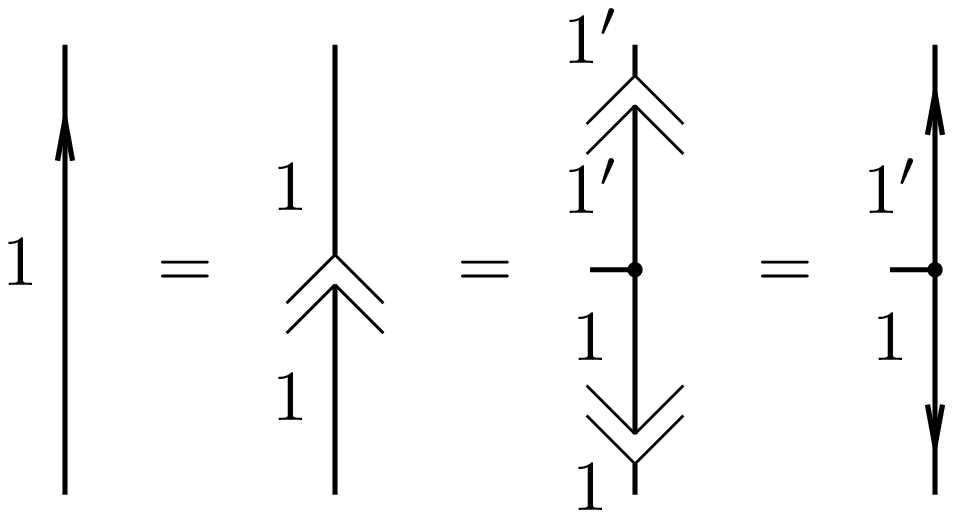}}
\end{center}
\caption[2jinsert]{\label{2jinsert} By inserting $2j$-symbols in each 
  edge of a Yutsis diagram, all $3j$-symbols can be put into standard
  form (purely contravariant).}
\end{figure}

We perform an operation on each edge of the spin network that is
illustrated for edge 1 in Fig.~\ref{2jinsert}.  After the second
equality edges are labeled by $1$ and $1'$.  These refer to two
distinct carrier spaces, say, $\CS_{j_1}$ and $\CS'_{j_1}$, with the
same value of $j$ (that is, $j_1$).  The introduction of such distinct
carrier spaces does not change the value of the $6j$-symbol, which is
just a number.  In the second equality we have expressed the lower ket
chevron as a bra chevron transformed by $K_1^{-1}$, as in
Fig.~\ref{raiseindex}.  In the final diagram the arrows are directed
toward both $3j$-nodes connected by the original edge, and a $2j$-node
has been inserted.  We do this on all six edges of the spin network in
Fig.~\ref{symm6j}.  The resulting diagram is somewhat busy so we do
not attempt to draw it, but each edge of the original diagram now
looks like the final diagram in Fig.~\ref{2jinsert}.

\begin{figure}[htb]
\begin{center}
\scalebox{0.5}{\includegraphics{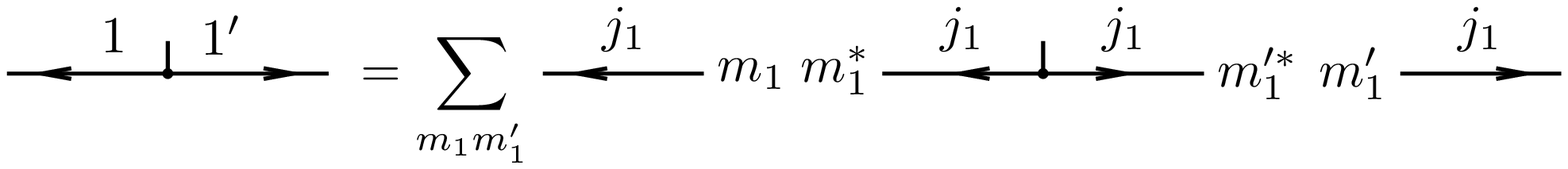}}
\end{center}
\caption{\label{edmonds} Breaking lines to represent $6j$-symbol as a
  sum over $2j$- and $3j$-symbols.}
\end{figure}

We now break up the final diagram in Fig.~\ref{2jinsert} in two
different ways.  The first way is illustrated in Fig.~\ref{edmonds},
in which the primed and unprimed lines are broken into summations over
primed and unprimed quantum numbers $m$.  This is done for all six
edges of the original spin network.  The $6j$-symbol is represented as a
product of six copies of a $2j$-symbol and four of a $3j$-symbol.
Using the definition (\ref{2jdef}) for the $2j$-symbol, the result is
    \begin{equation}
    \fl\eqalign{
    &\left\{\begin{array}{ccc}
        j_1 & j_2 & j_3 \\
        j_4 & j_5 & j_6
    \end{array}\right\} =
    \left(\prod_{r=1}^6 \sqrt{2j_r+1}\right)
    \sum_{\hbox{all $m$'s}} \\
    &\quad\times
    \left(\begin{array}{ccc}
        j_1 & j_2 & j_3 \\
        m_1 & m_2 & m_3
    \end{array}\right)
    \left(\begin{array}{ccc}
        j_1 & j_5 & j_6 \\
        m'_1 & m'_5 & m_6
    \end{array}\right)
    \left(\begin{array}{ccc}
        j_2 & j_6 & j_4 \\
        m'_2 & m'_6 & m_4
    \end{array}\right)
    \left(\begin{array}{ccc}
        j_3 & j_4 & j_5 \\
        m'_3 & m'_4 & m_5
    \end{array}\right) \\
    &\quad\times
    \left(\begin{array}{cc}
        j_1 & j_1 \\
        m_1 & m'_1
    \end{array}\right) \cdots
    \left(\begin{array}{cc}
        j_6 & j_6 \\
        m_6 & m'_6
    \end{array}\right)}
    \label{edmonds6j}
    \end{equation}
This formula may be compared to Edmonds (1960), eq.~(6.2.3).  Edmonds
uses what he calls a ``metric tensor'' (really the components of $K$
or $K^{-1}$, multiplied by $(-1)^{2j}$, see his eq.~(3.7.1)), which
relative to our $2j$-symbol introduces an overall phase of
$\prod_{r=1}^6 (-1)^{2j_r}$.  He also swaps $m_r$ and $m'_r$ for
$r=4,5,6$ relative to our definitions, which introduces further
phases.  The product of these phases is 1, showing that the formulas
agree.

\begin{figure}[htb]
\begin{center}
\scalebox{0.5}{\includegraphics{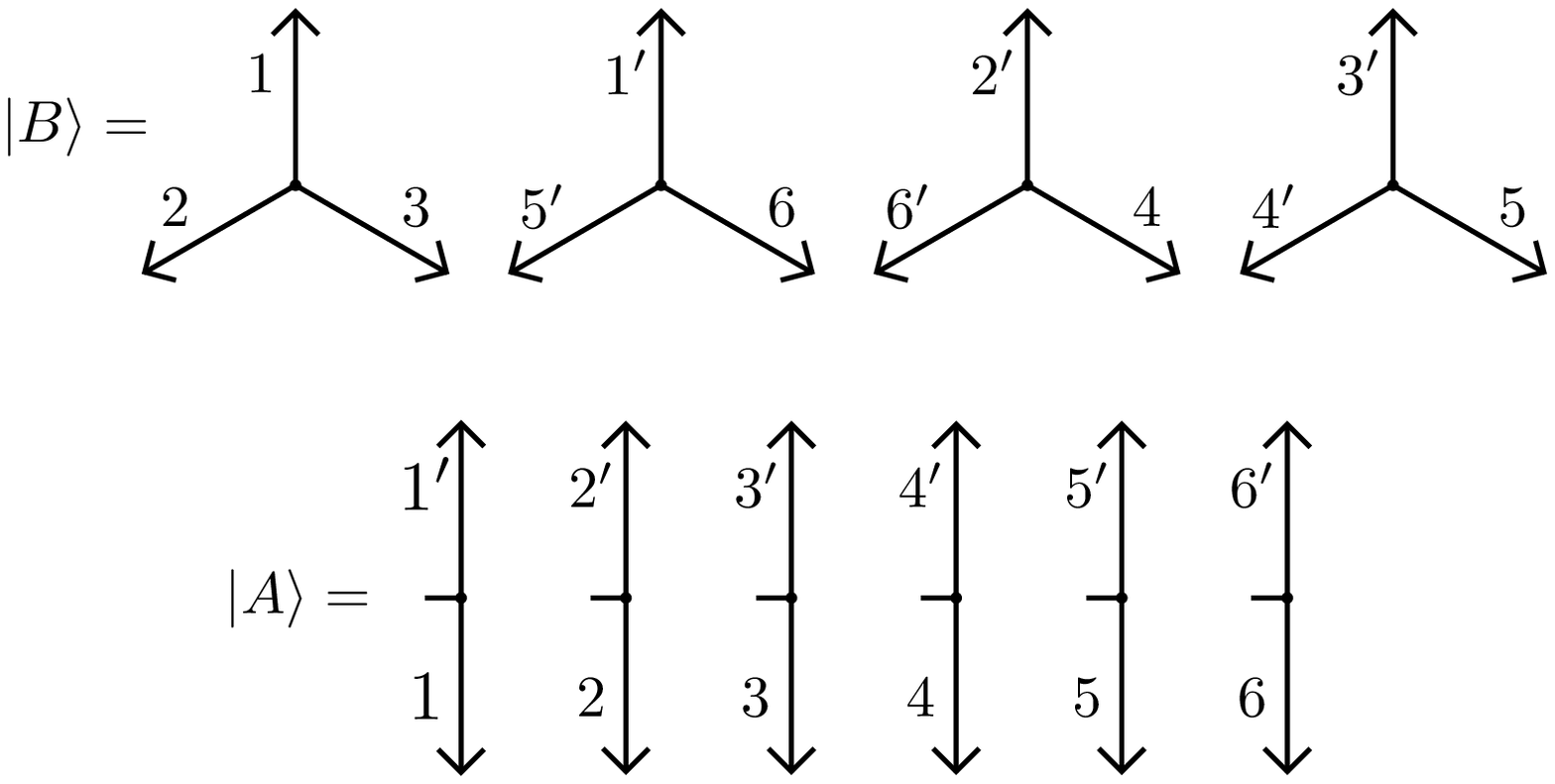}}
\end{center}
\caption{\label{robertsAB} The two states that appear in Roberts'
  (1999) expression for the $6j$-symbol.}
\end{figure}

Another way of breaking up the $6j$-symbol is to stop with the third
diagram of Fig.~\ref{2jinsert} in the transformation of the six edges
of the $6j$-symbol.  Again the resulting diagram is too busy to draw,
but it can be regarded as the complete contraction of two tensors, one
the tensor product of six $2j$-intertwiners, all terminating in ket
chevrons, the other the tensor product of four $3j$-intertwiners, all
terminating in bra chevrons.  There are twelve ket chevrons and twelve
bra chevrons altogether, which we think of as living in twelve carrier
spaces $\CS_{j_r}$ and $\CS'_{j_r}$ and their duals, where
$r=1,\ldots,6$.  These can be viewed as subspaces of twelve Schwinger
Hilbert spaces, $\SS_r$, $\SS'_r$, $r=1,\ldots,6$, and their duals.
Then the $6j$-symbol takes on the form $\braket{B}{A}$, where states
$\ket{A}$, $\ket{B}$ belong to the total Hilbert space $\HS_{12j} =
(\prod_r\otimes \SS_r)\otimes (\prod_r\otimes\SS'_r)$.  These states are
illustrated in Fig.~\ref{robertsAB}, where the state $\ket{B}$ has
been turned into ket form by Hermitian conjugation.

The usual custom in physics is to specify a state by the operators and
quantum numbers of which the state is a simultaneous eigenstate.  This
requires that the eigenstate be nondegenerate, so that it is
determined to within a normalization and phase.  This in turn
requires, in a certain sense, that the number of independent operators
should be equal to the number of degrees of freedom of the system.  We
will not attempt to be precise about this statement, but will
illustrate the principle in several examples.
  
\begin{figure}[htb]
\begin{center}
\scalebox{0.5}{\includegraphics{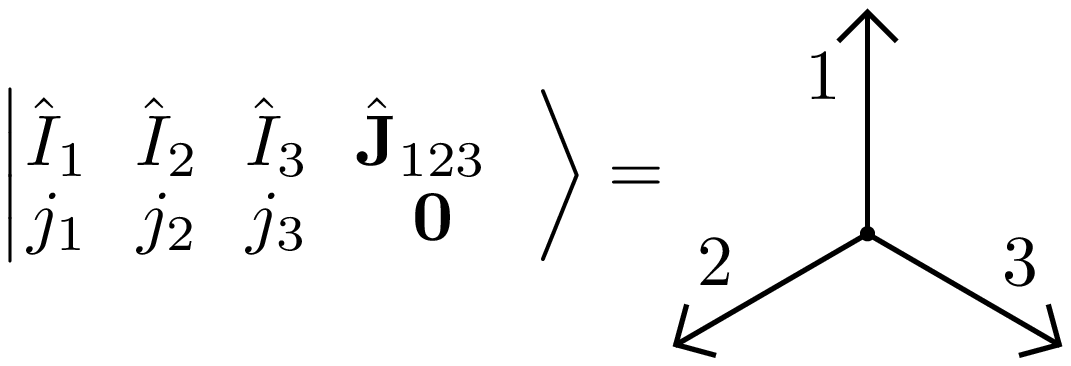}}
\end{center}
\caption{\label{3jket} Ket notation for invariant state of the $3j$-type.}
\end{figure}

One example is the first $3j$-state appearing in Fig.~\ref{robertsAB},
which lies in the Hilbert space $\SS_1 \otimes \SS_2 \otimes \SS_3$
and is a simultaneous eigenstate of $\Ihat_r$, $r=1,2,3$ with
eigenvalues $j_r$, $r=1,2,3$.  It is also an eigenstate of the vector
of operators
	\begin{equation}
	\Jvechat_{123} = \Jvechat_1 + \Jvechat_2 + \Jvechat_3,
	\label{Jvechat123def}
	\end{equation}
the total angular momentum on this Hilbert space, with eigenvalue
$\zerovec$.  That this simultaneous eigenstate is nondegenerate
follows from standard angular momentum theory; and the number of
operators (six) equals the number of degrees of freedom in the Hilbert
space (two for each $\SS_r$, $r=1,2,3$).  Thus we write this state as
illustrated in Fig.~\ref{3jket}, indicating both operators and
eigenvalues.  This is otherwise the state $\ket{W}$, illustrated in
bra form in Fig.~\ref{Wintertwiner}.  As for the normalization and
phase, these must be supplied by context.  For the $3j$-state
illustrated in Fig.~\ref{3jket}, these are given in terms of the
$3j$-symbol by Fig.~\ref{3jnetwork}; in particular, the state is
normalized.

\begin{figure}[htb]
\begin{center}
\scalebox{0.5}{\includegraphics{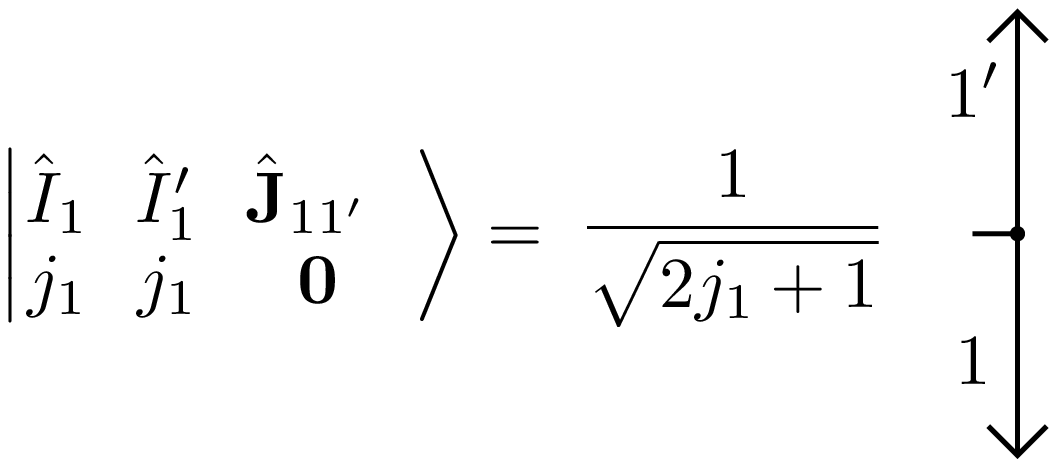}}
\end{center}
\caption{\label{2jket} Ket notation for invariant state of the $2j$-type.}
\end{figure}

Similarly, the first $2j$-state in Fig.~\ref{robertsAB} lies in the
Hilbert space $\SS_1\otimes\SS'_1$ and is a simultaneous eigenstate
of the operators $\Ihat_1$ and $\Ihat'_1$ with eigenvalues $j_1$ and
$j_1$, as well as of the total angular momentum operator on this
space,
	\begin{equation}
	\Jvechat_{11'} = \Jvechat_1 + \Jvechat'_1,
	\label{Jvechat11'}
	\end{equation}
with eigenvalue $\zerovec$.  This state is a simultaneous eigenstate
of five operators, but in a sense only two of the three components of
$\Jvechat_{11'}$ are independent, so we should count only four
independent operators, which agrees with the number of degrees of
freedom in the Hilbert space (again, two each for $\SS_1$ and
$\SS'_1$).  We write this state as illustrated in Fig.~\ref{2jket};
the normalization and phase are given by the components of $K^{-1}$
shown in Fig.~\ref{Kinvcomponents}.  In particular, with the square
root factor in Fig.~\ref{2jket}, this state is normalized.

In this notation we can write the equations of Fig.~\ref{robertsAB} in
the form
     \begin{equation}
       \fl\ket{B} = \Wket{\Ihat_1}{\Ihat_2}{\Ihat_3}{\Jvechat_{123}}
                      {j_1}{j_2}{j_3}{\zerovec}
                 \Wket{\Ihat'_1}{\Ihat'_5}{\Ihat_6}{\Jvechat_{1'5'6}}
                      {j_1}{j_5}{j_6}{\zerovec}
                 \Wket{\Ihat'_2}{\Ihat'_6}{\Ihat_4}{\Jvechat_{2'6'4}}
                      {j_2}{j_5}{j_6}{\zerovec}
                 \Wket{\Ihat'_3}{\Ihat'_4}{\Ihat_5}{\Jvechat_{3'4'5}}
                      {j_3}{j_4}{j_5}{\zerovec},
     \label{roberts3jstate}
     \end{equation}
and
     \begin{equation}
       \fl\ket{A} = \left(\prod_{r=1}^6 \sqrt{2j_r+1}\right)
                 \Kket{\Ihat_1}{\Ihat'_1}{\Jvechat_{11'}}
                      {j_1}{j_1}{\zerovec}
                 \cdots
                 \Kket{\Ihat_6}{\Ihat'_6}{\Jvechat_{66'}}
                      {j_6}{j_6}{\zerovec}.
       \label{roberts2jstate}
       \end{equation}
One can see that Edmonds' form of the $6j$-symbol (\ref{edmonds6j}) is
equal to $\braket{B}{A}$.

This scalar product was the starting point for Roberts' (1999)
analysis of the asymptotics of the $6j$-symbol.  We shall comment
below on further aspects of Roberts' calculation.

\subsection{The Triangle and Polygon Inequalities}
\label{polygoninequals}

We make a remark on a generalization of the triangle inequalities
before presenting the $4j$-model of the $6j$-symbol.  If
$(\ell_1,\ell_2,\ell_3)$ are three nonnegative lengths, the usual
triangle inequalities are $|\ell_i-\ell_j| \le \ell_k \le
\ell_i+\ell_j$, where $(i,j,k)=(1,2,3)$ and cyclic permutations.  We
generalize these as follows.  Let $\{\ell_i,i=1,\ldots,n\}$ be a set
of lengths, $\ell_i\ge0$, $i=1,\ldots,n$.  Then this set is said to
satisfy the ``polygon inequality'' if
	\begin{equation}
	\max\{\ell_i\} \le \frac{1}{2}\sum_{i=1}^n \ell_i.
	\label{polygoninequal}
	\end{equation}
This is equivalent to the triangle inequalities when $n=3$.  In
general, it represents the necessary and sufficient condition that
line segments of the given, nonnegative lengths can be fitted together
to form a polygon with $n$ sides (in $\Reals^N$, $N>0$).

\subsection{The $4j$-model of the $6j$-symbol}
\label{4jmodel}

\begin{figure}[htb]
\begin{center}
\scalebox{0.5}{\includegraphics{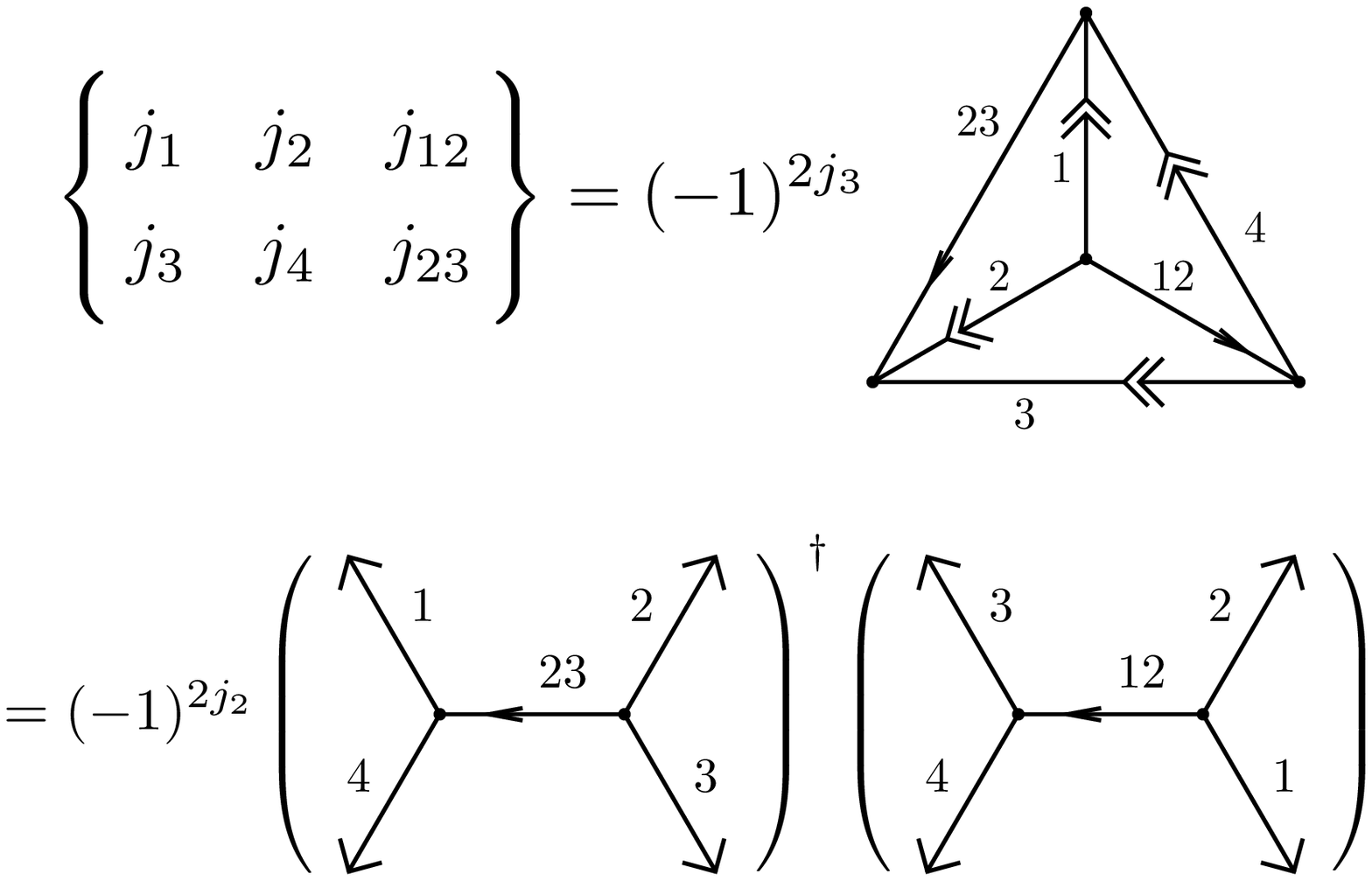}}
\end{center}
\caption[asym6j]{\label{asym6j} Asymmetric labeling of the $6j$-symbol, and
decomposition of spin network.}
\end{figure}

A different way of writing the $6j$-symbol as a scalar product begins
with Fig.~\ref{asym6j}, in which the $j$'s in the $6j$-symbol of
Fig.~\ref{symm6j} have been relabeled.  We will refer to the labeling
in Fig.~\ref{asym6j} as the ``asymmetric'' labeling, which is more
appropriate for the $4j$-model.  After the relabeling, we have
reversed the arrow on the edge $j_3$ of the spin network (labeled
simply by 3), incurring a phase $(-1)^{2j_3}$, and broken four edges
into scalar products of a bra and a ket.  Then, on the second line of
Fig.~\ref{asym6j}, we have unfolded the bra and the ket, written the
bra as the Hermitian conjugate of a ket, and adjusted phases.  The
result expresses the $6j$-symbol as a phase times a scalar product of
two states lying in the Hilbert space $\HS_{4j} =\prod_{r=1}^4 \otimes
\SS_r$.  It is understood that the ket terminating a line
labeled $r$ lies in the $j_r$-irreducible subspace of $\SS_r$.

These same states arise in the recoupling of four angular momenta with
a resultant of zero.  Let $\ZS$ be the subspace of $\CS_{j_1}
\otimes \CS_{j_2} \otimes\CS_{j_3} \otimes \CS_{j_4}$ upon which
	\begin{equation}
	\Jvechat_{\rm tot} = \sum_{r=1}^4 \Jvechat_r = 0.
	\label{Jtotdef}
	\end{equation}
This is the subspace of rotational invariants, that is, $\ZS$ is the
set of states $\ket{\psi }\in \CS_{j_1}\otimes \CS_{j_2}
\otimes\CS_{j_3} \otimes \CS_{j_4}$ that are invariant under
rotations.  According to the rules for addition of angular momenta,
subspace $\ZS$ is nontrivial ($\dim\ZS>0$) if $\sum_{r=1}^4 j_r={\rm
integer}$ and if the set $\{j_r\}$ satisfies the polygon inequality
(\ref{polygoninequal}).  In accordance with the remarks in
Sec.~\ref{intertwiners}, the subspace $\ZS$ can also be interpreted as
the space of 4-valent intertwiners, that is, $SU(2)$-invariant maps
$:\CS_{j_1}\otimes \CS_{j_2} \otimes\CS_{j_3} \otimes
\CS_{j_4}\to\Complexes$.  The notation $\ZS$ is a mnemonic for
``zero'' (the eigenvalue of $\Jvechat_{\rm tot}$).

As far as recoupling theory is concerned the spaces $\CS_{j_r}$ can be
any carrier spaces of $SU(2)$ for the given values of $j_r$, but in
our application we shall interpret the space $\CS_{j_r}$ as the
irreducible subspace $j_r$ of the $r$-th copy of the Schwinger Hilbert
space $\SS_r$.  Then $\ZS$ becomes a subspace of $\HS_{4j}$.  We shall
assume that the fixed values of the four $j_r$, $r=1,\ldots,4$ are
chosen such that $\dim\ZS>0$.  We should properly label $\ZS$ by the
four $j_r$ values since $\HS_{4j}$ contains as many subspaces of the
type $\ZS$ as there are choices of the four $j$'s.  For simplicity,
however, we will suppress this dependence in the notation, it being
understood that $j_r$, $r=1,\ldots,4$ are given.

\begin{figure}[htb]
\begin{center}
\scalebox{0.5}{\includegraphics{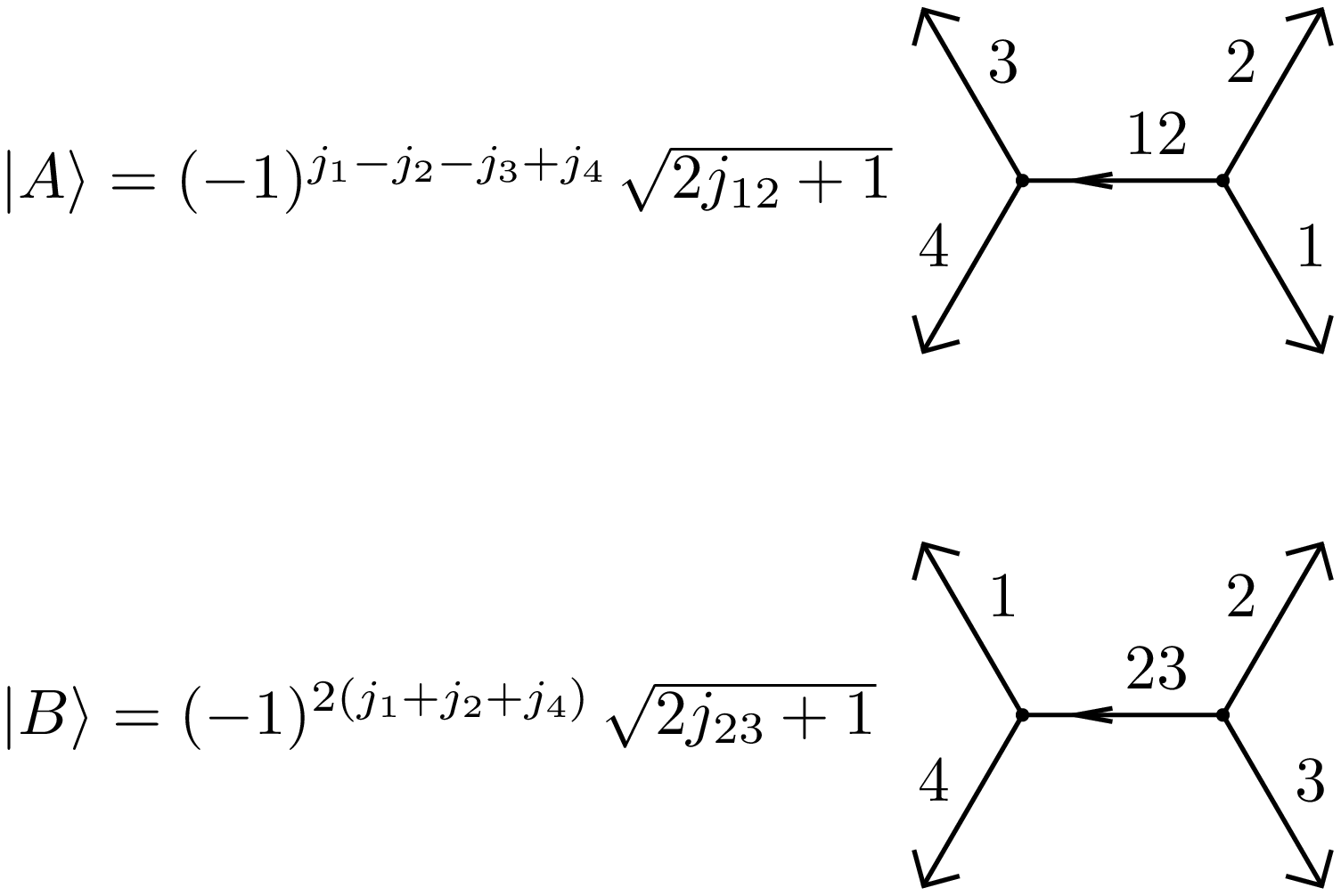}}
\end{center}
\caption[4jABstates]{\label{4jABstates} Two ways of recoupling four angular
momenta with a total of zero.}
\end{figure}

Standard recoupling theory gives three ways of constructing an
orthonormal basis in $\ZS$.  One uses Clebsch-Gordan coefficients to
couple angular momenta according to the pattern $1+2=12$, $12+3=123$,
$123+4=0$, resulting in the normalized state $\ket{B}$ lying in $\ZS$,
expressed in terms of a spin network in Fig.~\ref{4jABstates}.  A
second way couples according to the pattern $2+3=23$, $1+23=123$,
$123+4=0$, producing the normalized state $\ket{A}$ illustrated in
Fig.~\ref{4jABstates}.  A third way, which we will not consider
further, uses the intermediate coupling $1+3=13$.

The quantum numbers $j_{12}$ and $j_{23}$ of the intermediate angular
momenta range in integer steps between the bounds
	\begin{equation}
	\eqalign{
	j_{12,{\rm min}} &\le j_{12} \le j_{12,{\rm max}}, \\
	j_{23,{\rm min}} &\le j_{23} \le j_{23,{\rm max}},}
	\label{j12j23range}
	\end{equation}
where the maximum and minimum values are given in terms of the four
fixed $j_r$, $r=1,\ldots,4$ by
	\begin{equation}
	\eqalign{
	j_{12,{\rm min}}&=\max(|j_1-j_2|,|j_3-j_4|), \quad
	j_{12,{\rm max}}=\min(j_1+j_2,j_3+j_4), \\
	j_{23,{\rm min}}&=\max(|j_2-j_3|,|j_1-j_4|), \quad
	j_{23,{\rm max}}=\min(j_2+j_3,j_1+j_4).}
	\label{j12j23bounds}
	\end{equation}
Then the dimension of $\ZS$ is given by
	\begin{equation}
	D=\dim\ZS = j_{12,{\rm max}}-j_{12,{\rm min}}+1
	        = j_{23,{\rm max}}-j_{23,{\rm min}}+1.
	\label{dimZS}
	\end{equation}

An expression for $\dim\ZS$ can be given that is symmetrical in
$(j_1,j_2,j_3,j_4)$.  Using the fact that $|x|=\max(x,-x)$, the
difference between $j_{12,{\rm max}}$ and $j_{12,\rm{min}}$ becomes
the shortest distance between one set of four numbers, $\{j_1-j_2,
j_2-j_1, j_3-j_4, j_4-j_3\}$, and another set of two numbers,
$\{j_1+j_2, j_3+j_4\}$.  But this is the minimum of the distance
between all eight possible pairs taken from the two sets.  Thus
	\begin{equation}
	D=\dim\ZS = 2\min(j_1,j_2,j_3,j_4,s-j_1,s-j_2,s-j_3,s-j_4)
	+1,
	\label{symmdimZS}
	\end{equation}
where $s$ is the semiperimeter,
	\begin{equation}
	s=\frac{1}{2}(j_1+j_2+j_3+j_4).
	\label{sdef}
	\end{equation}
More precisely, if $D$ computed by (\ref{symmdimZS}) is $\le0$, then
subspace $\ZS$ is trivial ($\dim\ZS=0$); otherwise $D=\dim\ZS$.  The
formula (\ref{symmdimZS}) bears an interesting relationship to the
Regge symmetries of the $6j$-symbol (Varshalovich \etal\ 1981,
Eq.~(9.4.2.4)).
	
Now by comparing Figs.~\ref{asym6j} and \ref{4jABstates} we see that
the $6j$-symbol is proportional to a scalar product,
	\begin{equation}
	\braket{B}{A}=(-1)^{j_1+j_2+j_3+j_4}\,
	\sqrt{(2j_{12}+1)(2j_{23}+1)}
	\left\{\begin{array}{ccc}
	j_1 & j_2 & j_{12} \\
	j_3 & j_4 & j_{23}
	\end{array}\right\}.
	\label{6jme}
	\end{equation}
This is the scalar product that we shall use for the semiclassical
analysis of the $6j$-symbol in this paper.  In a different notation,
we can write $\braket{j_{23}}{j_{12}}$ instead of $\braket{B}{A}$,
which emphasizes the fact that this is a unitary matrix element
connecting two bases on $\ZS$.  The usual orthonormality relations
satisfied by the $6j$-symbol (see Edmonds eq.~(6.2.9)) are equivalent
to the unitarity of $\braket{j_{23}}{j_{12}}$.

Notice that in this $4j$-model, the angular momenta $\Jvechat_r$,
$r=1,2,3,4$ are independent operators, while the two remaining angular
momenta,
	\begin{equation}
	\Jvechat_{12} = \Jvechat_1 + \Jvechat_2,
	\qquad
	\Jvechat_{23} = \Jvechat_2 + \Jvechat_3,
	\label{Jvechat1223defs}
	\end{equation}
are not, rather they are functions of the first four.  As usual in the
Schwinger representation, the quantum number $j_r$, $r=1,\ldots,4$
specifies the eigenvalues of both $\Ihat_r$ and $\Jvechat_r^2$, that
is, $j_r$ and $j_r(j_r+1)$, respectively.  And the quantum numbers
$j_{12}$ and $j_{23}$ specify the eigenvalues of the operators
$\Jvechat_{12}^2$ and $\Jvechat_{23}^2$, that is, $j_{12}(j_{12}+1)$
and $j_{23}(j_{23}+1)$, respectively.  But there are no operators
$\Ihat_{12}$ or $\Ihat_{23}$.

The states $\ket{A}$ and $\ket{B}$ in Fig.~\ref{4jABstates} can be
expressed as eigenstates of complete sets of operators,
	\begin{equation}
	\ket{A}=\left|
	\begin{array}{@{\;}c@{\,}c@{\,}c@{\,}c@{\;}c@{\;}c@{}}
	\Ihat_1 & \Ihat_2 & \Ihat_3 & \Ihat_4 & \Jvechat_{12}^2
	        & \Jvechat_{\rm tot} \\
	j_1 & j_2 & j_3 & j_4 & j_{12} & \zerovec
	\end{array}\right>, 
	\qquad
	\ket{B}=\left|
	\begin{array}{@{\;}c@{\,}c@{\,}c@{\,}c@{\;}c@{\;}c@{}}
	\Ihat_1 & \Ihat_2 & \Ihat_3 & \Ihat_4 & \Jvechat_{23}^2
	        & \Jvechat_{\rm tot} \\
	j_1 & j_2 & j_3 & j_4 & j_{23} & \zerovec
	\end{array}\right>, 
	\label{4jABdefs}
	\end{equation}
in a notation like that used in (\ref{roberts3jstate}) and
(\ref{roberts2jstate}).  As mentioned, these states are normalized,
and their phases are specified by Fig.~\ref{4jABstates}.  

Notice that each state $\ket{A}$ and $\ket{B}$ has a list of eight
independent operators (counting the three components of $\Jvechat_{\rm
tot}$), corresponding to the eight degrees of freedom in the
$4j$-model.  We will call these lists of operators the $A$-list and
$B$-list, and write them collectively as
	\begin{equation}
	\eqalign{
		\Ahat &= (\Ihat_1,\Ihat_2,\Ihat_3,\Ihat_4,
		\Jvechat_{12}^2,\Jvechat_{\rm tot}), \\
		\Bhat &= (\Ihat_1,\Ihat_2,\Ihat_3,\Ihat_4,
		\Jvechat_{23}^2,\Jvechat_{\rm tot}),}
	\label{ABoplists}
	\end{equation}
We denote elements of these lists with subscripts, for example,
$\Ahat_i$ or $\Bhat_i$, $i=1,\ldots,8$.  The operators in the either
of the lists (\ref{ABoplists}) do not commute with one another
(because the components of $\Jvechat_{\rm tot}$ do not commute;
otherwise all commutators are zero), but they do possess simultaneous
eigenstates in $\ZS$, which are unique to within a phase, namely, the
states (\ref{4jABdefs}).

\subsection{The $8j$-model of the $6j$-symbol}
\label{8jmodel}

We obtain an $8j$-model of the $6j$-symbol by inserting $2j$-symbols
into edges 12 and 23 of the spin network of Fig.~\ref{asym6j} and
then treating them in the same way as the $2j$-symbols in the
$12j$-model.  The result is the Hilbert space $\HS_{8j}=
(\prod_{r=1}^4 \otimes\SS_r) \otimes\SS_{12} \otimes\SS'_{12} 
\otimes\SS_{23} \otimes\SS'_{23}$.  This model has
more symmetry than the $4j$-model but less than the $12j$-model.
Operators $\Ihat_{12}$ and $\Ihat_{23}$ exist in this model (as well
as operators $\Ihat'_{12}$ and $\Ihat'_{23}$), unlike the $4j$-model.
This gives the $8j$-model certain advantages over the $4j$-model.  We
shall not consider the $8j$-model further in this paper.

\section{The Classical Manifolds}
\label{classmanifolds}

In this section we study the classical mechanics that will be relevant
for the semiclassical analysis of the $6j$-symbol in a $4j$-model.  We
begin by presenting our notation for the Schwinger phase space and
products of it that are used to represent coupled, classical angular
momenta.  Other spaces that will be important are obtained by Poisson
and symplectic reduction.  Then we examine the geometry of the $A$-
and $B$-manifolds in phase space that support the states
(\ref{4jABdefs}), including a rather general analysis of why they are
Lagrangian.  Finally we discuss the Bohr-Sommerfeld quantization of
these manifolds.

\subsection{The Schwinger Phase Space and Other Spaces}
\label{schwingerphasespace}

The classical phase space for two harmonic oscillators (the Schwinger
phase space) is $\Phi=(\Reals^4,dp\wedge dx)$, with coordinates $(x,p)
\in \Reals^4$, for $x,p \in \Reals^2$.  See 
Sec.~I.\ISchwingermodel\ for more details on our use of the Schwinger
representation, as well as Cushman and Bates (1997) for a more
detailed discussion of the geometry of two harmonic oscillators and
the role played by the Hopf map.  Here $dp\wedge dx$ means
$\sum_{\mu=1}^2 dp_\mu \wedge dx_\mu$, and similarly for other obvious
contractions over $\mu=1,2$.  Complex coordinates $z_\mu=
(x_\mu+ip_\mu)/\sqrt{2}$, $\zbar_\mu = (x_\mu-ip_\mu)/\sqrt{2}$ are
also useful, allowing us to write $\Phi=(\Complexes^2,id\zbar\wedge
dz)$.

Interesting functions on $\Phi$ (classical observables) include
	\begin{equation}
	I = \frac{1}{2} \zbar z,
	\qquad
	J_i = \frac{1}{2} \zbar \sigma_i z,
	\label{IJidef}
	\end{equation}
where $i=1,2,3$, obviously the classical analogs of
(\ref{IhatJhatdef}), and where $z$ or $\zbar$ without indices
indicates a 2-component ``spinor,'' that is, an element of
$\Complexes^2$.  In comparison to (\ref{IhatJhatdef}) notice the
absence of the hats, indicating that these are classical observables.
These satisfy the Poisson bracket relations $\{I,J_i\}=0$ and
$\{J_i,J_j\} = \epsilon_{ijk}\,J_k$, as well as the identity $\Jvec^2
= I^2$.  The Hamiltonian flow of $I$ is a $U(1)$ action on $\Phi$,
while the flows of $J_i$, $i=1,2,3$, generate an $SU(2)$ action on
$\Phi$.  Both actions are easily expressed in the complex coordinates
$z$: that of $U(1)$ is $z\mapsto \exp(-i\psi/2) z$, where $\psi$ is
the variable conjugate to $I$, while that of $SU(2)$ is $z \mapsto
uz$ for $u\in SU(2)$.

The orbit of the $U(1)$ action generated by $I$ passing through any
point $z\ne0$ on $\Phi$ is a circle on which $\psi$ is a coordinate,
covered once when $0\le \psi < 4\pi$. The level set $I=J$ for $J>0$,
where $I:\Phi\to\Reals$ is the function and $J\in\Reals$ is the
contour value, is a 3-sphere to which the orbits of $I$ are confined.
The definition (\ref{IJidef}) of $J_i$ is interpreted as a map $\pi
:\Phi \to \Reals^3$, where the three components $J_i$ of $\Jvec$ are
coordinates on $\Reals^3$ (thus, this space is ``angular momentum
space'').  This map is a Poisson map (Marsden and Ratiu 1999), giving
$\Reals^3$ the Poisson structure $\{J_i,J_j\} = \epsilon_{ijk}\,J_k$.
We denote $\Reals^3$ with this Poisson structure by $\Lambda$, as
indicated by the diagram,
	\begin{equation}
	\begin{diagram}
	(I=J)    & \subset & \Phi \\
	\dTo>{\pi_H} &         & \dTo>\pi \\
	\Sigma   & \subset & \Lambda \\
	\end{diagram}
	\label{hopfmaps}
	\end{equation}
The map $\pi$ can also be interpreted as the momentum map (Abraham and
Marsden 1978) of the $SU(2)$ action on $\Phi$, so that $\Lambda$ or
angular momentum space is identified with ${\mathfrak s}{\mathfrak
u}(2)^*$.  When $\pi$ is restricted to a level set $I=J>0$ in $\Phi$
(a 3-sphere), it projects onto the 2-sphere $|\Jvec| = J$ in
$\Lambda$.  This is the projection map $\pi_H$ of the Hopf fibration,
in which the orbits of $I$ are the fibers (or ``Hopf circles'').  It
is also the map of symplectic reduction (Abraham and Marsden 1978) of
the level set $I=J$ in $\Phi$ by the $U(1)$ action, so that the
2-spheres $|\Jvec|=J$ in $\Lambda$ are symplectic manifolds.  The
symplectic form on one of these 2-spheres is $J\,d\Omega$, where
$d\Omega$ is the element of solid angle (to within a sign).  We
denote one of these spheres with its symplectic structure (for some
value $J>0$) by $\Sigma$; these spheres are also the symplectic leaves
of the Poisson structure in $\Lambda$.

In an $nj$-model of a spin network we take Cartesian products of the
Schwinger phase space $\Phi$ to obtain the phase space for $n$
independent classical angular momenta.  We will illustrate the
notation for the $4j$-model.  We write $\Phi_{4j} = \Phi^4$ for the
entire phase space; apart from the symplectic structure, this is
$\Complexes^8 = \Reals^{16}$.  Coordinates on $\Phi_{4j}$ are
$x_{\mu r}$, $p_{\mu r}$, $z_{\mu r}$, etc, $\mu=1,2$,
$r=1,\ldots,4$.   We denote the $r$-th copy of $\Phi$ by $\Phi_r$, and
define functions $I_r$, $J_{ri}$ on $\Phi_r$ on the pattern of
(\ref{IJidef}), that is, just by adding $r$ subscripts to all the
variables in those equations.  Naturally these can also be viewed as
functions on $\Phi_{4j}$.  The vector $\Jvec_r$ is the $r$-th
classical angular momentum.

We generalize the diagram (\ref{hopfmaps}) to the $4j$-model as follows,
	\begin{equation}
	\begin{diagram}
	           &         & (I_r=J_r)    & \subset & \Phi_{4j} \\
	           &         & \dTo>{\pi_H} &         & \dTo>\pi \\
   {\rm CL}        & \subset & \Sigma_{4j}  & \subset & \Lambda_{4j} \\
   \dTo>{\pi_{KM}} &         &              &         & \\
   \Gamma          &         &              &         & \\
	\end{diagram}
	\label{4jhopfmaps}
	\end{equation}
where $\pi$ means dividing $\Phi_{4j}$ by the $U(1)^4$ action
generated by $I_r$, $r=1,\ldots,4$.  Thus $\Jvec_r$, $r=1,\ldots,4$
are coordinates on $\Lambda_{4j}=(\Reals^3)^4$, in which the Poisson
bracket of two functions $f$ and $g$ is given by
	\begin{equation}
	\{f,g\} = \sum_{r=1}^4 \Jvec_r\cdot\left(
	\frac{\partial f}{\partial \Jvec_r} \times
	\frac{\partial g}{\partial \Jvec_r}\right).
	\label{Lambda4jPB}
	\end{equation}
Similarly, for four positive contour values $J_r>0$, $r=1,\ldots,4$,
the level set $I_r = |\Jvec_r| = J_r$, $r=1,\ldots,4$, in $\Phi_{4j}$
is $(S^3)^4$ (indicated simply by $I_r=J_r$ in the diagram).  The map
$\pi$ restricted to this space is a power of the Hopf map (simply
denoted $\pi_H$ in the diagram), which projects $(S^3)^4$ onto the
space $\Sigma_{4j} = (S^2)^4$, a symplectic manifold in
$\Lambda_{4j}$.  Notice that the radii of the spheres (the 3-spheres
in $\Phi_{4j}$ and the 2-spheres in $\Lambda_{4j}$) need not be equal
for different $r$.  The spaces CL and $\Gamma$ in (\ref{4jhopfmaps})
will be explained later.

Other important classical observables on $\Phi_{4j}$ are $\Jvec_{12}
=\Jvec_1 +\Jvec_2$ and $\Jvec_{23} =\Jvec_2 +\Jvec_3$, their squares,
$\Jvec_{12}^2$ and $\Jvec_{23}^2$, and the total angular momentum
$\Jvec_{\rm tot} = \sum_{r=1}^4 \Jvec_r$.
	
\subsection{Level Sets and Contour Values}
\label{levelsets}

The $A$- and $B$-lists of operators in (\ref{ABoplists}) correspond to
lists of classical observables (without the hats),
	\begin{equation}
	\eqalign{
	A &= (I_1,I_2,I_3,I_4,\Jvec_{12}^2,\Jvec_{\rm tot}), \\
	B &= (I_1,I_2,I_3,I_4,\Jvec_{23}^2,\Jvec_{\rm tot}).}
	\label{ABfunctionlists}
	\end{equation}
We will denote the members of these lists by $A_i$, $B_i$,
$i=1,\ldots,8$.  As discussed in I and in Littlejohn (1990), the
Lagrangian manifolds that support the semiclassical approximations to
the states $\ket{A}$ and $\ket{B}$ are the level sets in $\Phi_{4j}$
of these lists of classical observables, with quantized values of the
contour values.  We will defer the question of quantization to
Sec.~\ref{BSquant}, and for now just examine these level sets for some
suitable contour values.  We will call the level sets of the $A$- and
$B$-lists the $A$- and $B$-manifolds.

It will be convenient to distinguish notationally the functions in the
two lists, regarded as maps $:\Phi_{4j} \to \Reals$, from the contour
values, which are real numbers.  Our notation is summarized in
Table~\ref{fvtable}.  The vector of functions $\Jvec_{\rm tot}$ is
given the value $\zerovec$ because that is the only contour value we
will consider.  The other contour values are variable.  The
conventions in the table solve some notational problems in I, but the
notation requires care.  For example, the magnitude of the vector
$\Jvec_r$, regarded as a function $:\Phi_{4j}\to\Reals$, is not $J_r$
(because $J_r$ is a number, not a function), but on the level set
$I_r=J_r$ it is true that that $|\Jvec_r| = J_r$, in view of the
identity between functions, $\Jvec_r^2 = I_r^2$.  If we wish to refer
to the magnitude or the square of the vector $\Jvec_r$, regarded as a
function on $\Phi$, we will write $|\Jvec_r|$ or $\Jvec_r^2$ (in bold
face), not $J_r$ or $J_r^2$.  We will denote the lists of contour
values collectively by $a$ or $b$, so that
	\begin{equation}
	\eqalign{
	a&=(J_1,J_2,J_3,J_4,J_{12}^2,\zerovec), \\
	b&=(J_1,J_2,J_3,J_4,J_{23}^2,\zerovec),}
	\label{abdefs}
	\end{equation}
with components $a_i$, $b_i$, $i=1,\ldots,8$.

\begin{table}
\begin{center}
\begin{tabular}{|c|c|}\hline
function & value \\
\hline
$I_r$ & $J_r$ \\
$\Jvec_{12}^2$ & $J_{12}^2$ \\
$\Jvec_{23}^2$ & $J_{23}^2$ \\
$\Jvec_{\rm tot}$ & $\zerovec$ \\
\hline
\end{tabular}
\end{center}
\caption{\label{fvtable} Notation for functions $:\Phi_{4j} \to
\Reals$ and values (real numbers) of those functions.  In the first
row, $r=1,\ldots,4$.}
\end{table}

Consider now the conditions on the contour values $a$, $b$ such that
the $A$- and $B$-manifolds should exist (as nonempty sets of
points).  The question of the dimensionality of these manifolds will
be postponed to Sec.~\ref{ABproperties}, but it turns out that their
maximum (for any contour values) and generic dimensionality is 8, one
half the dimension of the $4j$ Schwinger phase space.  We work
with the $A$-manifold, since the conditions for the $B$-manifold are
completely analogous.

\begin{figure}[htb]
\begin{center}
\scalebox{0.5}{\includegraphics{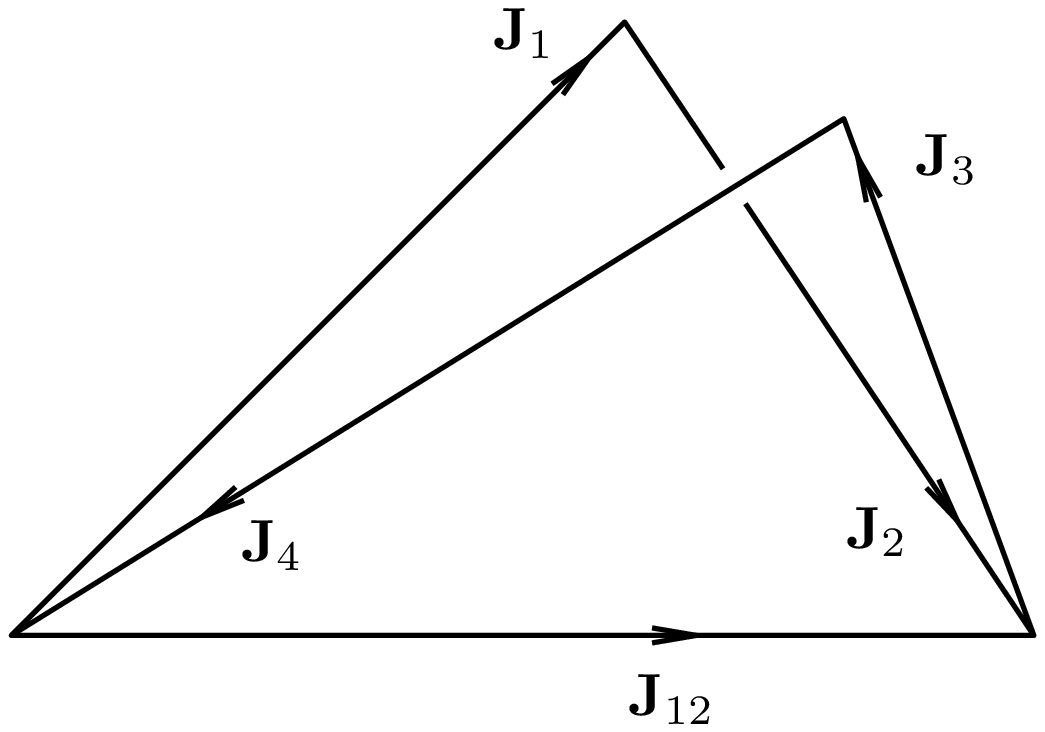}}
\end{center}
\caption[fourJ]{\label{fourJ} A point of the $A$- or $B$-manifolds 
in $\Phi_{4j}$
$={\mathbb C}^8$ projects under $\pi$ onto a point of $\Lambda_{4j}
=[{\mathbb R}^3]^4$ on which the four angular momenta $\Jvec_r$ add to
zero.  We visualize this point as four angular momentum vectors in a
single copy of ${\mathbb R}^3$ that add to zero.  In such diagrams,
the vectors need not be based at the origin; here we place them
end-to-end.  Also shown in the figure is the vector
$\Jvec_{12}=\Jvec_1+\Jvec_2$.}
\end{figure}

First, the $A$-manifold clearly does not exist unless $J_r\ge0$,
$r=1,\ldots,4$, so let us assume this condition.  Then if the
$A$-manifold exists, we can pick a point on it and evaluate the vector
functions $\Jvec_r$, $r=1,\ldots,4$, to determine the projected point
in $\Lambda_{4j}$.  We visualize this point as four vectors in a
single copy of $\Reals^3$.  In addition we compute the vector
$\Jvec_{12} =\Jvec_1 +\Jvec_2$ and plot it along with the others in
$\Reals^3$.  Next, we move the vectors $(\Jvec_1,\Jvec_2,-\Jvec_{12})$
end-to-end to create a triangle with sides $(J_1,J_2,J_{12})$.  We
move the vectors by parallel transport in $\Reals^3$, that is, without
rotating them.  Thus, the triangle inequalities are satisfied in the
triplet of lengths $(J_1,J_2,J_{12})$.  Also, since $\Jvec_{\rm
tot}=0$, we can form a second triangle out of vectors
$(\Jvec_3,\Jvec_4,\Jvec_{12})$, so the triangle inequalities are
satisfied in the triplet $(J_3,J_4,J_{12})$.  The two triangles have
the edge $J_{12}$ in common, creating a figure like Fig.~\ref{fourJ}.

Conversely, suppose that $J_r\ge0$, $r=1,\ldots,4$ and that the triangle
inequalities are satisfied in triplets $(J_1,J_2,J_{12})$ and
$(J_3,J_4,J_{12})$.  This means that two triangles can be constructed
in $\Reals^3$ with the given lengths.  By translating and/or rotating
the two triangles, we can bring the 12 edges into coincidence, thereby
creating a figure like Fig.~\ref{fourJ}, and hence a point of
$\Lambda_{4j}$.  But since the map $\pi:\Phi_{4j} \to \Lambda_{4j}$ is
onto, there exists an inverse image of this point in $\Phi_{4j}$,
hence the $A$-manifold exists. 

In summary, we have shown that the $A$-manifold exists iff $J_r\ge0$,
$r=1,\ldots,4$, and the triangle inequalities are satisfied in the
triplets $(J_1,J_2,J_{12})$ and $(J_3,J_4,J_{12})$.  A similar result
holds for the $B$-manifold, where the triplets are $(J_2,J_3,J_{23})$
and $(J_1,J_4,J_{23})$.  The triangle inequalities imply that $J_{12}$
and $J_{23}$ lie in the ranges,
	\begin{eqnarray}
	J_{12,{\rm min}} &\le J_{12} \le J_{12,{\rm max}}, 
	\label{J12range} \\
	J_{23,{\rm min}} &\le J_{23} \le J_{23,{\rm max}},
	\label{J23range}
	\end{eqnarray}
where
	\begin{eqnarray}
	\fl J_{12,{\rm min}} &= \max(|J_1-J_2|,|J_3-J_4|), 
	\qquad
	J_{12,{\rm max}} = \min(J_1+J_2,J_3+J_4), 
	\label{J12maxmin} \\
	\fl J_{23,{\rm min}} &= \max(|J_2-J_3|,|J_1-J_4|),
	\qquad
	J_{23,{\rm max}} = \min(J_2+J_3,J_1+J_4).
	\label{J23maxmin}
	\end{eqnarray}
These inequalities imply $J_{12}, J_{23 }\ge 0$.  Notice that the
value 0 is not always excluded, for example, if $J_1=J_2$ and
$J_3=J_4$ then $J_{12}=0$ is allowed.

It turns out that the $A$- and $B$-manifolds, when 8-dimensional, are
Lagrangian.  According to the Liouville-Arnold theorem (Arnold 1989)
the compact level sets of complete sets of Poisson commuting
observables are generically Lagrangian tori.  The sets of observables
$A$ or $B$ of interest in this paper are not commuting, however, so
the Liouville-Arnold theorem does not apply.  In a similar situation
in I we showed that the manifold in question (what we called the
``Wigner manifold'') was nevertheless Lagrangian.  In the following
section we present a general set of circumstances in which Lagrangian
manifolds are obtained, which cover not only the cases considered in
the Liouville-Arnold theorem but also all cases we know of in the
asymptotics of spin networks, including the manifolds studied in I and
the $A$- and $B$-manifolds of this paper.

\subsection{Level Sets, Orbits, and Lagrangian Manifolds}
\label{lsolm}

We use general notation in this section that differs somewhat from
that of the application to the $6j$-symbol in the rest of the
paper. The basic conclusion of this section is the following.  Let
$\{A_i, i=1,\ldots,m\}$ be a collection of classical observables on a
phase space (symplectic manifold) $P$ of dimension $2N$, that is,
$A_i: P\to\Reals$.  Let the set $\{A_i\}$ form a Lie algebra, that is,
the Poisson brackets $\{A_i,A_j\}$ are linear combinations of the
$A_i$.  Let $L$ be the level set $A_i=a_i$, for some contour values
$a_i$, and suppose that all Poisson brackets $\{A_i,A_j\}$ vanish on
$L$.  Finally, suppose $L$ is a smooth manifold of dimension $N$
(thus, $m\ge N$).  Then $L$ is Lagrangian.  The reader who is willing
to accept this conclusion can skip the remainder of this section.

We deal with classical Hamiltonian systems with symmetry.  Basic
references on this subject are Abraham and Marsden (1978), Marsden and
Ratiu (1999) and Cushman and Bates (1997).  We make quite a few
assumptions in this section, but most of them are generic.  The most
important one that is not is the assumption that the momentum, in the
general sense of the value of the momentum map, is a fixed point of
the coadjoint action of the group, that is, it is $G$-invariant.

Let $P$ be a symplectic manifold of dimension $2N$, let $G$ be a
connected Lie group of dimension $m$ with Lie algebra
${\Liealgebra{g}}$, dual ${\Liealgebra{g}}^*$, and symplectic action
on $P$.  Suppose the momentum map $M:P \to {\Liealgebra{g}}^*$ exists,
that is, the action of $G$ is generated by Hamiltonian flows of a set
of Hamiltonian functions.  Let $\{\xi_i, i=1,\ldots,m\}$ be a basis in
${\Liealgebra{g}}$, and let $c^k{}_{ij}$ be the structure constants,
so that $[\xi_i,\xi_j] = c^k{}_{ij}\, \xi_k$ (summation convention).
Define functions $A_i:P \to {\mathbb R}$ by $A_i(x) = \langle M(x),
\xi_i \rangle$ for all $x\in P$.  These functions form a Lie algebra
under the Poisson bracket, $\{A_i,A_j\} = c^k{}_{ij} \, A_k$, with the
same structure constants as the $\xi_i$ in ${\Liealgebra{g}}$.  Let
$X_i = \omega^{-1} dA_i$ be the Hamiltonian vector fields associated
with the $A_i$, where $\omega$ is the symplectic form on $P$, regarded
as a map from vector fields to 1-forms.

For some point $x_0 \in P$ let $a=M(x_0)$, so the level set $L$ of the
$A$'s passing through $x_0$ is given by $A=a$ (that is,
$L=M^{-1}(a)$).  For simplicity we assume that $L$ has only one
connected component, or else we restrict consideration to the
connected component passing through $x_0$.  Let $n$ be the rank of the
set of differential forms $\{dA_i, i=1,\ldots,m\}$, assumed to be
constant over $L$.  Then $n\le m$ and $\dim L = 2N-n$.  Also let the
orbit of the $G$-action through $x_0$ be $B$.  The vectors $\{X_i,
i=1,\ldots,m\}$ are tangent to $B$ and span its tangent space at each
point.  Also, $\rank \{X_i\} = \rank\{dA_i\} = n$, since $\omega$ is
nonsingular.  Thus, $\dim B=n$.

\begin{figure}[htb]
\begin{center}
\scalebox{0.5}{\includegraphics{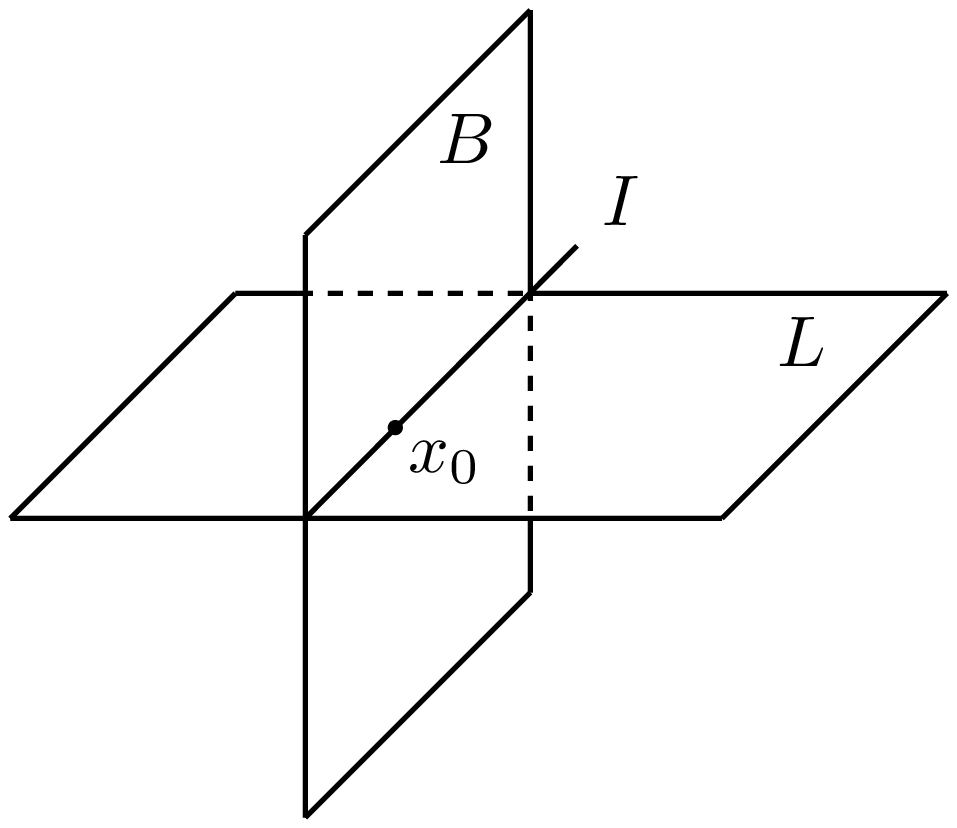}}
\end{center}
\caption[symred]{\label{symred} Schematic illustration of the basic 
geometry
of symplectic reduction.  $L$ is the level set of the momentum map
passing through $x_0$, $B$ is the orbit of the group action, and $I$
the intersection, also the orbit of the isotropy subgroup.}
\end{figure}

At this point we have the basic geometry for symplectic reduction, as
illustrated in Fig.~\ref{symred}.  Shown in the figure is the
intersection $I$ of $L$ and $B$, which is  the orbit of $x_0$ under
the isotropy subgroup of $a$ under the coadjoint action of the group.
Dividing $L$ by the isotropy subgroup produces the reduced symplectic
manifold.

Now suppose that $a \in {\Liealgebra{g}}^*$ (the generalized momentum)
is a fixed point of the coadjoint action of the group, that is,
$\Ad^*_g a =a$, $\forall g\in G$.  By differentiation and contraction
with an arbitrary element of ${\Liealgebra{g}}$ this implies $\langle
a, [\eta,\zeta]\rangle =0$, $\forall \, \eta,\zeta \in
{\Liealgebra{g}}$, or, by setting $\eta=\xi_i$, $\zeta=\xi_j$,
$c^k{}_{ij} a_k =0$.  But this implies $\{ A_i, A_j \}=c^k{}_{ij} \, A_k
=0$ on the level set $L$, where $A_k=a_k$.  In other words, the
functions $A_i$, which may form a nontrivial Lie algebra on $P$, have
vanishing Poisson brackets among themselves on $L$.  This in turn
implies that the $A_i$'s are constant along each other's flows on $B$,
that is, $X_i A_j = -\{A_j,A_i\} =0$, so $B \subset L$.  Therefore
$\dim B \le \dim L$, or $n \le 2N-n$, or $n\le N$.

Finally, let us assume that $n$ takes its maximum value $n=N$.  Then
$\dim B = \dim L = N$.  If $L$ is compact, then $B=L$, that is, the
level set and the orbit coincide.  Moreover $L$ is Lagrangian, since
the vectors $X_i$ span the tangent space to $L$ and $\omega(X_i,X_j) =
-\{A_i,A_j\}=0$ on $L$.  This in turn implies that $L$ supports
locally a solution of the simultaneous Hamilton-Jacobi equations
involved in finding a semiclassical eigenfunction of the operators with
principal symbols $A_i$.

In the case of an integrable system the group is Abelian, $G={\mathbb
R}^N$, the $N$ classical observables $A_i$ Poisson commute everywhere
in phase space, and on a generic level set $A=a$ the differentials
$dA_i$ are linearly independent everywhere.  Since the group is
Abelian, $a$ is automatically a fixed point of the coadjoint action,
and thus the level set is Lagrangian.  Moreover, if it is compact, it
is an $N$-torus.

In the case of the manifolds explored in I (the $jm$- and Wigner
manifolds) and the $A$- and $B$-manifolds of this paper, the isotropy
subgroup of the action of $G$ on $x_0\in P$ is zero-dimensional (a
discrete set of points), so $\dim G = \dim B = m = n$.  Moreover,
$n=N$, so the manifolds are Lagrangian.  

\subsection{Properties of the $A$- and $B$-manifolds}
\label{ABproperties}

We now apply the analysis of Sec.~\ref{lsolm} to the $A$-manifold.  In
the following we visualize a point of $\Phi_{4j}$ as a $2\times4$
matrix of complex coordinates $z_{\mu r}$, referring to the $r$-th
column as the $r$-th ``spinor.''

Here the symplectic manifold is $P=\Phi_{4j}$, and we take the group
to be $G= U(1)^5\times SU(2)$, where the action of $U(1)^5$ is
generated by $(I_1,I_2,I_3,I_4 ,\Jvec_{12}^2)$, and that of $SU(2)$ is
generated by $\Jvec_{\rm tot}$.  For given $r$, the observable $I_r$
generates a $U(1)$ action, the multiplication the $r$-th spinor by
$e^{-i\psi_r/2}$, where $\psi_r$ is the evolution variable conjugate
to $I_r$, with period $4\pi$.  See (I.\IIflow). Also, $\Jvec_{\rm tot}$
generates the diagonal $SU(2)$ action, that is, all four spinors are
multiplied by the same element $u\in SU(2)$ (the $2\times 4$ matrix is
multiplied on the left by $u$).  See (I.\IndotJeqns) and
(I.\IndotJflow).  As for the observable $\Jvec_{12}^2$, Hamilton's
equations are
	\begin{equation}
	\fl\frac{dz_{r\mu}}{dt} = \{ z_{r\mu}, \Jvec_{12}^2\}
	= 2 J_{12i} \{ z_{r\mu}, J_{12i} \} 
	= \cases{-i (\Jvec_{12} \cdot \bsigma)_{\mu\nu} \,
	z_{r\nu}, & if $r=1,2$, \\
	0, & if $r=3,4$,}
	\label{J12sqflow}
	\end{equation}
where $t$ is the variable of evolution.  See (I.\IndotJeqns) for a
similar calculation, and notice that $\Jvec_{12}$ is a constant of the
flow generated by $\Jvec_{12}^2$.  Thus, $\Jvec_{12}^2$ generates a
$U(1)$ action that rotates spinors 1 and 2 about the axis $\jvec_{12}
=\Jvec_{12} /|\Jvec_{12}|$, that is, it multiplies them by $u(\jvec_{12}
,\theta) \in SU(2)$ (in axis-angle notation for an element of $SU(2)$,
see (I.\Iudef)) while leaving spinors 3 and 4 invariant.  The period
is $\theta=4\pi$ or $t=2\pi/J_{12}$.  The $U(1)$ action generated by
$\Jvec_{12}^2$ is not a rotation about a fixed axis, since
$\Jvec_{12}$ depends on where we are in $\Phi_{4j}$, but it does
commute with the other $U(1)$ actions generated by the $I_r$, as well
as the $SU(2)$ action generated by $\Jvec$.  Similar statements can be
made about $\Jvec_{23}^2$.

Now we determine the maximum dimensionality of the $A$-manifold.
Since there are 8 functions in the $A$-list and $\dim\Phi_{4j}=16$,
the answer will be 8 if the functions are independent.  Functions are
independent at a point if their differentials, in this case, $\{dA_i,
i=1,\ldots,8\}$, are linearly independent.  In any case, the rank of
this set of differentials is the same as the rank of the set of
Hamiltonian vector fields $X_i = \omega^{-1}dA_i$.

The $A$-manifold cannot have its maximum dimension if any $J_r=0$, for
$r=1,\ldots,4$, since that implies $dI_r=0$.  So we assume $J_r>0$,
$i=1,\ldots,4$.  This means that all four spinors $z_r$ are nonzero,
and hence the orbit of the group $U(1)^4$ generated by the $I_r$,
$r=1,\ldots,4$ is a four-torus $T^4$.  This is also the fiber of the
projection $\pi :\Phi_{4j }\to\Lambda_{4j}$.

The condition $J_r>0$, $r=1,\ldots,4$ also means that all four vectors
$\Jvec_r$, $r=1,\ldots,4$ are nonzero.  We think of these vectors in a
single copy of $\Reals^3$.  When the group $SU(2)$, whose action is
generated by $\Jvec_{\rm tot}$, acts on $\Phi_{4j}$, the effect on the
vectors $\Jvec_r$, $r=1,\ldots,4$, is to rotate them by the
corresponding element of $SO(3)$ (see (I.\IRuproj)).  If any of these
vectors is moved by the $SO(3)$ rotation, then in $\Phi_{4j}$ we will
have moved off the initial fiber of the projection $\pi$, hence we
will have motions that are linearly independent of the $U(1)^4$
action.  To make all three independent directions of rotation in
$SO(3)$ give rise to linearly independent motions, we require that the
isotropy subgroup of the $SO(3)$ action on the set $\{\Jvec_r$,
$r=1,\ldots,4\}$ (that is, the diagonal action) be trivial.  This is
at a point at which $\Jvec_{\rm tot}=\sum_{r=1}^4 \Jvec_r=0$.  This
requires that the four vectors $\Jvec_r$, $r=1,\ldots,4$ (which are
nonzero) be noncollinear (otherwise the isotropy subgroup is $SO(2)$
or the whole group $SO(3)$).  This in turn requires that at least one
of the triangles, $(J_1,J_2,J_{12})$ or $(J_3,J_4,J_{12})$, have
nonzero area.

Next there is the $U(1)$-action on $\Phi_{4j}$, generated by
$\Jvec_{12}^2$.  This has the effect on the vectors $\Jvec_r$,
$r=1,\ldots,4$ of a rotation of vectors $\Jvec_1$ and $\Jvec_2$ about
axis $\jvec_{12}$, while leaving $\Jvec_3$ and $\Jvec_4$ invariant.
In order that this motion move us off the initial fiber, we require
that $\Jvec_1$ and $\Jvec_2$ be linearly independent, that is, that
triangle $(J_1,J_2,J_{12})$ have nonzero area.  And in order that this
motion be linearly independent of overall rotation generated by
$\Jvec_{\rm tot}$, we require that the triangle $(J_3,J_4,J_{12})$
also have nonzero area, in order that the rotation of $\Jvec_1$,
$\Jvec_2$ about the axis $\jvec_{12}$ should change the shape of the
figure created by the four vectors.  (The overall rotations generated
by $\Jvec_{\rm tot}$ do not change the shape of the figure.)

We conclude that at a point of the $A$-manifold, $\rank\{dA_i,
i=1,\ldots, 8\} = 8$ iff $J_r>0$, $r=1,\ldots,4$ and none of the
triangle inequalities in $(J_1,J_2,J_{12})$ or $(J_3,J_4,J_{12})$, or
equivalently, the inequalities in (\ref{J12range}), is saturated.
Notice in particular that this implies $J_{12}>0$.  Since these
conditions depend only on the contour values and not where we are on
the $A$-manifold, the $A$-manifold, under the stated conditions, is a
smooth, 8-dimensional manifold.  

In addition, the $A$-manifold is compact.  It is also connected, as
follows by consideration of its projection under $\pi$ (see
Sec.~\ref{project}).  This means that the action of the group $U(1)^5
\times SU(2)$ is transitive on the $A$-manifold (it is the orbit of
any point on it under the group action).  It then follows from the
discussion of Sec.~\ref{lsolm} that the $A$-manifold, when
8-dimensional, is Lagrangian.  Similar statements apply to the
$B$-manifold.

To find the topology of the $A$-manifold, we first find the isotropy
subgroup of the action of $U(1)^5\times SU(2)$ on a point $x_0$ on
this manifold.  If we denote coordinates on $U(1)^5 \times SU(2)$ by
$(\psi_1, \psi_2, \psi_3, \psi_4, \theta, u)$, where $u \in SU(2)$ and
where the five angles are the $4\pi$-periodic evolution variables
corresponding to $(I_1,I_2,I_3,I_4,\Jvec_{12}^2)$, respectively, then
the isotropy subgroup is generated by two elements, say, $x=(2\pi,
2\pi, 2\pi, 2\pi, 0, -1)$ and $y=(0,0,2\pi,2\pi,2\pi,-1)$. The
isotropy subgroup itself is the Viergruppe $\{e,x,y,xy\}=({\mathbb
Z}_2)^2$.  Thus the $A$-manifold is topologically $U(1)^5 \times SU(2)
/ ({\mathbb Z}_2)^2$.  This is the same logic used in I to find the
topology of the ``Wigner manifold'' of that paper.  The analysis is
the same for the $B$-manifold, which has the same topology.  The
isotropy subgroup would be larger in degenerate cases, for example,
when some triangle inequalities are saturated.

Now it is easy to find the invariant measure on the $A$- or
$B$-manifolds.  It is $d\psi_1\wedge d\psi_2\wedge d\psi_3\wedge
d\psi_4\wedge d\theta\wedge du$, where $du$ is the Haar measure on
$SU(2)$ ($du =\sin\beta\, d\alpha\wedge d\beta\wedge d\gamma$ in
Euler angles).  Thus, the volume of the $A$- or $B$-manifold with
respect to this measure is
	\begin{equation}
	V_A = V_B = \frac{1}{4} (4\pi)^5 \times 16\pi^2
	= 2^{12} \pi^7,
	\label{ABvolume}
	\end{equation}
where the $1/4$ compensates for the 4-element isotropy subgroup.

\subsection{Projections and tetrahedra}
\label{project}

\begin{figure}[htb]
\begin{center}
\scalebox{0.5}{\includegraphics{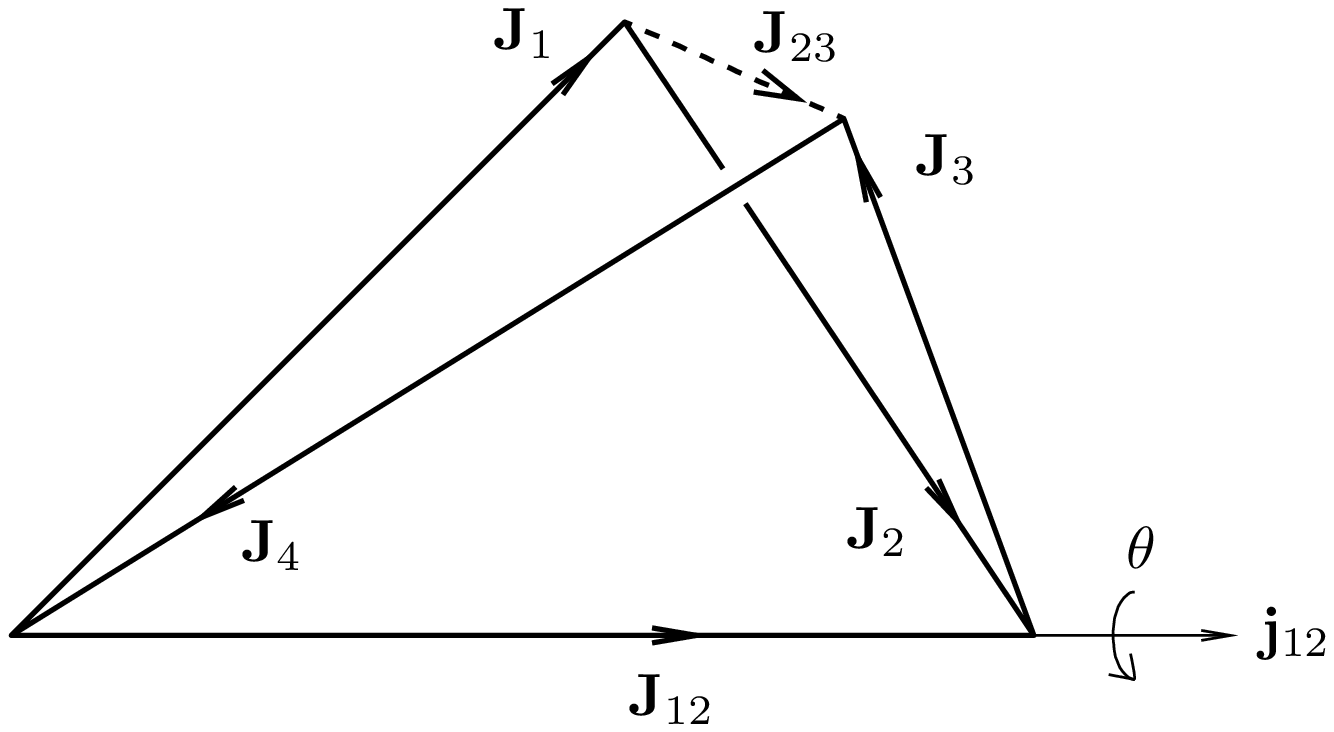}}
\end{center}
\caption[butterfly]{\label{butterfly} The function $\Jvec_{12}^2$ 
generates a
rotation about axis $\jvec_{12}$ of vectors $\Jvec_1$ and $\Jvec_2$,
rotating triangle 1-2-12 by an angle $\theta$ while leaving triangle
3-4-12 fixed.  The result is a family of tetrahedra with values of
$j_{23}$, the length of vector $\Jvec_{23}$, ranging from a minimum to
a maximum.}
\end{figure}

We now study the projection of the $A$-manifold onto $\Lambda_{4j}$,
for contour values such that the manifold is 8-dimensional.  The
projection consists of the set of four nonzero vectors in $\Reals^3$,
$\{\Jvec_r, r=1,\ldots, 4\}$ of given lengths $J_r>0$ with a sum of
zero, $\Jvec_{\rm tot}=0$, creating a closed chain of links as in
Fig.~\ref{fourJ}, such that $J_{12}$ has a given, positive value and
both triangles 1-2-12 and 3-4-12 have nonzero area.  We choose to
place the four vectors end-to-end in the order $(\Jvec_1 ,\Jvec_2
,\Jvec_3 ,\Jvec_4)$, as in Fig.~\ref{fourJ}; this is an arbitrary
choice, but convenient for studying observables $\Jvec_{12}^2$ and
$\Jvec_{23}^2$ (if we wished to examine $\Jvec_{13}^2$ we would choose
a different order).  By filling in the lines $\Jvec_{12}$ and
$\Jvec_{23}$, the closed chain of links becomes a tetrahedron, as
shown in Fig.~\ref{butterfly}.

The set of all such figures, modulo orientation, is the circle
$S^1=SO(2)$, whose angle is the dihedral angle about the edge
$\Jvec_{12}$.  Including the orientation, we see that the projection
of the $A$-manifold onto $\Lambda_{4j}$ is topologically $SO(2) \times
SO(3)$.  One might imagine that this should be described as an $SO(3)$
bundle over $SO(2)$, but in fact the bundle is trivial.  This is
connected, so the inverse image under $\pi$, which is the $A$-manifold
itself, is a 4-torus bundle over $SO(2)\times SO(3)$, and is also
connected.  An action of the group $SO(2)\times SO(3)$ on
$\Lambda_{4j}$ is generated by $\Jvec_{12}^2$ and $\Jvec_{\rm tot}$,
regarded as observables on $\Lambda_{4j}$ using the Poisson structure
(\ref{Lambda4jPB}).  This is the projection of the action of $U(1)^5
\times SU(2)$ on $\Phi_{4j}$.  The motion generated by $\Jvec_{12}^2$
is the ``butterfly'' motion illustrated in Fig.~\ref{butterfly}, in
which the butterfly flaps one of its wings.

As the triangle 1-2-12 rotates around the edge 12, the length
$|\Jvec_{23}|$ varies between a maximum and minimum value.  The
minimum value is reached when triangles 1-2-12 and 3-4-12 lie in the
same plane on the same side of line 12, and maximum when on opposite
sides of line 12; these are also the minimum and maximum of
$|\Jvec_{23}|$ on the $A$-manifold in $\Phi_{4j}$.  These extremal
values are reached when the tetrahedron is flat (its volume is zero),
and can be obtained in terms of the other five $J$'s as the roots of
the Cayley-Menger determinant (Berger 1987, Crippen and Havel 1988),
	\begin{equation}
	\det \left(\begin{array}{ccccc}
	0 & 1 & 1 & 1 & 1 \\
	1 & 0 & J_1^2 & J_{12}^2 & J_4^2 \\
	1 & J_1^2 & 0 & J_2^2 & x^2 \\
	1 & J_{12}^2 & J_2^2 & 0 & J_3^2 \\
	1 & J_4^2 & x^2 & J_3^2 & 0
	\end{array}\right)=0,
	\label{CMdet}
	\end{equation}
where $x=|\Jvec_{23}|$.  Such determinants were an important part of
the analysis of Ponzano and Regge (1968).  In this case the
determinant expresses the condition that the volume of the tetrahedron
is zero.  Expanding the determinant gives a quadratic equation in
$x^2$ in terms of the five $J$'s specifying the $A$-manifold.  The
same condition can be expressed in terms of a smaller ($3\times3$)
Gram matrix of dot products, as discussed in Littlejohn and Yu (2009).
	
A similar analysis applies to the $B$-manifold except that here we fix
the length $J_{23}$ of vector $\Jvec_{23}$, and create tetrahedra of
different shapes by varying the dihedral angle between triangles
4-1-23 and 2-3-23, that is, by rotating vectors $\Jvec_2$ and
$\Jvec_3$ about the axis $\jvec_{23} = \Jvec_{23}/J_{23}$.  In this
process, the length $|\Jvec_{12}|$ varies from some minimum to some
maximum, which can be obtained by replacing $x^2$ in (\ref{CMdet}) by
$J_{23}^2$, and then $J_{12}^2$ by $x^2$ where now $x=|\Jvec_{12}|$.

\subsection{Quantizing the Manifolds}
\label{BSquant}

The $A$- and $B$-manifolds can be subjected to Bohr-Sommerfeld
quantization, which selects certain contour values as quantized.  The
process uses the Weyl symbols or transforms (Weyl 1927, Wigner 1932,
Groenewold 1946, Moyal 1949, Berry 1977, Balazs and Jennings 1984,
Littlejohn 1986, Ozorio de Almeida 1998) of the lists
(\ref{ABoplists}) of operators, which are functions on $\Phi_{4j}$.
The Weyl symbols of selected operators are summarized in
Table~\ref{weylsymbols}.  It is important that the classical manifolds
be the level sets of the Weyl symbols of the operators whose
simultaneous eigenfunctions we seek (Littlejohn 1990); this holds in
the present case because the Weyl symbols in the table are always
equal to the corresponding classical observable (without the hat), to
within an additive constant.  Notationally we could have defined the
classical observable (without the hat) as the Weyl symbol of the
corresponding quantum observable, but the conventions in the table
make it easier to establish the connections with the usual quantum
numbers in physics.  In particular, the zero point energy has been
subtracted from the quantum observables $\Ihat_r$, but not from the
classical ones $I_r$, which explains the $1/2$ on the first row of the
table.

\begin{table}
\begin{center}
\begin{tabular}{|c|c|}\hline
operator & Weyl symbol \\
\hline
$\Ihat_r$ & $I_r-1/2$ \\
$\Jvechat_r$ & $\Jvec_r$ \\
$\Jvechat_{\rm tot}$ & $\Jvec_{\rm tot}$ \\
$\Jvechat_{12}$ & $\Jvec_{12}$ \\
$\Jvechat_{23}$ & $\Jvec_{23}$ \\
$\Jvechat_r^2$ & $\Jvec_r^2 - 3/8$ \\
$\Jvechat_{12}^2$ & $\Jvec_{12}^2 - 3/4$ \\
$\Jvechat_{23}^2$ & $\Jvec_{23}^2 - 3/4$ \\
\hline
\end{tabular}
\end{center}
\caption{\label{weylsymbols} Weyl symbols of selected operators.  In
rows containing operators $\Ihat_r$, $\Jvechat_r$ and $\Jvechat_r^2$,
$r=1,\ldots,4$.}
\end{table}

To quantize the $A$- or $B$-manifold, we first determine the homotopy
group, then we compute action integrals and Maslov indices along
generators of the group, then we require that the action plus Maslov
correction be an integer multiple of $2\pi$.  Only manifolds of full
dimensionality (8) can be quantized.  This is the procedure followed
in I, and the analysis is very similar in this case; in particular, in
both cases the homotopy group is Abelian.  

We just summarize the results, speaking of the $A$-manifold.  We find
that $J_r$, $r=1,\ldots,4$ are quantized in half-integer steps, which
we write in terms of the quantum numbers $j_r$ as
	\begin{equation}
	J_r=j_r+1/2,
	\label{Jrquant}
	\end{equation}
where the allowed values of $j_r$ are $0,\frac{1}{2},1,\ldots$.
Smaller values of $j_r$ are not allowed because for $j_r=-1/2$ the
manifolds do not have full dimensionality, while for $j_r<-1/2$ they
do not exist.  We choose the quantum number $j_r$ so that it agrees
with the usual notation in physics for the eigenvalues of various
operators, but that is not confirmed until we compute the
semiclassical eigenvalues in Sec.~\ref{sceigenvalues}.

Similarly, $J_{12}$ must be an integer or half-integer on quantized
manifolds, which we write in terms of a conventional quantum number by
$J_{12}=j_{12}+\frac{1}{2}$.  In addition, there is the condition that
$j_1+j_2+j_{12}$ and $j_3+j_4+j_{12}$ be integers, the requirement
(with the usual interpretation of the quantum numbers)
that the $3j$-symbols in Fig.~\ref{asym6j} should exist.  See also
eq.~(I.\Itwozquantone).  These imply that $j_1+j_2+j_3+j_4$ must be an
integer, part of the conditions that the subspace $\ZS$ defined in
Sec.~\ref{4jmodel} be nontrivial.

The range of the quantum number $j_{12}$ is determined by the
requirement that the $A$-manifold be 8-dimensional.  Looking first at
the upper limit, if $J_{12}$ is quantized we have
	\begin{equation}
	J_{12} = j_{12}+\frac{1}{2} < J_{12,{\rm max}}
	=\min(j_1+j_2+1,j_3+j_4+1).
	\label{j12upper}
	\end{equation}
Given the other conditions on $j_{12}$, this implies that the maximum
quantized value of $J_{12}$ is $J_{12,{\rm max}}-\frac{1}{2}$.
Similarly, we find that the minimum quantized value of $J_{12}$ is
$J_{12,{\rm min}}+\frac{1}{2}$.  The quantized values of $J_{12}$ are
separated from the maximum and minimum classical values (for given
$J_r$, $r=1,\ldots,4$) by a margin of $\frac{1}{2}$, and are spaced in
integer steps. These rules imply
	\begin{equation}
	j_{12,{\rm max}}=J_{12,{\rm max}}-1,
	\qquad j_{12,{\rm min}}=J_{12,{\rm min}}.
	\label{J12j12maxmin}
	\end{equation}
They also imply the bounds (\ref{j12j23bounds}) on the quantum number
$j_{12}$.  Similar results apply to the $B$-manifold and the quantized
values of $J_{23}$.

\begin{figure}[htb]
\begin{center}
\scalebox{0.5}{\includegraphics{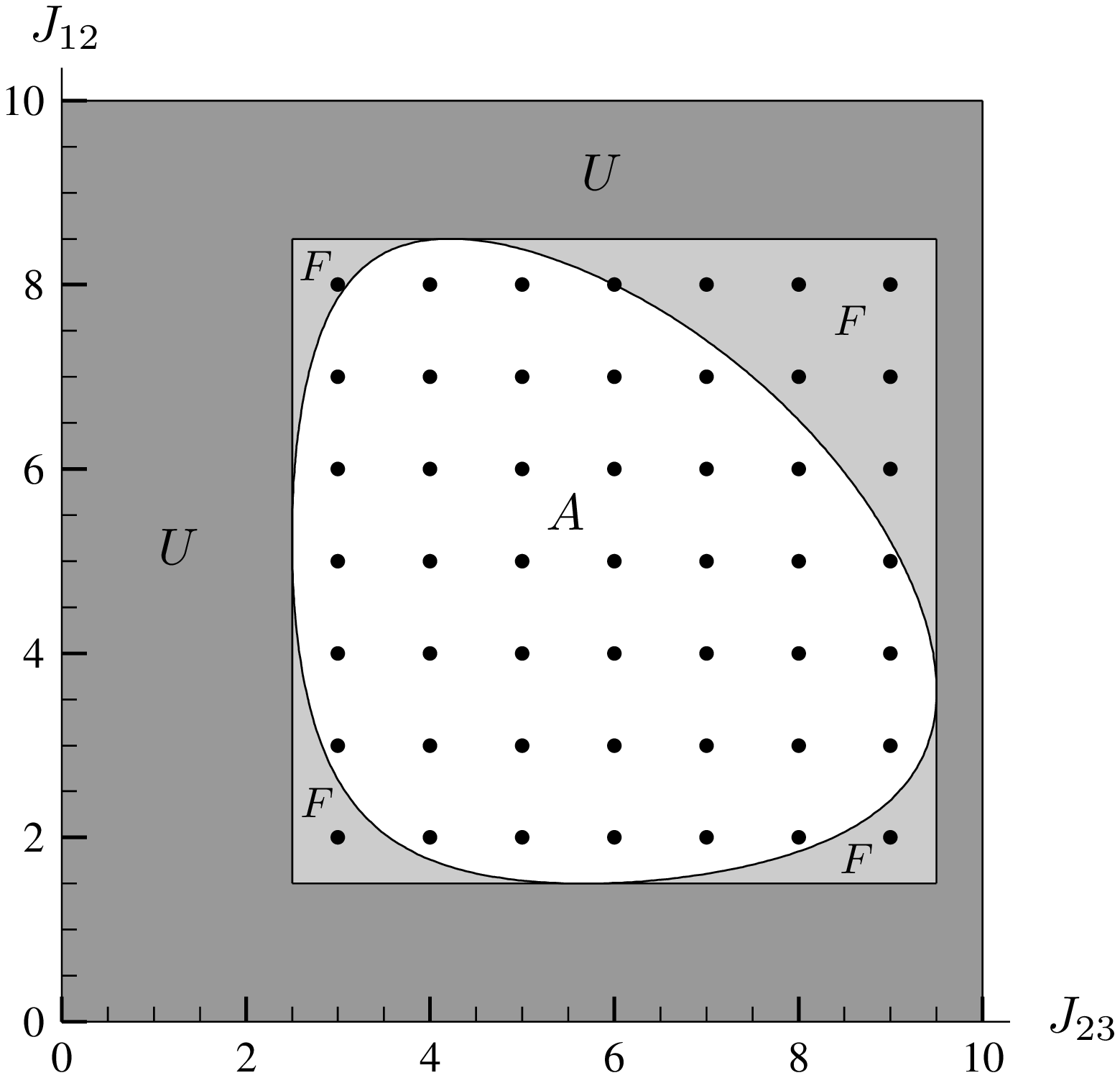}}
\end{center}
\caption[square]{\label{square} The $J_{12}$--$J_{23}$ plane for
$(J_1,J_2,J_3,J_4)=(5,\frac{7}{2},6,\frac{13}{2})$ (all quantized
values). Region $U$ is where either the $A$- or $B$- manifold does not
exist.  Inside the square both manifolds exist; $A$ and $F$ are the
classically allowed and classically forbidden regions, respectively.
Caustic curve is shown.}
\end{figure}

Figure~\ref{square} is a numerical example of these quantization rules.
The square in the figure is given by the bounds (\ref{J12range}) and
(\ref{J23range}), while the spots are the quantized values of $J_{12}$
and $J_{23}$.  The latter are related to the usual quantum numbers by
	\begin{equation}
	J_{12}=j_{12}+\frac{1}{2}, \qquad
	J_{23}=j_{23}+\frac{1}{2}.
	\label{J12J23quant}
	\end{equation}
The spots form a square array because $\braket{j_{23}}{j_{12}}$ is a
square matrix.  The size of the matrix (the number of rows or columns
of spots) is $\dim\ZS$, given by (\ref{dimZS}).  Other features of
this figure will be explained later.

Note that the minimum value of $J_r$, $r=1,\ldots,4$ and of $J_{12}$
on any quantized $A$-manifold is $\frac{1}{2}$, so the corresponding
vectors always have a positive length.  In addition, the triangles
1-2-12 and 3-4-12 always have a positive area.  Similar conclusions
apply to the quantized $B$-manifolds.

\subsection{Semiclassical Eigenvalues}
\label{sceigenvalues}

Once the classical manifolds are quantized, we find the semiclassical
approximations to the eigenvalues of the operators in the $A$- or
$B$-lists by evaluating the Weyl symbols of those operators on the
classical manifold.  Doing this for the operators $\Ihat_r$,
$r=1,\ldots,4$, we find the eigenvalue of $\Ihat_r$ is $j_r$; this is
the exact answer, something that was to be expected since $\Ihat_r$ is
a quadratic polynomial in the fundamental $\xhat$'s and $\phat$'s of
the system, and Weyl quantization of such operators is exact.  See
Littlejohn (1986) for more on classical and quantum quadratic
polynomials.  If we now compute the eigenvalue of $\Jvechat_r^2$ by
using the operator identity $\Jvechat_r^2 = \Ihat_r(\Ihat_r+1)$,
naturally we get the exact answer, $j_r(j_r+1)$.  On the other hand,
if we evaluate the Weyl symbol of the operator $\Jvechat_r^2$ on the
quantized level set, we get (according to the table) the eigenvalue
$(j_r+1/2)^2-3/8 = j_r(j_r+1) -1/8$, with an error of $-1/8$.  There
is an error because $\Jvechat_r^2$ is a quartic polynomial in the
fundamental $\xhat$'s and $\phat$'s of the system, so Weyl
quantization is not exact.  The error is, however, of relative order
$\hbar^2$, that is, $1/j^2$, which is the error expected in lowest
order semiclassical approximations.  The operators $\Jvechat_r^2$ do
not appear in the $A$- or $B$-lists, but operators $\Jvechat_{12}^2$
and $\Jvechat_{23}^2$ do, and here again there is an error at order
$1/j^2$.  Moreover, in this case there are no operators $\Ihat_{12}$
or $\Ihat_{23}$ which could be used to obtain the exact eigenvalues.
This is a drawback of the $4j$-model in comparison to the $8j$- or
$12j$-models, where such operators exist.  It does not, however,
change any of the subsequent analysis, which depends only on using the
quantized manifolds to carry out the stationary phase calculation.

\subsection{Time Reversal}
\label{trevclass}

The action of the antilinear time reversal operator $\Thetahat$ on a
carrier space $\CS_j$ can be defined by
	\begin{equation}
	\Thetahat\ket{jm} = (-1)^{j-m} \ket{j,-m}
	\label{Thetahatdef}
	\end{equation}
(Messiah 1966).  This is equivalent to $\Thetahat=K_1^{-1} \circ G$,
where $G$ is the antilinear metric or map of Hermitian conjugation,
and $K_1^{-1}$ is defined in Sec.~\ref{brastokets}.  The composition
of the antilinear $G$ with the linear $K_1^{-1}$ is the antilinear
time reversal map $\Thetahat$.  The map $\Thetahat$ is easily extended
to the full Schwinger Hilbert space $\SS$ and tensor products thereof
such as $\HS_{4j}$.

Classically the antilinear $\Thetahat:\HS_{4j}\to\HS_{4j}$ corresponds to an
antisymplectic map $\Theta :\Phi_{4j }\to\Phi_{4j}$.  In the complex
coordinates, its action on all four spinors is given by
	\begin{equation}
	\Theta \left(\begin{array}{c}
	z_{1r} \\ z_{2r}
	\end{array}\right)
	=\left(\begin{array}{c}
	-\zbar_{2r} \\ \zbar_{1r}
	\end{array}\right),
	\label{classThetadef}
	\end{equation}
that is, $\Theta:z_r \mapsto v \zbar_r$, where $z_r\in\Complexes^2$
and where 
	\begin{equation}
	v=\exp(-i\sigma_2\pi/2)=\left(\begin{array}{cc}
	0 & -1 \cr
	1 & 0
	\end{array}\right).
	\label{vdef}
	\end{equation}
We recall that in quantum mechanics, it is time reversal, not parity,
that reverses the direction of angular momenta.  At the classical
level, this means that $\Jvec_r\bigl (\Theta(x )\bigr)= -\Jvec_r(x)$,
where $x\in\Phi_{4j}$.

Time reversal can be projected via $\pi$ onto $\Lambda_{4j}$, where
its effect on the coordinates is $\Theta:\Jvec_r \mapsto -\Jvec_r$. It
is an anti-Poisson map on $\Lambda_{4j}$.  

\section{Intersections and Actions}
\label{intact}

In this section we consider the intersections of the $A$- and
$B$-manifolds, assuming that $J_r>0$, $r=1,\ldots,4$ are given.  For
now we treat this as a classical problem in which the $J$'s (including
$J_{12}$ and $J_{23}$) are continuous variables, but we note that if
$J_r$, $r=1,\ldots,4$ are quantized then they are automatically
positive.  The motivation, however, is to find the stationary phase
points of the scalar product (\ref{6jme}), which are the intersections
of the quantized manifolds.

\subsection{Classically allowed and forbidden regions}
\label{classallowforbid}

A simple analogy will help to understand the results.  Consider a
one-dimensional harmonic oscillator, $H=(1/2)(x^2+p^2)$ (classical or
quantum).  For a given value of the energy $E$, the classically
allowed region is the interval of the $x$-axis between the turning
points, given by $x=\pm\sqrt{2E}$, while the classically forbidden
region is outside this interval.  The classically allowed region can
also be defined as the region of the $x$-axis where the two curves in
phase space, $x={\rm const}$ and $H=E$, have intersections.  These
curves are level sets of the observables $x$ and $H$, which appear on
the two sides of the matrix element when we write the energy
eigenfunction as $\psi_E(x) =\braket{x}{E}$.  Inside the classically
allowed region the intersections between the two curves consist of two
points, with opposite momentum values.  These are related by
time reversal ($p\to -p$).  In the classically forbidden region the
two curves have no real intersections, but if we complexify phase
space and the curves $x={\rm const}$ and $H=E$ (maintaining real
contour values $x$ and $E$), then they do have complex intersections
that are related to the exponentially decaying wave function in the
classically forbidden region.

\begin{figure}[htb]
\begin{center}
\scalebox{0.5}{\includegraphics{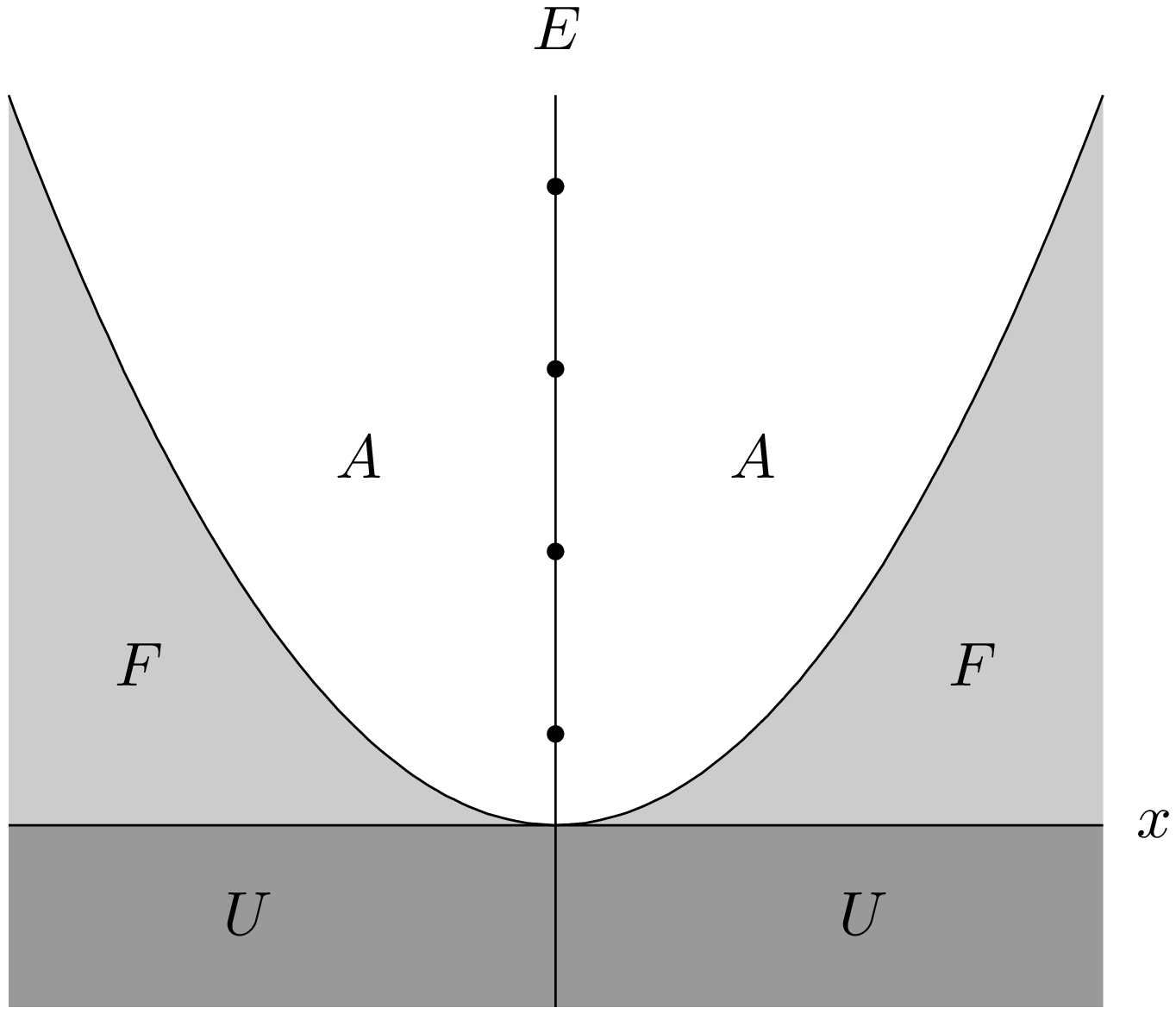}}
\end{center}
\caption[horegions]{\label{horegions} Regions of the $x$-$E$ plane 
for the 
harmonic oscillator: $UU$ where curves $H=E$ do not exist, and $AA$
and $FF$, the classically allowed and forbidden regions,
respectively. The parabola is the caustic curve.}
\end{figure}

We can view the classically allowed and forbidden regions in the
$x$-$E$ plane, in which both $x$ and $E$ are variables.  See
Fig.~\ref{horegions}.  The darkly shaded region in the figure, labeled
$UU$, is the region $E<0$, for which the level set $H=E$ does not
exist.  Above the line $E=0$, both level sets, $x={\rm const}$ and
$H=E$ exist.  The unshaded region, labeled $AA$, is the classically
allowed region, while the lightly shaded region, labeled $FF$, is the
classically forbidden region.  The quantized values of the energy
($E=n+\frac{1}{2}$) are indicated as spots on the $E$-axis.  There is
only a one-dimensional array of spots because the observable $x$ has a
continuous spectrum.  The parabola $E=x^2/2$ is the caustic curve,
separating the classically allowed from the classically forbidden
region.

\subsection{Intersections of the $A$- and $B$-manifolds}
\label{ABintersect}

Similarly, for positive $J_r$, $r=1,\ldots,4$, either the $A$- or
$B$-manifold does not exist if $J_{12}$ or $J_{23}$ lies outside the
bounds (\ref{J12range}) and (\ref{J23range}).  Those inequalities
define a square region of the $J_{23}$-$J_{12}$ plane, as illustrated
in Fig.~\ref{square}.  A figure like this was first given by Neville
(1971).  The darkly shaded region outside the square, labeled $U$, is
where either the $A$- or $B$-manifold does not exist.  In the interior
of the square both manifolds exist and have full dimensionality.
On the boundary of the square they exist, but have less than full
dimensionality, since some triangle inequality is saturated.  If the
$A$- and $B$-manifolds exist and intersect, then we are in the
classically allowed (unshaded) region in the interior of the square.
If they exist but do not intersect, then we are in the classically
forbidden (lightly shaded) region.  The caustic curve is the oval
curve in the figure, separating the classically allowed from the
classically forbidden regions.  Notice that it touches the square
boundary at four points.

If we fix a value of $J_{12}$, we can regard the classically allowed
and forbidden regions as intervals of the $J_{23}$-axis.  The interval
within which the $B$-manifold exists is given by (\ref{J23range});
this is the interval of allowed $J_{23}$ values, given $J_r$,
$r=1,\ldots,4$.  The classically allowed region, on the other hand, is
the interval of allowed $J_{23}$ values, given $J_r$, $r=1,\ldots,4$
and $J_{12}$ (equivalent to the statement that the manifolds
intersect).  Since this is a more restrictive condition, the
classically allowed region must be a subset of the interval
(\ref{J23range}).  Moreover, since the $A$-manifold is connected, the
subset must be a connected interval, since this subset is the range of
$J_{23}$ values that occur on a given $A$-manifold.  This is just what
we see in Fig.~\ref{square}: for most values of $J_{12}$, the interval
inside the square consists of a classically allowed region, surrounded
on both sides by classically forbidden regions, outside of which are
the regions in which manifolds of the given $J_{23}$ values do not
exist.  The bounds of the classically allowed region are given by the
roots of (\ref{CMdet}), that is, they correspond to tetrahedra of zero
volume.  The caustic curve in Fig.~\ref{square} is the contour of
$\det M=0$, where $M$ is the Cayley-Menger determinant (\ref{CMdet}).

The volume vanishes if the tetrahedron is flat.  For most values of
the parameters, this does not require that any of the triangular faces
have zero area.  But for certain values of $J_{12}$, the volume
vanishes at a place where the face 2-3-23 has zero area (thus,
$\Jvec_2$, $\Jvec_3$ and $\Jvec_{23}$ are linearly dependent).  This
is the point where the caustic curve touches the boundary of the
square on the left or right.  Similarly, the caustic curve touches the
boundary of the square on the top and bottom of the square, where not
only does the volume vanish, but also the area of triangle 1-2-12.

The caustic curve never lies outside the bounds (\ref{J23range})
defined by the triangle inequalities.  Biedenharn and Louck (1981)
appear to claim the contrary, but there is the question of whether one
is talking about the classical or quantum triangle inequalities.  That
is, the caustic curve does pass outside the bounds given by the square
array of quantized spots.  It also seems to us that the interpretation
of Biedenharn and Louck of Fig.~6 from Ponzano and Regge is incorrect.

If the values of $J_r$, $r=1,\ldots,4$ are quantized, then we can plot
the quantized values of $J_{12}$ and $J_{23}$ as a square array of
spots, as in the figure.  See other comments on this array in
Sec.~\ref{BSquant}.  In Littlejohn and Yu (2009) we incorrectly stated
that the quantized values of $J_{12}$ and $J_{23}$ can fall exactly on
a caustic, citing the theory of Brahmagupta quadrilaterals (Sastry
2002).  That theory shows that plane quadrilaterals with integer sides
and integer diagonals exist, that is, flat tetrahedra with all
integer edges.  However, to represent a $6j$-symbol, the sums of the
integers around the faces of triangles must be odd, and this condition
cannot be met.  (The integers divided by 2 are the values of
$J_r=j_r+\frac{1}{2}$, and the sum of the $j_r$ around the faces must
be an integer.)  A correct proof of the nonexistence of flat, quantized
tetrahedra, credited to Adler, is given in a brief citation by Ponzano
and Regge (1968).

We now examine the intersections of the $A$- and $B$-manifolds in
greater detail.  First, the $A$- and $B$-manifolds (assumed to exist)
have an intersection in $\Phi_{4j}$ iff their projections onto
$\Lambda_{4j}$ intersect.  Furthermore, the intersection in
$\Phi_{4j}$ is the lift of the intersection of the projections in
$\Lambda_{4j}$, with a $T^4$ fiber over every point.  These statements
follow from the fact that over every point $x\in\Lambda_{4j}$ there is
a 4-torus fiber in $\Phi_{4j}$, and that if $x$ lies on the
intersection of the projections of the $A$- and $B$-manifolds, then
the 4-torus belongs to both the $A$- and $B$-manifolds in $\Phi_{4j}$.
This is the same logic used in I under similar circumstances.

To find the intersections of the projections in $\Lambda_{4j}$ we
require four vectors $\Jvec_r$, $r=1,\ldots,4$ that satisfy
	\begin{equation}
	\eqalign{
	&|\Jvec_r| = J_r, \qquad
	\sum_{r=1}^4 \Jvec_r=0, \\
	&|\Jvec_1+\Jvec_2| = J_{12}, \qquad
	|\Jvec_2+\Jvec_3| = J_{23},}
	\label{ABconditions}
	\end{equation}
for the given values of $J_r>0$, $r=1,\ldots,4$ and of $J_{12}$ and
$J_{23}$.  A nice way of constructing these vectors is given in
Appendix~A of Littlejohn and Yu (2009), which uses the singular value
decomposition of the Gram matrix of dot products associated with the
Cayley-Menger determinant (\ref{CMdet}).  This method not only gives
an explicit solution for these vectors at any point in the classically
allowed region, it also shows that they are unique to within the
overall action of $O(3)$.  It is obvious in any case that if
$\Jvec_r$, $r=1,\ldots,4$ is a solution of (\ref{ABconditions}), then
so is $S\Jvec_r$ for any $S\in O(3)$.  This method was generalized to
the $9j$-symbol in Haggard and Littlejohn (2010).

The group $O(3)$ is conveniently decomposed into proper rotations in
$SO(3)$ and spatial inversion, which is time reversal in the present
case.  It is a basic fact of geometrical figures in $\Reals^3$ that
spatial inversion is not equivalent to any proper rotation unless the
figure is planar.  Another fact is that the orbit of a geometrical
figure under $SO(3)$ is diffeomorphic to $SO(3)$ itself, unless the
dimension of the figure is $\le 1$.  These issues are discussed by
Littlejohn and Reinsch (1995, 1997) in the context of molecular
configurations.  They imply that except at the caustics, where the
tetrahedron is flat, two tetrahedra related by time reversal are not
related by any proper rotation.  Therefore, except at the caustics,
the solution set of (\ref{ABconditions}) in $\Lambda_{4j}$ consists of
two disconnected subsets, each diffeomorphic to $SO(3)$, related by
time reversal.  Each subset consists of tetrahedra of nonzero volume
related by proper rotations.  At a generic point of the caustic curve,
where the tetrahedron is flat but still 2-dimensional, the two subsets
merge into one, which is still diffeomorphic to $SO(3)$.

\begin{figure}[htb]
\begin{center}
\scalebox{0.5}{\includegraphics{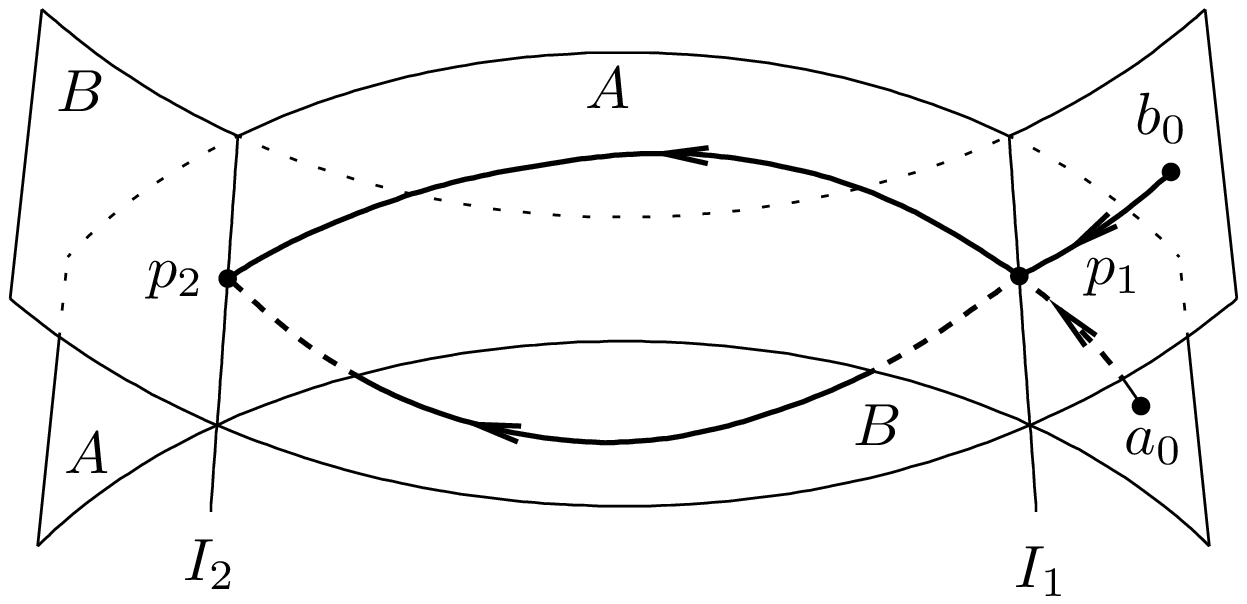}}
\end{center}
\caption[ABintsect]{\label{ABintsect} A schematic illustration of 
the intersection of the two 8-dimensional $A$- and $B$-manifolds in
$\Phi_{4j}$.  The two intersections $I_1$ and $I_2$ are 7-dimensional,
and are related by time reversal.  Paths for computing action
functions $S_A$ and $S_B$ relative to initial points $a_0$ and $b_0$
on the two manifolds are shown.}
\end{figure}

The intersections in $\Phi_{4j}$ are the lifts of these intersections
in $\Lambda_{4j}$.  Therefore, except at the caustics, the
intersection of the $A$- and $B$-manifolds consists of two
disconnected subsets, related by time reversal, where each subset is a
$T^4$-bundle over $SO(3)$.  These subsets are 7-dimensional, so the
$A$- and $B$-manifolds, which are 8-dimensional, intersect in two
7-dimensional submanifolds.  The situation can be visualized as in
Fig.~\ref{ABintsect}, where $A \cap B = I_1 \cup I_2$ and where $I_1$
and $I_2$ are the connected intersection sets, related by $\Theta$
(see (\ref{classThetadef})).  Each intersection set is an the orbit of
the group $U(1)^4 \times SU(2)$, where $U(1)^4$ represents the phases
of the four spinors and $SU(2)$ is the diagonal action (thus, the
group is generated by $I_r$, $r=1 ,\ldots,4$ and $\Jvec_{\rm tot}$).

The isotropy subgroup of this group is ${\mathbb Z}_2$, generated by
element $(2\pi,2\pi, 2\pi,2\pi, -1)$, in coordinates $(\psi_1, \psi_2,
\psi_3, \psi_4 ,u)$ for group $T^4\times SU(2)$, where $u\in SU(2)$.
The volume of intersection manifold $I_1$ or $I_2$ with respect to the
measure $d\psi_1 \wedge d\psi_2 \wedge d\psi_3 \wedge d\psi_4 \wedge
du$ is
	\begin{equation}
	V_I = \frac{1}{2} (4\pi)^4 \times 16\pi^2 = 2^{11} \pi^6.
	\label{volumeI}
	\end{equation}

An interesting aspect of the method of Appendix~A of Littlejohn and Yu
(2009) for finding the vectors $\Jvec_r$, $r=1,\ldots,4$ is that it
also works in the classically forbidden region, where it produces
complex 3-vectors that satisfy (\ref{ABconditions}).  These solutions
are determined modulo the action of spatial inversion and
$SO(3,\Complexes)$.  In fact, as discussed in that reference, the
vectors can be chosen so that two components are real and one purely
imaginary, so that the symmetry group of the solution set is the
Lorentz group $SO(2,1) \subset SO(3,\Complexes)$.  Roberts (1999)
and others have referred to the tetrahedra in the classically
forbidden region as living in $\Reals^3$ with a Minkowski metric,
while those in the classically allowed region live in $\Reals^3$ with
a Euclidean metric.  This is a correct interpretation of the situation
for the $6j$-symbol, but we do not think it is appropriate for
generalizations to other spin networks.  For example, in the
$9j$-symbol (Haggard and Littlejohn 2010) the vectors in the
classically forbidden region cannot be chosen so that two components
are real, instead they belong to $\Complexes^3$, and the symmetry
group of the solution set is $SO(3,\Complexes)$, not some Lorentz
subgroup thereof.  This simply means that to explore the complex
Lagrangian manifolds, complex Euler angles must be used when following
the Hamiltonian flows generated by $\Jvec_{\rm tot}$.  The
complexified $A$- and $B$-manifolds in the $6j$-symbol support the
asymptotic forms given by Ponzano and Regge (1968) in the classically
forbidden region.

\subsection{Actions and phases on the $A$- and $B$-manifolds}
\label{ABactions}

Let the $x$-space wave functions associated with states $\ket{A}$ and
$\ket{B}$ of (\ref{4jABdefs}) be denoted $\psi_A(x) =\braket{x}{A}$
and $\psi_B(x) =\braket{x}{B}$, where $x\in\Reals^8$.  The
semiclassical approximations to these wave functions involve phases
$e^{iS_A(x)}$ and $e^{iS_B(x)}$, where actions $S_A(x)$ and $S_B(x)$
are integrals of $p\,dx =\sum_{r\mu} p_{r\mu} \, dx_{r\mu}$ from
some initial points on the two manifolds to some final point.  The
initial points on the $A$- and $B$-manifolds are denoted $a_0$ and
$b_0$, respectively, in Fig.~\ref{ABintsect}; they determine the
overall phases of the states $\vert A\rangle$ and $\vert B\rangle$.

As explained in I, the branches of the stationary phase evaluation of
the matrix element $\langle B \vert A \rangle$ are associated with the
intersections of the $A$- and $B$-manifolds, in this case the
manifolds $I_1$ and $I_2$, so the asymptotic form of the $6j$-symbol
has two branches.  Moreover, the phase associated with each branch is
$S_A - S_B$ evaluated on the corresponding intersection manifold (see,
for example, (I.\Iabmatrixelement) or (I.\Iabmatrixelementone)), and
is independent of where we evaluate it on that manifold.
Figure~\ref{ABintsect} illustrates two points $p_1$ and $p_2$, on
intersection manifolds $I_1$ and $I_2$, respectively, with paths that
may be used for computing the actions $S_A$ and $S_B$.

In the following we shall be interested in the relative phase between
the two branches.  Let us define
	\begin{equation}
	S_1 = S_{A1}-S_{B1}, \qquad S_2 = S_{A2} - S_{B2},
	\label{S1S2defs}
	\end{equation}
and
	\begin{equation}
	S=S_2-S_1=\int_{a_0}^{p_2} -\int_{b_i}^{p_2} - \int_{a_0}^{p_1} 
	+ \int_{b_0}^{p_1} p\,dx = \oint p \, dx,
	\label{relphase}
	\end{equation}
where the final integral is taken along the path that goes from $p_1$
to $p_2$ along the $A$-manifold and then back to $p_1$ along the
$B$-manifold.  The relative phase is independent of the initial points
$a_0$ and $b_0$, and is moreover a symplectic invariant.  The relative
phase is easier to determine than the absolute phases of either
branch, which are related to the overall phase convention for the
$6j$-symbol.  In computing the relative phase, we note that the loop
integral in (\ref{relphase}) can be evaluated with respect to any
symplectic 1-form, such as the complex one used in I (see
(I.\Izaction)),
	\begin{equation}
	\oint \sum_{r\mu} p_{r\mu} \, dx_{r\mu} =
	\Im \oint \sum z_{r\mu} \, d\zbar_{r\mu}.
	\label{1forms}
	\end{equation}
The loop integral can be transformed by Stokes' theorem into an
integral of the symplectic form over the enclosed area, since on
$\Phi_{4j}={\mathbb C}^8$ all cycles are boundaries.

\subsection{Closing the loop in $\Lambda_{4j}$}
\label{closeloopAMS}

\begin{figure}[htb]
\begin{center}
\scalebox{0.5}{\includegraphics{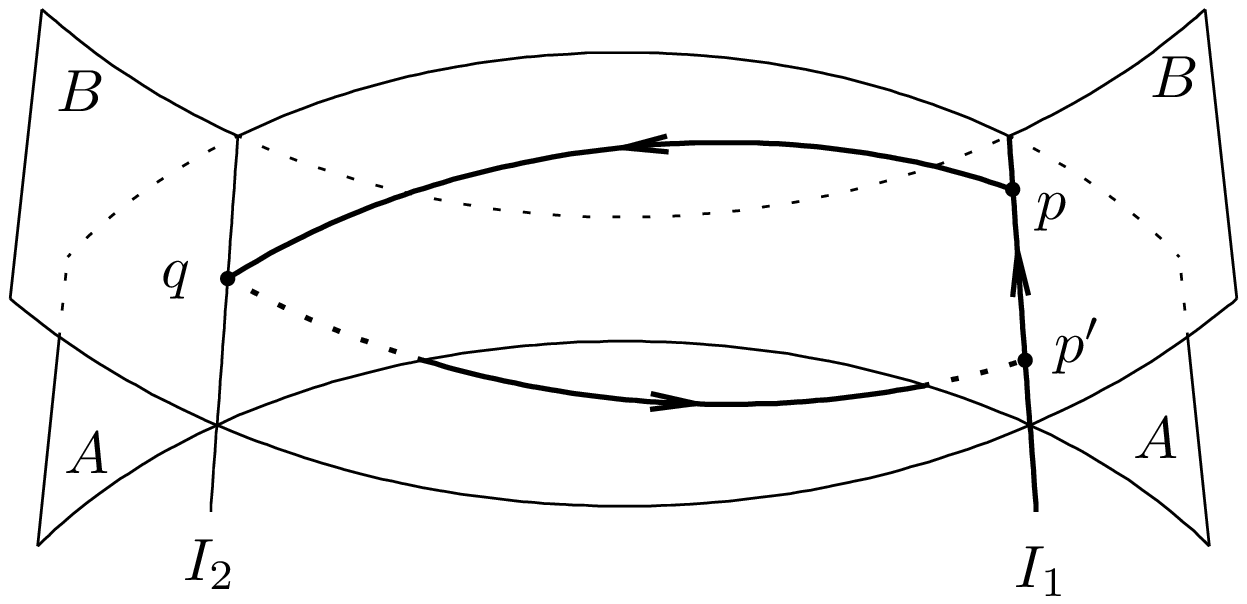}}
\end{center}
\caption[ABintsect1]{\label{ABintsect1} The $\Jvec_{12}^2$-flow takes 
us from a point $p$ of intersection manifold $I_1$ along the
$A$-manifold to point $q$ of manifold $I_2$; and then the
$\Jvec_{23}^2$-flow takes us back to point $p'$ of $I_1$.  To close
the loop it is necessary to connect $p'$ to $p$ inside $I_1$.}
\end{figure}

We shall construct the closed loop giving the relative phase between
the branches according to (\ref{relphase}) by following the
Hamiltonian flows of various observables.  Let us define the signed
volume of the tetrahedron as $V=(1/6) \Jvec_1 \cdot(\Jvec_2 \times
\Jvec_3)$, and let us take manifold $I_1$ to be the one on which
$V<0$, so that $V>0$ on $I_2$.  Time reversal changes the sign of the
volume when mapping $I_1$ into $I_2$ and vice versa.  Let us start at
a point $p$ of $I_1$, as in Fig.~\ref{ABintsect1}.  Then by following
the $\Jvec_{12}^2$-flow we trace out a path that takes us along the
$A$-manifold to a point $q$ of $I_2$.  We cannot use any of the other
seven observables defining the $A$-manifold for this purpose, namely,
$(I_1,I_2,I_3,I_4,\Jvec)$, since their flows confine us to the
intersection manifold $I_1$.

\begin{figure}[htb]
\begin{center}
\scalebox{0.7}{\includegraphics{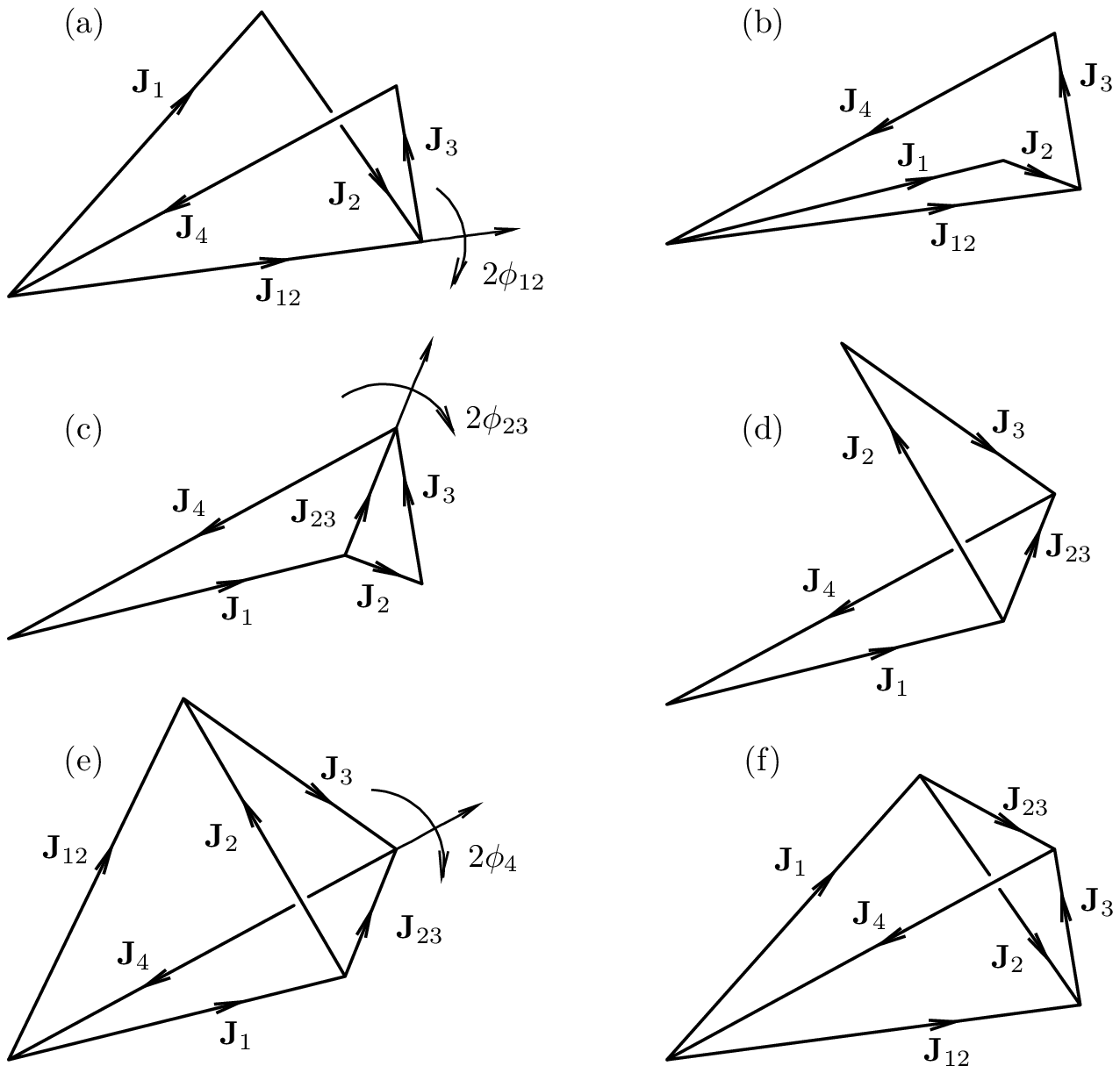}}
\end{center}
\caption[cycle]{\label{cycle} A cycle of rotational transformations that
takes a tetrahedron in ${\mathbb R}^3$ back into itself.}
\end{figure}

We see that the $\Jvec_{12}^2$-flow actually does take us to manifold $I_2$ by
considering the projection of the path $p$-$q$ in
Fig.~\ref{ABintsect1} onto $\Lambda_{4j}$.  We visualize the projected
path as a transformation applied to a set of four vectors in ${\mathbb
R}^3$ that define a tetrahedron.  The situation is illustrated in
Fig.~\ref{cycle}.  In part (a) of that figure, we have four vectors
$\Jvec_r$ that sum to zero, defining a tetrahedron of negative volume.
The lengths $J_r>0$, $r=1,\ldots,4$, $J_{12}$ and $J_{23}$ are assumed
to have the prescribed values, and vector $\Jvec_{12}$ is drawn (but
not $\Jvec_{23}$).  We take the point $p$ of Fig.~\ref{ABintsect1} to
lie on the $T^4$ fiber above this tetrahedron.  The
$\Jvec_{12}^2$-flow rotates the 1-2-12 triangle about the 12-axis by
the right-hand rule while leaving the 3-4-12 triangle fixed (see
(\ref{J12sqflow})), that is, it rotates the 1-2-12 triangle into
the foreground.  Let the angle of rotation be $2\phi_{12}$, where
$\phi_{12}$ is the interior dihedral angle of the tetrahedron along
edge 12 in its original configuration.  This brings triangle 1-2-12
through triangle 3-4-12 to the opposite side, creating a new
tetrahedron with the same lengths (the new $J_{23}$ is the same as the
old one), hence the same dihedral angles, but with the opposite signed
volume.  The result is illustrated in part (b) of Fig.~\ref{cycle}, a
tetrahedron that is the projection of a point $q
\in I_2$ in $\Phi_{4j}$, as illustrated in Fig.~\ref{ABintsect1}.  Thus 
we see that the $\Jvec_{12}^2$-flow does take us from $I_1$ to $I_2$
along the $A$-manifold, as claimed.

Having reached point $q \in I_2$, we can go back to $I_1$ along the
$B$-manifold by following the $\Jvec_{23}^2$-flow, reaching point
$p'\in I_1$ as illustrated in Fig.~\ref{ABintsect1}.  The
transformation in $\Lambda_{4j}$ is illustrated in Fig.~\ref{cycle}.
Part (c) of that figure is the same as part (b), except that vector
$\Jvec_{23}$ is drawn and $\Jvec_{12}$ is suppressed.  The
$\Jvec_{23}^2$-flow rotates triangle 2-3-23 about the axis $\jvec_{23}
=\Jvec_{23} /J_{23}$, while leaving triangle 1-4-23 fixed.  Let the
angle of rotation be twice the interior dihedral angle along edge 23,
that is, $2\phi_{23}$, as illustrated in part (c) of Fig.~\ref{cycle}.
The result is part (d) of that figure, a tetrahedron in which the
volume has been inverted a second time, taking us back to the original
(negative) volume in part (a).  We arrive at point $p' \in I_1$, as in
Fig.~\ref{ABintsect1}.

It is clear that $p'$ is not the same as the original point $p$,
because if it were, the orientation of the tetrahedron in part (d) of
Fig.~\ref{cycle} would be the same as that in part (a).  Thus to
create a closed loop in $\Phi_{4j}$, we must follow some path in $I_1$
taking us from $p'$ to $p$, as in Fig.~\ref{ABintsect1}.  

We create this path from $p'$ to $p$ in $I_1$ in two steps.  First we
apply an $SU(2)$ transformation to all four spinors at point $p'$,
that is, a diagonal transformation, whose projection onto
$\Lambda_{4j}$ is an $SO(3)$ transformation of the tetrahedron in part
(d) of Fig.~\ref{cycle}, returning it to the original orientation in
part (a).  This is a proper rotation of all four vectors $\Jvec_r$,
that is, a rigid rotation of the entire tetrahedron, and it is the
final step in a cycle of rotations that transform the original
tetrahedron in part (a) into itself.  That figures (a) and (d) must be
related by some proper rotation is clear, since the lengths of the
sides are the same and the signed volume is the same.  In fact, it is
easy to see that the axis of the final rotation is $\Jvec_4$, since
this vector is left invariant by both the $\Jvec_{12}^2$- and
$\Jvec_{23}^2$-flows, and is the same in parts (a) and (d).  As it
turns out, the final rotation has axis $-\jvec_4 = -\Jvec_4/j_4$
(notice the minus sign) and angle $2\phi_4$, twice the internal
dihedral angle along edge 4.  This is illustrated in part (e) of
Fig.~\ref{cycle}, which is the same as part (d) except that all
vectors are drawn.  The effect of the final rotation is illustrated in
part (f), which is the same as part (a) except that all vectors are
drawn.

\subsection{Angle of the final rotation}
\label{finalangle}

\begin{figure}[htb]
\begin{center}
\scalebox{0.5}{\includegraphics{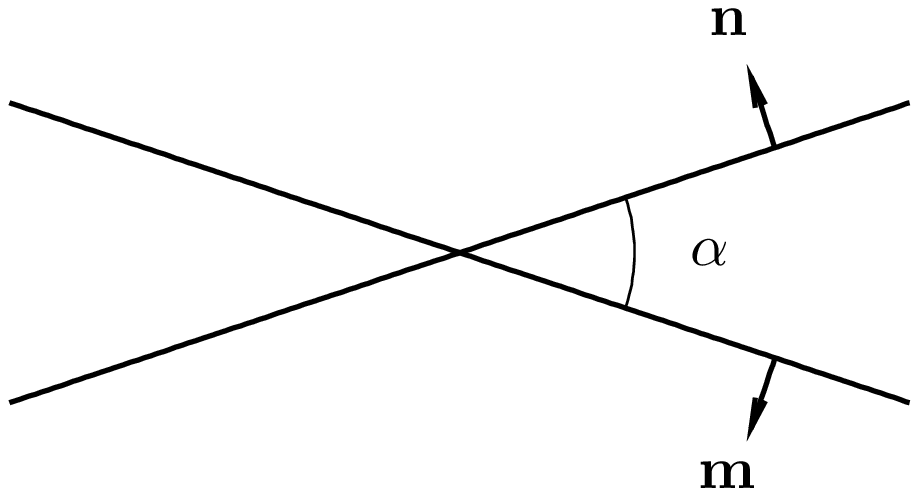}}
\end{center}
\caption{\label{2refl} The product of two reflections is a rotation.}
\end{figure}

To obtain the angle of the final rotation about axis $-\jvec_4$, we
use the fact that the product of two reflections is rotation.  Let a
reflection about a plane $P$ be $Q(P)$.  There are four ways to draw
the angle between the planes, one of which is denoted by $\alpha$ in
Fig.~\ref{2refl}.  For a given choice of dihedral angle $\alpha$, let
the outward pointing normals of the two planes be $\nvec$ and $\mvec$,
as in the figure.  Then
	\begin{equation}
	Q(\nvec)Q(\mvec) = R(\avec,2\alpha),
	\label{QQR}
	\end{equation}
where we use axis-angle notation for the rotation $R$ and where the
axis $\avec$ of the rotation is along the line of intersection of the
two planes and is given by
	\begin{equation}
	\avec=\frac{\nvec \times \mvec}{|\nvec\times\mvec|}
	=\frac{\nvec \times \mvec}{\sin\alpha}.
	\label{QQRaxis}
	\end{equation}
Although the angle $\alpha$, the normals $\mvec$ and $\nvec$, and the
axis $\avec$ depend on which of the four choices is made for the
dihedral angle, the resulting rotation does not (although it does
depend on the order in which the reflections are applied).

Let us denote the first rotation taking us from part (a) to part (b)
of Fig.~\ref{cycle} in axis-angle form by $R_{12}(\jvec_{12},
2\phi_{12})$, where the $12$-subscript on $R$ indicates that this
rotation only affects vectors $\Jvec_1$ and $\Jvec_2$.  It is obvious
from Fig.~\ref{cycle} that the effect of this rotation on all the
angular momentum vectors is to reflect them in the plane 3-4-12, that
is,
	\begin{equation}
	\Jvec'_r = Q(\hbox{\rm 3-4-12})\Jvec_r,
	\label{firstQ}
	\end{equation}
where the prime refers to the values of the vectors after the first
rotation and where $r=1,\ldots,4$.  This applies for $r=1,2$ because
the reflection in the 3-4-12 plane has the same effect as the
rotation, and for $r=3,4$ because the reflection does nothing to these
vectors and the rotation does not apply to them.  Note that the 3-4-12
plane is the same in parts (a) and (b) of the figure (the plane is not
affected by the first rotation).  

\begin{figure}[htb]
\begin{center}
\scalebox{0.5}{\includegraphics{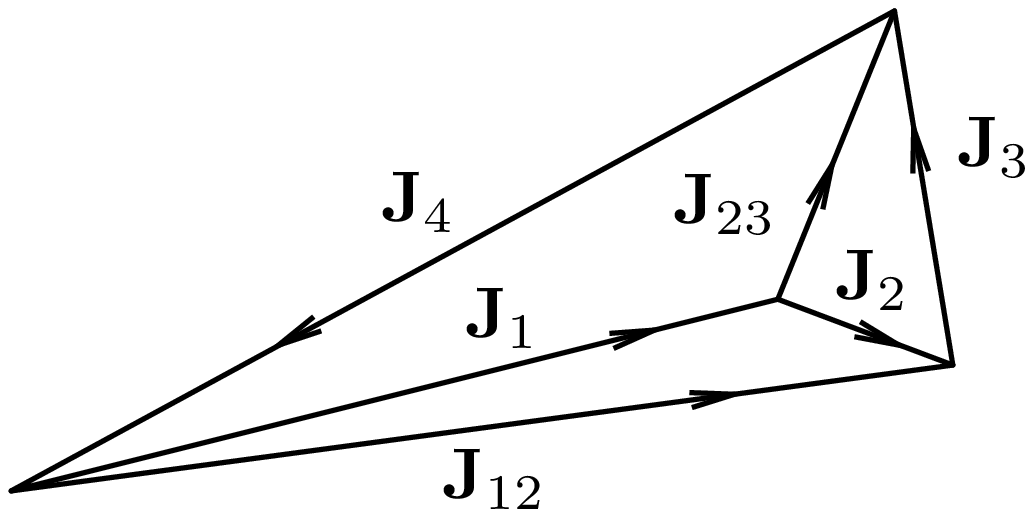}}
\end{center}
\caption{\label{cyclec} The plane of the first reflection is the back
plane, 3-4-12, while that of the second is the plane $1'$-4-$2'3$
(primes suppressed in the figure).}
\end{figure}

Similarly, we denote the second rotation, taking us from part (c) to
part (d) of Fig.~\ref{cycle}, by $R_{23}(\jvec'_{23},2\phi_{23})$,
where the prime on $\jvec'_{23}$ indicates that the axis is the
$23$-direction after the first rotation.  Then the effect of the
second rotation on all four $\Jvec'_r$ is the same as a reflection in
the plane $1'$-4-$2'3$, that is,
	\begin{equation}
	\Jvec''_r = Q(\hbox{\rm $1'$-4-$2'3$})\Jvec'_r.
	\label{secondQ}
	\end{equation}
The planes of the two rotations intersect in edge 4 of the
tetrahedron, which can be seen more clearly in Fig.~\ref{cyclec},
which is the same as part (c) of Fig.~\ref{cycle} except that all
vectors are drawn.  In Fig.~\ref{cyclec}, plane 3-4-12 is the back
plane, and is also the plane of the first reflection.  Plane
$1'$-4-$2'3$ (primes are omitted in the figure) is the plane of the
second reflection.  In comparison to (\ref{QQR}), if we identify
$\alpha$ with the interior dihedral angle $\phi_4$ at edge 4, then
$\mvec$ is the outward normal to plane 3-4-12, while $\nvec$ is the
outward normal to plane $1'$-4-$2'3$.  Their cross product is in the
direction $\jvec_4$, so we have
	\begin{equation}
	\Jvec''_r = R(\jvec_4,2\phi_4)\Jvec_r, \qquad r=1,\ldots,4.
	\label{R1R2}
	\end{equation}
This is the rotation taking us from part (a) to part (e) of
Fig.~\ref{cycle}; to undo that rotation, we apply $R(-\jvec_4,
2\phi_4)$ to pass from part (e) to part (f) of that figure.

To summarize the rotational history, we have applied the rotations
	\begin{equation}
	R(-\jvec_4,2\phi_4) R_{23}(\jvec'_{23},2\phi_{23})
	R_{12}(\jvec_{12},2\phi_{12})
	\label{3rots}
	\end{equation}
to the tetrahedron in part (a) of Fig.~\ref{cycle}, taking it through
a cycle of tetrahedra and returning it to its original shape and
orientation.  The corresponding $SU(2)$ rotations, with the same axes
and angles, are applied to point $p$ in Fig.~\ref{ABintsect1}, taking
us along a path $p \to q \to p' \to p''$.  Point $p''$ is not shown in
Fig.~\ref{ABintsect1}, but it is a point of $I_1$ that projects onto
the same tetrahedron as point $p$, since the projected path in
$\Lambda_{4j}$ is closed.  Points $p$ and $p''$ differ by the phases of
the four spinors, that is, by transformations generated by
$(I_1,I_2,I_3,I_4)$.  Thus, there is a final segment $p'' \to p$
needed to close the path in $\Phi_{4j}$, which runs along the $T^4$
fiber over the initial configuration in $\Lambda_{4j}$. 

\subsection{Closing the loop in $\Phi_{4j}$}
\label{closeloopLPS}

There are several ways to compute the final four phases, but we will
discuss just one.  We start with vector $\Jvec_1$.  The action of the
rotations (\ref{3rots}) on this vector can be written
	\begin{equation}
	R(-\jvec_4,2\phi_4)R(\jvec_{12},2\phi_{12})\Jvec_1 = \Jvec_1,
	\label{J1xfm}
	\end{equation}
where we omit the subscripts on the $R$'s because it is understood
that only vector $\Jvec_1$ is being acted upon, and where we omit the
middle rotation in (\ref{3rots}) since it does not act on $\Jvec_1$.
The product of the two rotations in (\ref{J1xfm}) is not the identity,
but it is a rotation about axis $\jvec_1$ since it leaves $\Jvec_1$
invariant.

\begin{figure}[htb]
\begin{center}
\scalebox{0.5}{\includegraphics{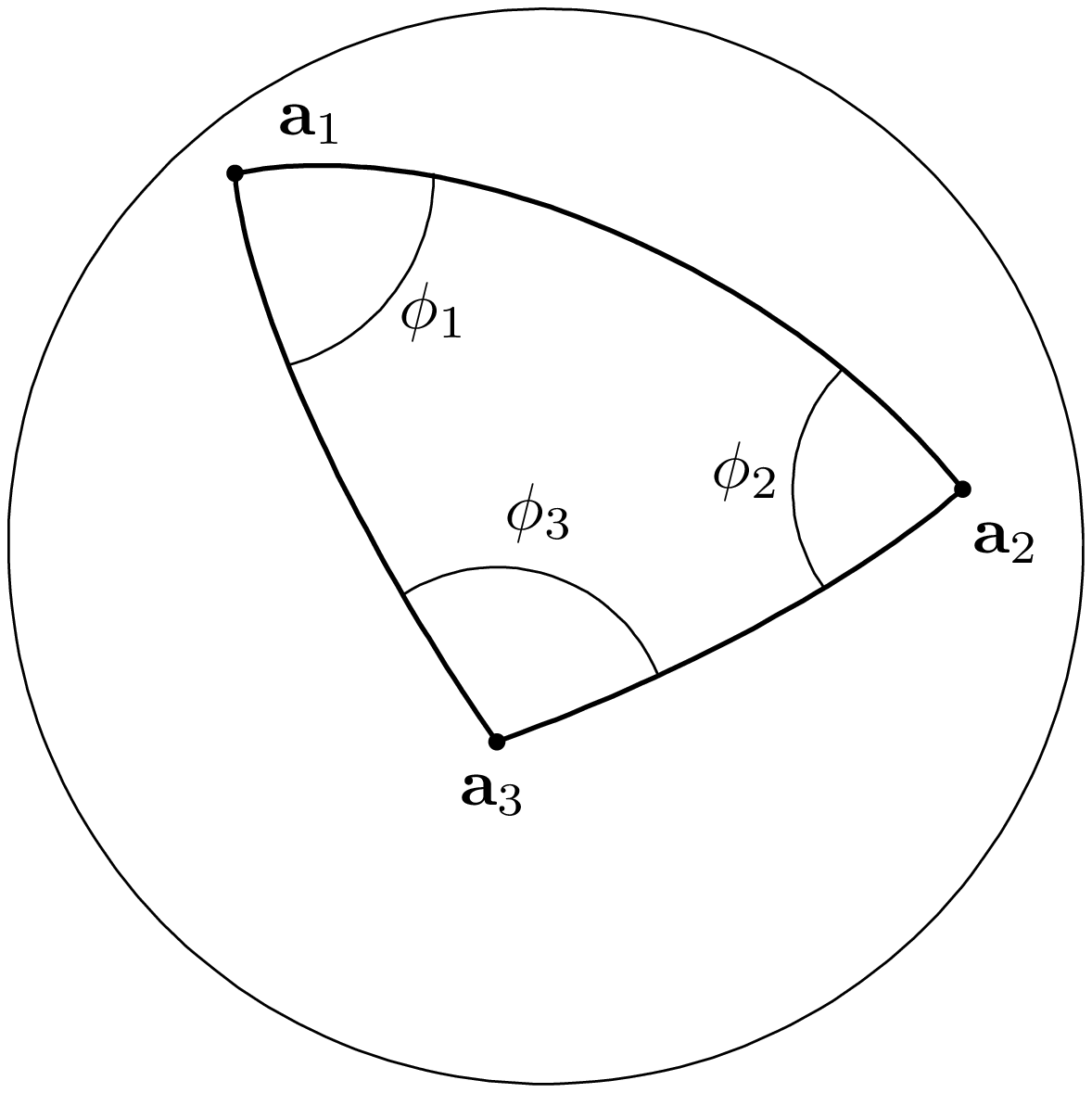}}
\end{center}
\caption{\label{hrformula} Axes $\avec_i$ and angles $\phi_i$, 
$i=1,2,3$ for the product of three rotations.}
\end{figure}

To find the angle of this rotation, we use the Rodrigues-Hamilton
formula (Whittaker 1960) for the product of two rotations in
axis-angle form. Let $\avec_i$, $i=1,2,3$ be three unit vectors, which
we can plot on the unit sphere as in Fig.~\ref{hrformula}.  We join
the three points on the unit sphere by arcs of great circles.  On
following the path $1\to 2\to 3\to 1$ we regard the region to our
right as the interior of the spherical triangle formed by the arcs.
This gives meaning to the interior angles of the triangle, labeled
$\phi_i$, $i=1,2,3$ in the figure.  On going from point $i$ to point
$i+1$ we can go either the long way or short way around the great
circle; the interior of the triangle and the definitions of the
interior angles depend on which way we go, but the formula is valid in
any case.  If we follow the arcs the short way around, we obtain a
spherical triangle such as that shown in Fig.~\ref{hrformula}.  Then
the formula of Rodrigues and Hamilton is
	\begin{equation}
	R(\avec_3,2\phi_3)R(\avec_2,2\phi_2)R(\avec_1,2\phi_1)=I.
	\label{prodrule}
	\end{equation}
The proof is obtained by using (\ref{QQR}) to write each rotation as a
product of reflections, that is, $R(\avec_1,2\phi_1)=Q(31)Q(12)$ and
cyclic permutations, where for example $Q(12)$ means reflection in the
plane defined by axes $\avec_1$ and $\avec_2$.

To apply (\ref{prodrule}) to the rotations in (\ref{J1xfm}) we imagine
the vertex of the original tetrahedron (part (a) of Fig.~\ref{cycle})
where edges 1, 4 and 12 meet as at the center of the sphere in
Fig.~\ref{hrformula}, and we identify axes $(\avec_1 ,\avec_2
,\avec_3)$ with $(\jvec_{12} ,-\jvec_4 ,\jvec_1)$.  Then the $\phi$'s
of Fig.~\ref{hrformula} become the interior dihedral angles of the
tetrahedron, and we have
	\begin{equation}
	R(\jvec_1,2\phi_1)R(-\jvec_4,2\phi_4)
	R(\jvec_{12},2\phi_{12})=I.
	\label{RloopJ1}
	\end{equation}
Thus, the product of the two rotations in (\ref{J1xfm}) is $R(
-\jvec_1, 2\phi_1)$.

The third rotation found in this manner can be regarded as a holonomy.
As the first two rotations in (\ref{J1xfm}) are applied to $\Jvec_1$,
that vector traces out a closed curve on $S^2$ which is composed of
the arcs of two small circles (the curve is a ``small lune'').  The
sphere in question can be regarded as $\Sigma_1$, the symplectic
manifold denoted $\Sigma$ in (\ref{hopfmaps}) (the 1 subscript
indicates that we are dealing with the first of the four copies of
$\Sigma$ in $\Sigma_{4j}$).  The two rotations themselves in
(\ref{J1xfm}) can be regarded as the lift of that curve on $S^2$ into
$SO(3)$.  The lift is an open curve starting at the identity in
$SO(3)$ and ending at the product of the two rotations shown in
(\ref{J1xfm}).  To close this curve in $SO(3)$, we apply the third
rotation shown in (\ref{RloopJ1}).

This closed curve in $SO(3)$ may then be lifted (a second time) into
$SU(2)$, by replacing each $SO(3)$ rotation by an $SU(2)$ rotation
with the same axis and angle, that is, the lift is specified by the
product $u(\jvec_1,2\phi_1) u(-\jvec_4,2\phi_4)
u(\jvec_{12},2\phi_{12})$.  This product is either $+1$ or $-1$,
depending on the homotopy class of the closed loop in $SO(3)$.  To
find this class, we continuously deform the tetrahedron (see part (a)
of Fig.~\ref{cycle}), bringing dihedral angle $\phi_{12}$ to zero, so
that the tetrahedron becomes flat.  At the end of this deformation,
$\phi_{12}=0$ and one of $\phi_1$ and $\phi_4$ is 0 and the other is
$\pi$.  Thus, the closed loop in $SO(3)$ becomes an element of the
noncontractible homotopy class of the homotopy group ${\mathbb Z}_2$
of $SO(3)$, so the product of the three $SU(2)$ matrices is $-1$.
Therefore to close the loop in $SU(2)$ we apply a final rotation by an
angle $-2\pi$, to obtain
	\begin{equation}
	u(\jvec_1,-2\pi)u(\jvec_1,2\phi_1) u(-\jvec_4,2\phi_4) 
	u(\jvec_{12},2\phi_{12})=+1.
	\label{uloopJ1}
	\end{equation}

The final rotation by angle $-2\pi$ could have been chosen about any
axis if all we wanted to do was to close the loop in $SU(2)$, but we
choose axis $\jvec_1$ for the following reason.  The first two spin
rotations in (\ref{uloopJ1}), when applied to a spinor in ${\mathbb
C}^2$ over vector $\Jvec_1$, produce another spinor that projects onto
$\Jvec_1$ again, that is, it differs from the initial spinor only by
an overall phase.  This overall phase is the $U(1)$ holonomy mentioned
above.  The final step in closing the loop in Fig.~\ref{ABintsect1} is
to follow the $I_r$-flows to adjust the phases of the four spinors, in
particular, we must follow the $I_1$-flow to adjust the phase of the
first spinor.  But the observable $\Jvec_1^2=I_1^2$ is a function of
$I_1$, so we can follow its flow just as well.  But $\Jvec_1^2$
generates a rotation about the direction $\Jvec_1$, that is, it is
equivalent to multiplying the first spinor by an $SU(2)$
transformation with axis $\jvec_1$ and some angle.  The angle required
is seen in (\ref{uloopJ1}): it is $2\phi_1-2\pi$.

We remark that the final rotation in (\ref{uloopJ1}) could have been
any angle $-2\pi+4n\pi$.  Adding a multiple of $4\pi$ to this angle is
equivalent to going around some closed contour on the $A$-manifold (in
fact, within $I_1$), which, when the manifolds are quantized and the
Maslov phase is taken into account, changes the relative
phase of the two branches by a multiple of $2\pi$.

Next we turn to vector $\Jvec_3$ and the phase needed to bring the
third spinor back its original value after the rotations
(\ref{3rots}).  The action of these rotations on $\Jvec_3$ is given by
	\begin{equation}
	R(-\jvec_4,2\phi_4) R(\jvec'_{23},2\phi_{23}) \Jvec_3 
	= \Jvec_3,
	\label{J3xfm}
	\end{equation}
where we omit the first one since it does not act on $\Jvec_3$.  The
product of the two rotations in (\ref{J3xfm}) is a third rotation
about axis $\jvec_3$, which can be computed with the help of
(\ref{prodrule}).  This time we identify the 3-4-23 vertex of part (c)
of Fig.~\ref{cycle}, seen more clearly in Fig.~\ref{cyclec}, with the
origin of the sphere in Fig.~\ref{hrformula}.  We use the version of
the tetrahedron after the first rotation (part (c)) since the middle
rotation in (\ref{J3xfm}) involves the rotated axis $\jvec'_{23}$.
Thus we find
	\begin{equation}
	R(-\jvec'_{23},2\phi_{23})R(\jvec_4,2\phi_4)
	R(-\jvec_3,2\phi_3)=I.
	\label{3rotsJ3}
	\end{equation}
Now using $R(-\avec,\alpha)=R(\avec,\alpha)^{-1}$ and taking the
inverse of (\ref{3rotsJ3}), we obtain
	\begin{equation}
	R(\jvec_3,2\phi_3) R(-\jvec_4,2\phi_4) 
	R(\jvec'_{23},2\phi_{23})=I,
	\label{RloopJ3}
	\end{equation}
which specifies a closed loop in $SO(3)$.  We find the homotopy class
of this loop by taking $\phi_{23} \to 0$, which makes one of $\phi_4$
and $\phi_3$ zero and the other $\pi$, so the loop in $SO(3)$ belongs
to the noncontractible class.  Thus, the lift into $SU(2)$ is not
closed, but we can close it by appending a final spin rotation
about axis $\jvec_3$ by angle $-2\pi$.  Thus the closed loop in
$SU(2)$ is specified by
	\begin{equation}
	u(\jvec_3,-2\pi) u(\jvec_3,2\phi_3) u(-\jvec_4,2\phi_4) 
	u(\jvec'_{23},2\phi_{23})=+1,
	\label{uloopJ3}
	\end{equation}
which when applied to the third spinor at point $p$ in
Fig.~\ref{ABintsect1} returns it to its initial value.  The final
$U(1)$ holonomy of spinor 3, after the application of the three
rotations (\ref{3rots}), is $2\phi_3 -2\pi$.

As for $\Jvec_2$, its cycle on the sphere is specified by
	\begin{equation}
	R(-\jvec_4,2\phi_4) R(\jvec'_{23},2\phi_{23})
	R(\jvec_{12},2\phi_{12}) \Jvec_2 = \Jvec_2.
	\label{J2xfm}
	\end{equation}
The product of the three rotations shown must be a rotation with axis
$\jvec_2$.  To find the angle, we first use (\ref{RloopJ3}) to obtain
	\begin{equation}
	R(-\jvec_4,2\phi_4) R(\jvec'_{23},2\phi_{23})
	= R(-\jvec_3,2\phi_3).
	\label{RprodJ2}
	\end{equation}
Substituting this into (\ref{J2xfm}) we obtain a product that we can
evaluate with the help of (\ref{prodrule}) and with reference to the
2-3-12 vertex of the original tetrahedron (part (a) or (f) of
Fig.~\ref{cycle}).  The result is
	\begin{equation}
	R(\jvec_2,2\phi_2)
	R(-\jvec_3,2\phi_3) R(\jvec_{12},2\phi_{12}) = I.
	\label{RprodJ2a}
	\end{equation}
Thus the closed loop in $SO(3)$ associated with the loop traced by
$\Jvec_2$ on the sphere is specified by 
	\begin{equation}
	R(\jvec_2,2\phi_2) R(-\jvec_4,2\phi_4) 
	R(\jvec'_{23},2\phi_{23}) R(\jvec_{12},2\phi_{12}) = I.
	\label{Rloop2}
	\end{equation}
To find the homotopy class of this loop we deform the tetrahedron into
a planar shape as before, and find that two of the four angles
$(\phi_2,\phi_4, \phi_{23}, \phi_{12})$ are 0 and two are $\pi$.  The
loop in $SO(3)$ thus becomes the product of two rotations with angles
$2\pi$, which belongs to the contractible homotopy class.  Therefore
the lift into $SU(2)$ is closed,
	\begin{equation}
	u(\jvec_2,2\phi_2) u(-\jvec_4,2\phi_4) 
	u(\jvec'_{23},2\phi_{23}) u(\jvec_{12},2\phi_{12}) = 1,
	\label{uloop2}
	\end{equation}
which when applied to the second spinor at point $p$ in
Fig.~\ref{ABintsect1} returns it to its initial value.  The final
$U(1)$ holonomy of spinor 2, after the application of the three
rotations (\ref{3rots}), is $2\phi_2$. 

Finally, we treat vector $\Jvec_4$ and the corresponding $U(1)$
holonomy.  The effect of (\ref{3rots}) on $\Jvec_4$ is simply
	\begin{equation}
	R(-\jvec_4,2\phi_4)\Jvec_4 = \Jvec_4,
	\label{J4xfm}
	\end{equation}
since the first two rotations do not act on $\Jvec_4$.  This specifies
an open curve in $SO(3)$ that can be closed (trivially) by multiplying
by a rotation about axis $\jvec_4$,
	\begin{equation}
	R(\jvec_4,2\phi_4) R(-\jvec_4,2\phi_4) = I.
	\label{Rloop4}
	\end{equation}
The closed loop in $SO(3)$ belongs to the contractible homotopy class,
so its lift into $SU(2)$ is closed and is specified by
	\begin{equation}
	u(\jvec_4,2\phi_4) u(-\jvec_4,2\phi_4) = 1.
	\label{uloop4}
	\end{equation}
When applied to the fourth spinor at point $p$ in
Fig.~\ref{ABintsect1} this sequence of spin rotations returns it to
its initial value.  The final $U(1)$ holonomy of spinor 4, after the
application of the three rotations (\ref{3rots}), is $2\phi_4$.

To summarize, we have succeeded in constructing the closed loop
$p$-$q$-$p'$-$p$ illustrated in Fig.~\ref{ABintsect1} as the product
of a sequence of seven spin rotations, each one generated by the
Hamiltonian flow of one of the observables in the $A$- or $B$-list (in
the $A$-list while we move on the $A$-manifold, and in the $B$-list
while we move on the $B$-manifold).  Most of these can be regarded as
being generated by the square of some angular momentum vector; for
example, the first and second spin rotations, specified by the axes
and angles of the right-most rotations in (\ref{3rots}), are generated
by $\Jvec_{12}^2$ and $\Jvec_{23}^2$, respectively, while the last
four rotations can be regarded as being generated by $\Jvec_r^2$,
$r=1,\ldots,4$.  The third rotation in (\ref{3rots}), about axis
$\jvec_4$, is generated by $\jvec_4 \cdot \Jvec$, that is,
$u(-\jvec_4,2\phi_4) = \exp(i\phi_4 \jvec_4 \cdot \bsigma)$ is applied
to all four spinors.

\subsection{The Ponzano-Regge phase}
\label{PRphase}

The actions associated with these spin rotations are easily computed,
using the complex 1-form (\ref{1forms}) and the methods of
Sec.~I.\Icomputingactions. To summarize the results, let $\Jvec_p$ be
a partial or total sum of the four angular momentum vectors (some
range of $r=1,\ldots,4$ is summed over), with magnitude $|\Jvec_p| =
J_p$.  Then the action along the path generated by $\nvec \cdot
\Jvec_p$ with elapsed angle $\theta$ is simply $(\nvec \cdot
\Jvec_p)\theta$.  In particular, the third rotation in (\ref{3rots}),
the overall rotation of the tetrahedron taking us from part (e) to
part (f) in Fig.~\ref{cycle}, does not contribute to the action since
in this case $\Jvec_p =\Jvec_{\rm tot}$ which vanishes on the $A$- and
$B$-manifolds.  See also (I.\Irotaction). As for the rotations
generated by some $\Jvec_p^2$, in this case $\nvec=\jvec_p =
\Jvec_p/J_p$, so the action is simply $J_p\theta$.  Thus the first and
second spin rotations specified by (\ref{3rots}) contribute
$J_{12}(2\phi_{12}) + J_{23}(2\phi_{23})$ to the total action.  As for
the last four rotations, in this case $\Jvec_p$ is one of the
$\Jvec_r$, $r=1,\ldots,4$, and the four angles are summarized in
Sec.~\ref{closeloopLPS}.  These rotations therefore contribute
$J_1(2\phi_1-2\pi) + J_2(2\phi_2) + J_3(2\phi_3-2\pi) + J_4(2\phi_4)$
to the total action.  Altogether, we have
	\begin{equation}
	S = \oint p\, dx = 2\sum_{r=1}^6 J_r \phi_r
	-2\pi(J_1+J_3),
	\label{totalaction}
	\end{equation}
where index $r=5$ means $r=12$ and $r=6$ means $r=23$.  This can be
written
	\begin{equation}
	S = -2\Psi + 2\pi(J_2 + J_4 + J_{12} + J_{23}),
	\label{thephase}
	\end{equation}
where on quantized manifolds the final term is an integer multiple of
$2\pi$, and where
	\begin{equation}
	\Psi = \sum_{r=1}^6 J_r(\pi-\phi_r).
	\label{Psidef}
	\end{equation}
The angle $\pi-\phi_r$ is the exterior dihedral angle, so $\Psi$ is
the phase of Ponzano and Regge.

\section{The amplitude determinant and reduced phase space}
\label{ampdet}

Amplitude determinants are notorious for the trouble they cause in
semiclassical approximations, for example, Gutzwiller's amplitude
determinant (Gutzwiller 1967, 1969, 1970, 1971) has that reputation
and in several studies of asymptotic approximations to spin networks
the authors have resorted to numerical calculations for the amplitude
determinant.  In fact, amplitude determinants can be expressed in
terms of Poisson brackets, which aids considerably in their
evaluation.  For example, Wigner's (1959) amplitude for the
$6j$-symbol is a single Poisson bracket, while the
amplitude determinant for the $9j$-symbol is a $2\times2$ matrix of
Poisson brackets (Haggard and Littlejohn 2010), the derivation of
which was the easiest part in the asymptotic formula for the
$9j$-symbol.  Similarly it is easy to obtain tractable expressions for
the amplitude determinant for the $15j$-symbol and other more
complicated cases of interest in quantum gravity.  

In Sec.~I.\Iampdet\ we presented a coordinate-based discussion of
amplitude determinants in the $3j$-symbol.  For a more geometrical
treatment of some of the issues discussed there we refer to the
literature on the ``quantization commutes with reduction'' theorems of
Guillemin and Sternberg (1982).  Here we will simply review the
results of I and discuss their geometrical content.

The discussion involves symplectic reduction, which in the case of the
$4j$-model of the $6j$-symbol leads to the reduced phase space of the
$6j$-symbol, a 2-sphere denoted by $\Gamma$ in (\ref{4jhopfmaps}).
This space is essential for understanding the semiclassical mechanics
of the $6j$-symbol, for example, it is the phase space that underlies
the 1-dimensional WKB methods used by Schulten and Gordon (1975a,b),
and it played an important role in the derivation of the uniform
asymptotic approximation of Littlejohn and Yu (2009), as well as in
the semiclassical studies of the volume operator in quantum gravity by
Bianchi and Haggard (2011).

\subsection{Densities and amplitude determinants}
\label{densities}

In this section we adopt a general notation, as in Sec.~\ref{lsolm},
so that our results can be applied to the $4j$-model of the
$6j$-symbol, the $12j$-model of the $6j$-symbol, the $9j$-symbol, and
many other examples.  The phase space is $P=\Reals^{2N}$ with
symplectic form $\omega=dp\wedge dx$.  The theory of symplectic
reduction usually begins with a symplectic group action on phase
space, but a more natural starting point in our applications is the
components of the momentum map of the group, which is a list of
classical observables.  Actually, there are two lists.

We assume that there exist on $P$ an $A$- and a $B$-list of
observables, $(A_1 ,\ldots, A_N)$ and $(B_1 ,\ldots, B_N)$, each of
which forms a Lie algebra under the Poisson bracket.  We denote
contour values by $a_i$ and $b_i$, so that level sets are specified by
$A_i=a_i$ and $B_i=b_i$.  We denote these level sets by $L_a$ and
$L_b$ (these are the $A$- and $B$-manifolds).  We assume that the
$A_i$'s and $B_i$'s are functionally independent at most places in
phase space, so that the generic dimensionality of $L_a$ and $L_b$ is
$N$, although it may be less in exceptional cases.  The $A$- and
$B$-lists of observables correspond to two groups with symplectic
actions on phase space, which we denote by $G_a$ and $G_b$, with Lie
algebras $\Liealgebra{g}_a$ and $\Liealgebra{g}_b$.  We assume the
groups are connected and have dimensionality $N$, as in
Sec.~\ref{lsolm}.  We denote the structure constants of one or the
other group by $c^k{}_{ij}$, as in Sec.~\ref{lsolm}.

We are exclusively interested in contour values $a$ and $b$ that are
fixed points of the coadjoint actions of the respective groups, since
this implies that $L_a$ and $L_b$, if $N$-dimensional, are Lagrangian.
Only Lagrangian manifolds can support semiclassical wave functions.
The restriction on the contour values means, however, that we do not
have a Lagrangian foliation of $P$.

We will be interested in the WKB or semiclassical $x$-space wave
function associated with the $A$- or $B$-manifold.  We will work with
the $A$-manifold, since the $B$-case is similar.  For modern
treatments of WKB theory, see for example Martinez (2002) or
Mishchenko \etal (1990).  The variables $x$ are half of the
coordinates $(x,p)$ on $P$; we denote ``$x$-space'' by $Q=\Reals^N$, which
abstractly is best seen as the quotient space when $P$ is divided by
the foliation into vertical Lagrangian planes, $x={\rm const}$.  We
write this wave function in the form
	\begin{equation}
	\psi_a(x) = \braket{x}{a}= K \sum_{\rm br} |\Omega(x)|^{1/2}
	\exp\{i[S(x)-\mu\pi/2]\},
	\label{psiaofx}
	\end{equation}
where $K$ is a normalization; where the sum is over branches of the
projection $\pi:L_a\to Q$ from the Lagrangian manifold to $Q$, assumed
to be locally invertible; where the branch index is suppressed in the
sum; where $S$ is the action computed as in Sec.~\ref{ABactions}; and
where $\mu$ is the Maslov index.

We associate the function $\Omega(x)$ with an $N$-form $\Omega$ on $Q$
by
	\begin{equation}
	\Omega = \Omega(x)\, dx^1 \wedge \ldots \wedge dx^N.
	\label{Omegadef}
	\end{equation}
It follows from WKB theory that $\Omega(x)$ satisfies a set of
amplitude-transport equations on $Q$, one for each observable $A_i$.
These are conveniently expressed in terms of an $N$-form $\sigma$ on
$L_a$, defined by $\sigma = \pi^* \Omega$. Then the amplitude
transport equations on $Q$ are equivalent to $\Liederiv_{X_i}
\sigma=0$ on $L_a$, where $\Liederiv$ is the Lie derivative and where
$X_i$, $i=1 ,\ldots,N$ are the Hamiltonian vector fields on $L_a$
associated with the $A_i$, that is, $X_i =\omega^{-1} dA_i$.  This means that
$\sigma$ is invariant under the infinitesimal action of the group,
hence under any finite action (recall that $L_a$ is an orbit of the
group action).  These are left actions.

We express the solution $\sigma$ as follows.  The vector fields $X_i$
may be chosen as a (generally non-coordinate) basis in each tangent
space at each point of $L_a$.  Let $\lambda^i$, $i=1 ,\ldots, N$ be
the dual 1-forms on $L_a$, that is, $\lambda^i(X_j) =\delta^i_j$.
Then, in the case of compact groups, the solution $\sigma$ is given by
	\begin{equation}
	\sigma =\lambda^1\wedge\ldots\wedge\lambda^N,
	\label{sigmadef}
	\end{equation}
to within a multiplicative constant.  The reason is that $\sigma$ is a
version of the right-invariant Haar measure on the group, and for
compact groups the left and right Haar measures are equal.  Thus,
$\sigma$ is invariant under the (left) action of the group, and
$\Liederiv_{X_i }\sigma=0$.

The action of $G_a$ on $L_a$ provides two ways in which geometric
structures on the group can be transferred to $L_a$.  First, the group
action implies a linear map $:\Liealgebra{g}_a \to T_xL_a$ for each
$x\in L_a$, which, under our assumptions, is invertible.  This map can
be used to push forward a standard $N$-form on $\Liealgebra{g}_a$ to
$L_a$.  Let the basis in $\Liealgebra{g}_a$ be $\{\xi_i, i=1 ,\ldots,
N\}$, corresponding to the observables $A_i$ and vector fields $X_i$,
and let $\{\alpha^i, i=1,\ldots,N\}$ be the dual basis in $\Liealgebra{g
}_a^*$, that is, $\alpha^i(\xi_j) =\delta^i_j$.  Then if the $N$-form
$\alpha^ 1\wedge\ldots\wedge\alpha^N$ is pushed forward to $L_a$ in this
manner, we obtain $\sigma$, defined by (\ref{sigmadef}).

A second way involves picking a point $x_0\in L_a$ to serve as an
``origin'' in $L_a$, and then identifying points $x$ of $L_a$ by the
group element $g$ such that $x=gx_0$.  This creates a diffeomorphism
between a neighborhood of the identity in $G_a$ and a neighborhood of
$x_0$ in $L_a$, which can be used to push forward differential
geometric structures from $G_a$ to $L_a$.  It then turns out that
$\sigma$ given by (\ref{sigmadef}) is the push forward of the
right-invariant $N$-form on $G_a$ associated with $\alpha^
1\wedge\ldots\wedge\alpha^N$.  Thus, $\sigma$ is naturally invariant
only under a right action of the group; the only reason it is also
invariant under the left action is that for the groups we consider,
the left and right Haar measures are identical.

To see this from another standpoint, a short calculation shows that
$\Liederiv_{X_i }\lambda^j = c^j{}_{ik }\,\lambda^k$, from which follows
$\Liederiv_{X_i }\sigma = c^j{}_{ij }\,\sigma$.  To derive this we recall that
although the Poisson brackets $\{A_i, A_j\}$ vanish on $L_a$, the Lie
brackets of the corresponding Hamiltonian vector fields do not,
instead we have $[X_i, X_j] = -X_k\,c^k{}_{ij}$.  But if the left and
right Haar measures are equal, then the adjoint representation $\Ad_g$
is volume-preserving, so the structure constants are traceless, and
$\Liederiv_{X_i }\sigma=0$.  In our examples we deal only with compact groups,
so this condition is met.

That the solution is unique to within a multiplicative constant can be
seen by supposing that $\sigma'$ is another $N$-form on $L_a$ such that
$\Liederiv_{X_i }\sigma'=0$.  Since all $N$-forms are proportional, we must
have $\sigma' = f\sigma$ where $f$ is a function on $L_a$.  This 
implies $X_i(f)=0$, or $f={\rm const}$ on $L_a$.

Now given that (\ref{sigmadef}) is the solution we want, we write
$\sigma = \pi^*\Omega$ in the form,
	\begin{equation}
	\Omega(x)\, dx^1 \wedge \ldots \wedge dx^N =
	\lambda^1 \wedge \ldots \wedge \lambda^N,
	\label{Omegasigmaeqn}
	\end{equation}
where we write simply $dx^i$ for $\pi^*(dx^i)$, and we evaluate both
sides on the set of vectors $(X_1 ,\ldots ,X_N)$.  This gives
	\begin{equation}
	\Omega(x) \det \{x^i,A_j\} = 1,
	\label{OmegaPB}
	\end{equation}
where we use $dx^i(X_j) =\{ x^i,A_j\}$.  This reproduces
Eq.~(I.\IOmegasoln).

Finally, as shown in I, the normalization integral, evaluated in the
stationary phase approximation, implies $K=1/\sqrt{V_a}$, where $V_a$
is the volume of $L_a$ with respect to $\sigma$.

\subsection{The amplitude of $\braket{b}{a}$}
\label{scmeba}

A coordinate-based derivation of the amplitude of the semiclassical
matrix element $\braket{b}{a}$ was presented in I for the case of the
$3j$-symbol.  Here we discuss the results from a geometrical point of
view, using the general notation of Sec.~\ref{lsolm} and the previous
section.  The main issue is that the stationary phase set $L_a \cap
L_b$ in the evaluation of the integral $\braket{b}{a} =\int dx
\braket{b}{x} \braket{x}{a}$ is not a set of isolated points, as
expected on the basis of a naive dimensionality count, but rather a
set of manifolds of dimensionality $\ge1$. These nontrivial
intersection manifolds are due to the existence of a common
``intersection group'' of $G_a$ and $G_b$, which may be defined as
follows.

The basic idea is that the $A$- and $B$-lists of observables may have
some observables in common, which generate the intersection group.
This is obviously the case in (\ref{ABfunctionlists}), for example.
But the specific observables that occur in the $A$- and $B$-lists
depend on the bases chosen in the Lie algebras $\Liealgebra{g}_a$ and
$\Liealgebra{g}_b$, and if a basis is changed then the $A$- or
$B$-observables are replaced by linear combinations of themselves.  So
we need a precise definition of the observables ``in common.''

The actions of groups $G_a$ and $G_b$ on $P$ provide Lie algebra
anti-homomorphisms between $\Liealgebra{g}_a$ and $\Liealgebra{g}_b$
and the Lie algebra of (globally) Hamiltonian vector fields on $P$.
By our assumptions these anti-homomorphisms have full rank, so the
images of these two anti-homomorphisms are two $N$-dimensional Lie
algebras of Hamiltonian vector fields.  These two Lie algebras have an
intersection which itself is a Lie algebra.  Let the dimension of the
intersection be $p$, and let $p+q=N$.  The intersection Lie algebra
is generated by a set of Hamiltonian functions, call them $(C_1
,\ldots, C_p)$.  These can be regarded as the functions common to the
original $A$- and $B$-lists, that is, by a change of basis in
$\Liealgebra{g}_a$ and $\Liealgebra{g}_b$ we can bring the $A$- and
$B$-lists into the form $A=(C,D)$ and $B=(C,E)$, where $D=(D_1
,\ldots, D_q)$ and $E=(E_1 ,\ldots, E_q)$ are sets of observables that
the $A$- and $B$-lists do not have in common.  The common observables
$C$ generate the action of an ``intersection group'' $G_0$, that is,
they are the components of the momentum map of the action of $G_0$ on
$P$.  The group $G_0$ is not uniquely determined by the $C$
observables, only its action on $P$ is.  The same applies to $G_a$ and
$G_b$, which are generated by the $A$- and $B$-lists of observables.
But in practice there are convenient ways of choosing all these groups
so that the kernels of their actions are discrete.  This means that
$\dim G_a = \dim G_b = N$ and $\dim G_0 = p$.  We may assume moreover
that $G_0$ is connected (as we have already for $G_a$ and $G_b$),
since we are free to take the connected identity component of any of these
groups.

The level set of the momentum map of the intersection group $G_0$
plays an important role in what follows.  We denote this level set by
$L_0$; its equation is $C_i=c_i$, $i=1 ,\ldots ,p$, for some contour
values $c_i$.  We choose the $c_i$ such that $L_0$ has its generic
dimension $2N-p=N+q>N$.

The level set $L_0$ is foliated into $A$- and $B$-manifolds, where the
$A$-manifolds are parameterized by the values $d_i$ of the observables
$D_i$, and the $B$-manifolds by the values $e_i$ of the observables
$E_i$, $i=1 ,\ldots, q$.  We will assume that the Poisson brackets
$\{A_i, A_j\}$ and $\{B_i, B_j\}$ vanish on $L_0$, so that the generic
$A$- and $B$-manifolds in $L_0$ are Lagrangian.  This is an extension
of our earlier assumption, that these Poisson brackets vanish on a
specific pair of $A$- and $B$-manifolds.  Thus, after removing
exceptional manifolds of less than generic dimensionality, $L_0$ is
foliated into Lagrangian manifolds in two different ways.  These
Poisson bracket relations imply $\{C_i, C_j\}=0$ on $L_0$, so that $c$
is a fixed point of the coadjoint action of $G_0$.  This does not mean
that $L_0$ is Lagrangian (the dimension is wrong), but it is
coisotropic (Abraham and Marsden 1978).  It also means that the
isotropy subgroup of the coadjoint action is the whole group, so the
reduced phase space of level set $L_0$ under the $G_0$ action is the
space $L_0/G_0$.  

This draws attention to the orbits of $G_0$, which generically have
dimension $p$.  Not only is $L_0$ foliated into orbits of $G_0$, so
is each $A$- and $B$-manifold, since $G_0$ is a simultaneous subgroup
of $G_a$ and $G_b$.  Thus, the intersections of $L_a$ and $L_b$, which
are the stationary phase sets, are also foliated into orbits of
$G_0$.  We will assume that $L_a \cap L_b$ is a union of a discrete
set of orbits of $G_0$.  Based on a dimensionality count, this is the
generic case.  It holds for example in the $4j$-model of the
$6j$-symbol, where $L_a \cap L_b$ consists of two orbits of $G_0$, the
sets $I_1$ and $I_2$ in Fig.~\ref{ABintsect}.  It also holds in the
$12j$-model used by Roberts (1999) and in our own work on the
$9j$-symbol (Haggard and Littlejohn 2010).

The orbits of $G_0$ appear in the stationary phase evaluation of
	\begin{eqnarray}
	\braket{b}{a} &= \int dx\,\braket{b}{x}\braket{x}{a}
	= \frac{1}{\sqrt{V_aV_b}}
	\int dx \sum_{\rm br} 
	{1\over\sqrt{|\det\{x,A\}\det\{x,B\}|}} \nonumber \\
	&\qquad\times\exp\{i[S_A(x)-S_B(x) -\mu\pi/2]\},
	\label{abintegral}
	\end{eqnarray}
where we have inserted the normalized $x$-space wave functions for the
states $\ket{a}$ and $\ket{b}$, where the sum is over all pairs of
branches of both wave functions, and where $\mu$ is the cumulative
Maslov index.  The stationary phase set is the projection of $L_a \cap
L_b$ onto $Q$ ($x$-space); it is a union of the projections of the
discrete set of orbits of $G_0$ that make up $L_a\cap L_b$.

The projection of each orbit is a subset of $Q$ that is locally
$p$-dimensional.  As in I, we introduce a local change of coordinates
$x \to (y,z)$, where $(y^1,\ldots,y^p)$ are coordinates along the
projected orbits and $(z^1,\ldots,z^q)$ are coordinates transverse,
with $z=0$ being the projected orbit itself.  Then the integral
becomes (suppressing normalization and branch sums)
	\begin{equation}
	\fl\int \frac{d^py}{|\det\{y,C\}|}
	\int \frac{d^qz}{\sqrt{|\det\{z,D\} \det\{z,E\}|}}
	\exp\{i[S_A(y,z)-S_B(y,z)-\mu\pi/2]\}.
	\label{yzintegral}
	\end{equation}
When the exponent is expanded to second order in $z$, the leading term
$S_A(y,0) - S_B(y,0)$ is independent of $y$, that is, it is constant
along the orbits of $G_0$.  The second derivative matrix of $S_A-S_B$
with respect to $z$ does depend on $y$, but after doing the Gaussian
integral and combining with the determinants in the denominator of the
$z$-integration, the result is expressed purely in terms of the
Poisson brackets $\{E_i,D_j\}$, which are independent of $y$.  

The fact that the $z$-integral is independent of $y$, that is, our
location on the orbit of $G_0$, is noteworthy.  It means that the
integral does not depend on the detailed nature of the
$z$-coordinates, for example, it is invariant under a coordinate
transformation of the form $z'=z'(z,y)$ such that $z=0$ implies
$z'=0$.  It suggests that we are dealing with a quotient operation in
which we divide by the orbits of $G_0$.

Another remark is that there is nothing special about the
$x$-representation in which the integral (\ref{abintegral}) is carried
out.  The $x$-coordinates were introduced as half of the $(x,p)$
coordinates on $P=\Reals^{2N}$, but any representation related to this
one by a metaplectic transformation (Littlejohn 1986) would work as
well.  This amounts to foliating phase space, not by the vertical
Lagrangian planes $x={\rm const}$, but rather by other Lagrangian
planes related to this one by any linear, symplectic map.  In this
manner one can divide the orbits of $G_0$ into segments and do the
integral over each segment in a representation in which the projection
of the orbit onto the representation space has full rank.  The
segments into which the orbit of $G_0$ is divided can even be
infinitesimal, effectively making the representation for the integral
a function of where we are along those orbits.  Similarly, one can
avoid the caustics of the wave functions $\psi_a$ and $\psi_b$.  This
is the old idea underlying the Maslov method (Maslov and Fedoriuk
1981) in WKB theory.

In any case, once the $z$-integral is done and is recognized to be
independent of $y$, the remaining $y$-integral can be lifted to the
orbit of $G_0$ in phase space whereupon it becomes just the integral
of the Haar measure of $G_0$, giving the volume of the orbit.  This
Haar measure is normalized as in the previous section, that is, we
start with the observables $C_i$, $i=1 ,\ldots, p$, we associate these
with Hamiltonian vector fields $X_i =\omega^{-1} dC_i$ (a change of
notation from above, where the $X$'s were associated with the $A$'s),
we define form $\lambda^i$ dual to the $X_i$ on the orbits of $G_0$,
and then the Haar measure on the orbit is taken to be $\lambda^
1\wedge\ldots\wedge \lambda^p$.  The final result is
	\begin{equation}
	\braket{b}{a} = (2\pi i)^{q/2} \frac{V_I}{\sqrt{V_aV_b}}
	\sum_{\rm br} \frac{1}{\sqrt{|\det\{D_i,E_j\}|}}
	\exp[i(S_I -\mu\pi/2)],
	\label{scabme}
	\end{equation}
where now the branches refer to the discrete set of orbits of $G_0$
that make up $L_a \cap L_b$, where $V_I$ is the volume of intersection
manifold $I$ (an orbit of $G_0$), where $S_I$ means $S_A-S_B$ evaluated
on intersection manifold $I$, and where $\mu$ is a cumulative Maslov
index (not necessarily the same as the previous ones).  The branch
index could otherwise be written as $I$, a label of the intersection
manifold, and $V_I$ is taken out of the sum because it does not
depend on which intersection manifold is taken.  The volume $V_I$
differs from the volume of $G_0$ because in general there is a
discrete isotropy subgroup, as in (\ref{volumeI}) (one is really
computing the volume of a coset space).

The result (\ref{scabme}) contains a $q\times q$ matrix of Poisson
brackets, $\{E_i, D_j\}$, whose geometrical content may be understood
in terms of a variation of the discussion of densities in
Sec.~\ref{densities}.  First we recall that the $A$- and $B$-lists are
decomposed according to $A=(C,D)$, $B=(C,E)$.  Next, we fix the
contour values $c_i$, $i=1 ,\ldots, p$, so that we have a definite
level set $L_0$ of the momentum map of $G_0$.  Then we let
``$b$-space'' be $\Reals^q$ with coordinates $(e_1 ,\ldots, e_q)$ or
the region of $\Reals^q$ that is the projection of $L_0$ onto
$\Reals^q$, where coordinates $e_i$ are interpreted as the contour
values in $E_i=e_i$.  The matrix element $\braket{b}{a}$ can be
thought of as a wave function on $b$-space for fixed values of the
$a$'s, that is, of the $c$'s and $d$'s.  We write the amplitude of the
semiclassical approximation to this wave function as
$|\Omega(e)|^{1/2}$, where $\Omega =\Omega(e)\, de_
1\wedge\ldots\wedge de_q$ is the associated density (a $q$-form on
$b$-space).  This density is the projection of the natural density on
the $A$-manifold, in the following sense.  Let the $A$- or
$(C,D)$-observables be associated with vector fields $X_i
=\omega^{-1}dC_i$, $i=1 ,\ldots, p$, and $Y_i =\omega^{-1} dD_i$, $i=1
,\ldots, q$, with dual 1-forms $\lambda^i$, $i=1 ,\ldots, p$ and
$\mu^i$, $i=1 ,\ldots, q$.  These induce a density $\sigma_
0\wedge\mu^ 1\wedge\ldots\wedge\mu^q$ on $L_a$, where $\sigma_0
=\lambda^ 1\wedge\ldots\wedge\lambda ^p$ is the Haar measure on $G_0$.
Then $\Omega$ satisfies
	\begin{equation}
	\sigma_0 \wedge \pi^*\Omega = \sigma_0 \wedge \mu^1 \wedge
	\ldots \wedge \mu^q,
	\label{Omegaofeeqn}
	\end{equation}
where $\pi$ is the projection from $L_0$ or $L_a \subset L_0$ onto
$b$-space.  Now evaluating both sides on the set of vectors $(X_1
,\ldots, X_p, Y_1 ,\ldots, Y_q)$, we obtain
	\begin{equation}
	\Omega(e) \det dE_i(Y_j) =1.
	\label{Omegaofedet}
	\end{equation}
This gives the amplitude shown in (\ref{scabme}), since $dE_i(Y_j) =
\{E_i, D_j\}$.

This discussion has treated the $A$- and $B$-manifolds asymmetrically,
projecting from the $A$-manifold onto $b$-space, but we could have
projected the density on the $B$-manifold onto $a$-space with the same
result.  There are really four densities, two {\it on} the $A$- and
$B$-manifolds, and two {\it of} the $A$- and $B$-manifolds.

This discussion leads us to consider the reduced phase space $\Gamma =
L_0/G_0$, which is parameterized by the contour values $c_i$, $i=1
,\ldots, p$.  As is standard in symplectic reduction, the symplectic
form on $\Gamma$ is obtained by pulling back vectors from $\Gamma$ to
$L_ 0\subset P$ and evaluating them on the symplectic form on $P$;
this is meaningful because the answer does not depend on where on the
orbit of $G_0$ they are pulled back to, nor on the component of the
pulled-back vectors along the orbit.  A consequence is that the
projections of the $A$- and $B$-manifolds onto $\Gamma$, which are
$q$-dimensional since the $G_0$ orbits are $p$-dimensional, are
Lagrangian on $\Gamma$.  Since we are assuming that $L_a \cap L_b$ is
a discrete union of $G_0$ orbits, the projected manifolds on $\Gamma$
intersect in a discrete set of isolated points.  This would be the
generic case on a symplectic manifold of dimension $2q$.  Also,
$G_0$-invariant functions on $L_0$ project onto functions on $\Gamma$,
whose Poisson brackets on $\Gamma$ are the same as the Poisson
brackets of the original functions on $P$.  Such functions in the
present discussion include the observables $D_i$ and $E_i$, $i=1
,\ldots, q$, so the Poisson brackets $\{E_i,D_j\}$ of (\ref{scabme})
are naturally interpreted as living on $\Gamma$.

\subsection{The case of the $6j$-symbol}
\label{caseof6j}

It is straightforward to apply the general theory of
Secs.~\ref{densities} and \ref{scmeba} to the case of the
$6j$-symbol.  The phase space is $\Complexes^8 =\Reals^{16}$ so
$N=8$.  The common observables $C$ are $I_r$, $r=1 ,\ldots,4$,
and $\Jvec_{\rm tot}$, so $p=7$ and $q=1$.  The group $G_0$ is $T^4
\times SU(2)$.  The level set $L_0$ of the momentum map of $G_0$, for
$I_r = J_r$, $r=1,\ldots,4$, and $\Jvec_{\rm tot}=\zerovec$, is the
subset of $\Phi_{4j}$ upon which the four angular momenta have
specified lengths and their vector sum is zero.  It is logical that
this would be the subset of the classical phase space $\Phi_{4j}$ that
corresponds to the space $\ZS$ of four-valent intertwiners, introduced
in Sec.~\ref{4jmodel}, on which $\Ihat_r =j_r$, $r=1,\ldots,4$ and
$\Jvechat_{\rm tot} =\zerovec$.

The contour values of the $C$'s, that is, of $I_r$, $r=1 ,\ldots,4$
and $\Jvec_{\rm tot}$, must be chosen so that $L_0$ has its maximum
dimensionality, namely, 9.  The condition is $J_r>0$, $r=1 ,\ldots, 4$
and the polygon inequality,
	\begin{equation}
	\max\{J_1,J_2,J_3,J_4\} < \frac{1}{2}
	\sum_{r=1}^4 J_r,
	\label{polygonJr}
	\end{equation}
which as indicated must not be saturated.  This is the condition that
it is possible to make a noncollinear polygon in $\Reals^3$ out of
vectors of four given positive lengths.  If this condition is
satisfied, then $L_0$ is foliated into $A$- and $B$-submanifolds whose
generic dimensionality is 8.

There is only one observable of the $D$- and $E$-type; according to
(\ref{ABfunctionlists}) we should make the identifications
$D=\Jvec_{12}^2$ and $E=\Jvec_{23}^2$.  However, since we used
$d\theta$ in the volume form on the $A$- and $B$-manifolds when
computing the volume (\ref{ABvolume}), we should use instead
$D=|\Jvec_{12}|$ and $E=|\Jvec_{23}|$, since these are conjugate to
$\theta$ (really $\theta_a$ and $\theta_b$, since there are two of
them).  Then the Poisson bracket for the amplitude is
	\begin{equation}
	\fl\{E,D\}=
	\{|\Jvec_2+\Jvec_3|,|\Jvec_1+\Jvec_2|\}=
	\frac{\Jvec_2\cdot[(\Jvec_2+\Jvec_3)\times(\Jvec_1+\Jvec_2)]}
	{|\Jvec_{12}||\Jvec_{23}|}
	= \frac{6V}{J_{12} J_{23}},
	\label{6jPB}
	\end{equation}
where $V$ is the signed volume of the tetrahedron, $6V =\Jvec_ 1\cdot
(\Jvec_ 2\times\Jvec_3)$, where we have used (\ref{Lambda4jPB}) to
evaluate the Poisson bracket, and where in the final step we have
evaluated the Poisson bracket on $A$- and $B$-manifolds with contour
values $J_{12}$ and $J_{23}$.  We see the appearance of Wigner's
volume (Wigner 1959).  The volume changes sign between the two
intersection manifolds $I_1$ and $I_2$, but it appears with an
absolute value sign in (\ref{scabme}) so both stationary phase points
in the $6j$-symbol have the same amplitude.  

The amplitude $|\Omega|^{1/2}$ contains the factor 
	\begin{equation}
	\sqrt{J_{12}J_{23}} = 
	\frac{1}{2}\sqrt{(2j_{12}+1)(2j_{23}+1)},
	\label{sqrtfacs} 
	\end{equation}
which, when evaluated as shown on quantized manifolds, reproduces the
square roots seen in (\ref{6jme}).  Thus, based on the pieces of the
formula we have determined so far, we can write
	\begin{equation} 
	\left\{\begin{array}{ccc} j_1 & j_2 &
	j_{12} \\ j_3 & j_4 & j_{23} \end{array}\right\}
	=\frac{e^{i\pi/4}}{\sqrt{12\pi|V|}} 
	\frac{1}{2}
	\left[e^{i(S_1-\mu_1\pi/2)} + 
	e^{i(S_2-\mu_2\pi/2)} \right],
	\label{PRv1}
	\end{equation}
where $S_1$ and $S_2$ are given by (\ref{S1S2defs}) (they are the
phases $S_A-S_B$, evaluated on the intersection manifolds $I_1$ and
$I_2$ of Fig.~\ref{ABintsect}).  These phases contain the phase
conventions for the states $\ket{A}$ and $\ket{B}$ shown in
Fig.~\ref{4jABstates}, which appear semiclassically as the origins for
action integrals on the Lagrangian manifolds, denoted $a_0$ and $b_0$
in Fig.~\ref{ABintsect}.  Since we have not considered these phase
conventions yet, we cannot say what $S_1$ and $S_2$ are, but the
difference $S=S_2-S_1$ is given in terms of the Ponzano-Regge phase by
(\ref{thephase}).  

To return to the 9-dimensional space $L_0$, it projects onto what was
called ``$b$-space'' in Sec.~\ref{scmeba}, which in the present case
is the $J_{23}$-axis, producing the interval $[J_{23,{\rm min}},
J_{23,{\rm max}}]$, given by (\ref{J23maxmin}).  The inverse image of
a point on this interval is a $B$-manifold with a given value of
$J_{23}$.  Similarly $L_0$ projects onto ``$a$-space'', that is, the
$J_{12}$-axis, producing the interval (\ref{J12maxmin}),  the inverse
image of a point of which is an $A$-manifold with given $J_{12}$
value.  The space $L_0$ projects onto the $J_{12}$-$J_{23}$ plane producing
the classically allowed region of Fig.~\ref{square}; the inverse image
of a point of this region is an intersection of an $A$- and a
$B$-manifold of given $J_{12}$ and $J_{23}$ values.  If the point does
not lie on the caustic curve, the intersection consists of two
disconnected components $I_1$ and $I_2$, each an orbit of $G_0$, as in
Fig.~\ref{ABintsect}.  On the caustic curve these components merge
into one.  Finally, by dividing by $G_0$, $L_0$ projects onto the
reduced phase space $\Gamma$, which has dimensionality $9-7=2$.  We
now turn to this space.
	
\subsection{The reduced phase space $\Gamma$}
\label{evalrps}

Spaces of the type $\Gamma$ seem to have appeared first in the work of
Kapovich and Millson (1995, 1996).  Those authors showed that the
space of polygons of a given number of sides with fixed lengths in
$\Reals^3$, modulo overall rotations, is a symplectic manifold.  In
fact, for quadrilaterals the space is precisely $\Gamma$.  This space
was recently subjected to direct geometric quantization by Charles
(2008), who connected it with the $6j$-symbol and used it for a new
derivation of the Ponzano-Regge asymptotic formula.  The space of
five-sided polygons is the analog of $\Gamma$ for the $9j$-symbol; it
was used by Haggard and Littlejohn (2010) in their study of the
asymptotics of the $9j$-symbol.

To visualize $\Gamma=L_0/G_0$ it helps to carry out the reduction in
two steps.  In the first step, we choose contour values $J_r>0$, $r=1
,\ldots,4$, for which the level set $I_r=J_r$ in $\Phi_{4j}$ is the
product of 3-spheres $(S^3)^4$, as shown in (\ref{4jhopfmaps}).  The
$I_r$, $r=1 ,\ldots,4$, generate the action of the group $U(1)^4$,
corresponding to the phases of the four spinors.  Dividing this level
set by $U(1)^4$, we obtain the symplectic manifold $\Sigma_{4j}$ shown
in (\ref{4jhopfmaps}), which topologically is $(S^2)^4$, and which
consists of all sets of four vectors $\Jvec_r$ in $\Reals^3$, $r=1
,\ldots,4$, with fixed lengths, $|\Jvec_r| = J_r$.  The space
$\Sigma_{4j}$ is 8-dimensional.  In the second step, we consider the
submanifold in $\Sigma_{4j}$ upon which $\Jvec_{\rm tot}=0$, which is
a level set of the momentum map of the action of $SO(3)$ on
$\Sigma_{4j}$.  This manifold consists of sets of four vectors
$\Jvec_r$ in $\Reals^3$ of fixed lengths $J_r$ such that $\sum_r
\Jvec_r=\zerovec$.  The vectors can be placed end-to-end to
form a ``closed link,'' that is, a four-sided polygon in $\Reals^3$.
The set of closed links is denoted CL in (\ref{4jhopfmaps}).  We
assume the polygon inequality (\ref{polygonJr}) is satisfied, so the
space CL has dimensionality $8-3=5$.  Since this is the level set
$\Jvec_{\rm tot}=0$, the isotropy subgroup of the $SO(3)$-action is
$SO(3)$ itself, so the reduced phase space is ${\rm CL}/SO(3)$, which
is the space $\Gamma$ in (\ref{4jhopfmaps}).  This has $5-3=2$
dimensions.

The phase space $\Gamma$ is parameterized by the four fixed, positive
values $J_r$, $r=1 ,\ldots,4$ that satisfy the polygon inequality
(\ref{polygonJr}).  A point of this space specifies a quadrilateral in
$\Reals^3$ of the given lengths, modulo overall proper rotations.
Once the quadrilateral has been determined, one can draw in the two
remaining edges, of lengths $J_{12}$ and $J_{23}$, to fill in a
tetrahedron.  Thus, $\Gamma$ can be thought of as the shape space for
a set of tetrahedra, four of whose edges have fixed, positive lengths.
The two lengths that are variable are on opposite sides of the
tetrahedron.  Here we define ``shape'' as a configuration modulo
proper rotations, as in Littlejohn and Reinsch (1997); two shapes
related by spatial inversion are generally distinct.  The lengths
$|\Jvec_{12}|$ and $|\Jvec_{23}|$ are variable functions on $\Gamma$.
In fact, any rotationally invariant quantity associated with the
tetrahedron, such as the dihedral angles, the areas of the faces, the
signed volume, etc, is also a function on $\Gamma$.

\begin{figure}[htb]
\begin{center}
\scalebox{0.7}{\includegraphics{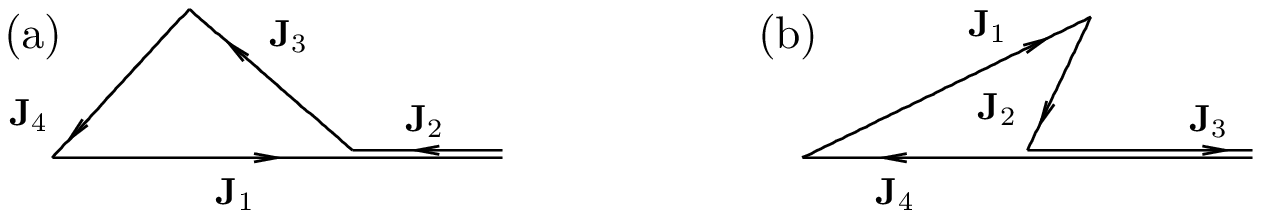}}
\end{center}
\caption[jlimits]{\label{jlimits} When $J_{12}$ is at its lower limit 
for given $(J_1,J_2,J_3,J_4)$, the shape of the tetrahedron is not
changed under the $\Jvec_{12}^2$-action, so the lower limit
corresponds to only one point of the reduced phase space.}
\end{figure}

It is easy to find coordinates on $\Gamma$.  We may take one
coordinate to be $J_{12}$, which varies between the bounds
(\ref{J12maxmin}).  (From this point on we drop the distinction
between $|\Jvec_{12}|$ and $J_{12}$, and similarly for $J_{23}$ and
$J_{13}$.)  For a fixed value of $J_{12}$, the allowed set of shapes
is generated by executing the ``butterfly'' motion about the axis
$\Jvec_{12}$, that is, rotating the triangle 1-2-12 about this axis,
relative to the 3-4-12 triangle.  We recall this motion is the
Hamiltonian flow generated by $\Jvec_{12}^2$ or the magnitude
$|\Jvec_{12}|$ (see (\ref{J12sqflow})).  Thus, coordinates can be
taken to be $(J_{12},\phi_{12})$, where $\phi_{12}$ is the interior
dihedral angle about the 12-edge.  For each value of $J_{12}$ in the
interior of the range (\ref{J12maxmin}), a circle of shapes is
generated as $\phi_{12}$ goes from 0 to $2\pi$; but at the endpoints
there is only a single shape.  For example, at the lower limit of
(\ref{J12maxmin}), the case $J_{12,{\rm min}} = |J_1-J_2|$ is
illustrated in part (a) of Fig.~\ref{jlimits}. In this case the
rotation of vectors $\Jvec_1$ and $\Jvec_2$ about the axis
$\Jvec_{12}$ does not change the shape.  Similarly, part (b)
illustrates the case $J_{12}= J_{12,{\rm min}} =|J_3-J_4|$.  In this
case, the $\Jvec_{12}^2$-action rotates the 1-2-12 triangle, but does
not change the shape since the new configurations that result are
related to the original ones by an overall $SO(3)$ transformation.  A
similar analysis applies at the upper limit of (\ref{J12maxmin}).

We see that the set of shapes generated in this manner for $J_{12}$ in
the interior of the interval (\ref{J12maxmin}) is a cylinder, but the
two endpoints are single points that pinch the cylinder at the two
ends, creating a topological sphere.  Topologically, $\Gamma=S^2$.

The symplectic form on $\Gamma$ may be obtained by projecting
$\omega =dp\wedge dx$ on $\Phi_{4j}$ as described in Sec.~\ref{scmeba}
, but it is easier just to notice that $\phi_{12}$ is the parameter of
evolution along the flow generated by $J_{12}$, so
$(\phi_{12},J_{12})$ form a canonically conjugate $(q,p)$ pair on the
sphere.  The same obviously applies to $J_{23}$ and $J_{31}$ and their
conjugate (dihedral) angles, so we have
	\begin{equation}
	dJ_{12} \wedge d\phi_{12} =
	dJ_{23} \wedge d\phi_{23} =
	dJ_{31} \wedge d\phi_{31},
	\label{RPSomegas}
	\end{equation}
indicating three sets of canonical coordinates on $\Gamma$, related by
canonical transformations.   These are examples of the action-angle
variables discovered by Kapovich and Millson (1995, 1996), which in
all cases are closely related to the recoupling schemes used in
angular momentum theory.  The length $J_{31}$ and the associated
dihedral angle $\phi_{31}$ do not appear in the coupling scheme we
described in Sec.~\ref{4jmodel} or in the tetrahedra we have
discussed so far, but would appear in a different tetrahedron in which
the four vectors are placed end-to-end in a different order.

So far we have described the construction of $\Gamma$ as a purely
classical problem, but if the $J_r$ are quantized, $J_r=j_r+1/2$, $r=1
,\ldots 4$, then one can speak of a quantized level set $L_0$ and
quotient space $\Gamma$.  

The area of the sphere $\Gamma$ with respect to the form $dJ_{12
}\wedge d\phi_{12}$ is obviously just $2\pi(J_{12,{\rm max}} -
J_{12,{\rm min}})$.  If $\Gamma$ is quantized, then by
(\ref{J12j12maxmin}) and (\ref{dimZS}) the area is $2\pi D$, where
$D=\dim\ZS$.  When quantized, $\Gamma$ contains one Planck cell of
area $2\pi$ for every state in the Hilbert space $\ZS$.  Obviously we
obtain the same area if we use either of the other symplectic forms in
(\ref{RPSomegas}).

\subsection{Quantized curves in $\Gamma$}
\label{quantrps}

States in $\ZS$, such as the $A$- and $B$-states given by
(\ref{4jABdefs}), are represented semiclassically by Lagrangian
manifolds in $\Gamma$, which are quantized curves in that space.  For
example, the $A$-states are represented by quantized levels sets of
$J_{12}$.
 
To plot these we map $\Gamma$ into a unit 2-sphere in ${\mathbb
R}^3$ with standard coordinates $(x,y,z)$ by associating
$(\phi_{12},J_{12})$ with a standard set of spherical angles
$(\theta,\phi)$, where $\phi=\phi_{12}$ and
	\begin{equation}
	J_{12} = J_{12,{\rm min}} + \frac{D}{2}
	(1+\cos\theta),
	\label{RPSthetadef}
	\end{equation}
where $D=J_{12,{\rm max}} - J_{12,{\rm min}}$ (generally), or
$D=\dim\ZS$ (when $\Gamma$ is quantized).  Here $x
=\sin\theta\cos\phi$, $y =\sin\theta\sin\phi$, $z =\cos\theta$ are
understood.  Then the symplectic form on the sphere is
$(-1/2)D\sin\theta\, d\theta\wedge d\phi$.  These coordinates make the
$J_{12}$ orbits look nice (they are small circles), but not the orbits
of the other variables, so they should not be thought of as having any
privileged role.  We use them mainly for plotting figures.  The
$x$-$z$ plane in these coordinates does have an invariant meaning,
however; it is the plane upon which the tetrahedra are planar (the
volume vanishes).  Moreover, time reversal is a reflection in this
plane ($y \to -y$).

\begin{figure}[htb]
\begin{center}
\scalebox{0.7}{\includegraphics{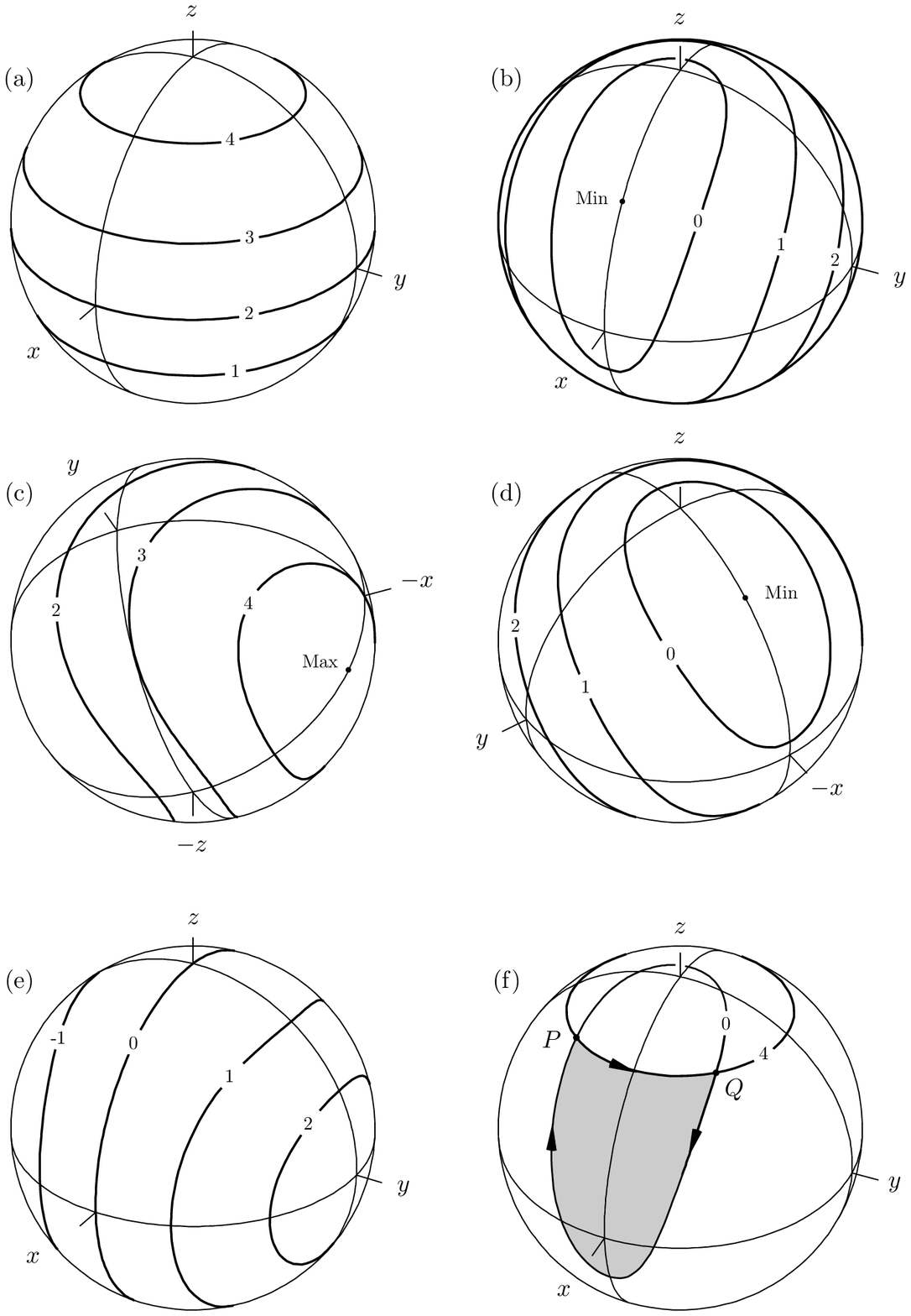}}
\end{center}
\caption[rps]{\label{rps} Orbits (or level sets) on the reduced phase
space.  In (a), orbits of $J_{12}$; in (b) and (c), of $J_{23}$; in
(d), of $J_{13}$; in (e), of the volume $V$.  Part (f) shows the
orbits relevant to the $6j$-symbol.}
\end{figure}

The quantized orbits of $J_{12}$, on a space $\Gamma$ with five Planck
cells of area, are illustrated in part (a) of Fig.~\ref{rps}.  The
orbits are just small circles, as noted.  The orbits are numbered $n=0
,\ldots,4$, in order of increasing $J_{12}$, and enclose area
$(n+\frac{1}{2})2\pi$.  The minimum and maximum values of $J_{12}$ are
at the south and north poles, respectively, because of the choice
(\ref{RPSthetadef}) of coordinates, but the minimum and maximum
quantized values are separated from the values at the poles by
$\frac{1}{2}$, as indicated by (\ref{J12J23quant}), since the polar
caps defined by the last quantized orbit before the poles must have
area $\frac{1}{2}$.

The quantized orbits of $J_{23}$ are illustrated in parts (b) and (c)
of that figure.  The function $J_{23}$ has a minimum on the $x$-$z$
plane with $x>0$, and increases monotonically toward a maximum on the
same plane with $x<0$.  The orbits are time reversal symmetric, but
have no symmetry under inversion through the origin in the coordinates
used (there is no reason why they should).  Also shown are some
quantized orbits of $J_{13}$, in part (d) of Fig.~\ref{rps}.

Another interesting observable on the reduced phase space is the
signed volume $V=(1/6)\Jvec_1 \cdot (\Jvec_2 \times \Jvec_3)$.  This
has been suggested by Chakrabarti (1964) and by L\'evy-Leblond and
L\'evy-Nahas (1965) as a ``democratic'' alternative to the usual
choices $J_{12}^2$, $J_{23}^2$, $J_{13}^2$ for the intermediate
variable in the quantum problem of the recoupling of three angular
momenta.  More recently, Carbone \etal (2002) have reconsidered the
use of this observable, and have derived both recursion relations and
asymptotic formulas for the matrix elements connecting the
$J_{12}$-basis and the $V$-basis.  The level sets (orbits) of $V$ are
plotted in part (e) of Fig.~\ref{rps}.  The orbits plotted are not
quantized, but are evenly spaced in $V$ about $V=0$, and labelled
$n=-2 ,\ldots, +2$.  The $V=0$ contour is the $x$-$z$ plane, which is
also a quantized orbit when $\dim\ZS$ is odd.  The orbit $V=v$ is
mapped by time reversal into the orbit $V=-v$.  

The Bohr-Sommerfeld condition for the quantized orbits of $V$ can be
expressed in terms of complete elliptic integrals of third kind.  We
will report on this and other results on the spectrum of $V$ and on
the wave functions $\braket{j_{12}}{V}$ in future publications.
Preliminary results are reported in Bianchi and Haggard (2011).

\subsection{The $6j$-symbol on the reduced phase space}
\label{6jrps}

As noted below (\ref{6jme}) the $6j$-symbol is proportional to the
matrix element $\braket{j_{23}}{j_{12}}$, so on $\Gamma$ it is
represented semiclassically by the quantized curves, $J_{23} =j_{23}
+\frac{1}{2}$ and $J_{12} =j_{12} +\frac{1}{2}$.  This is illustrated
in part (f) of Fig.~\ref{rps} for the case $\braket{j_{23}} {j_{12}}
=\braket {0}{4}$, where the numbers refer to the labeling of the
quantized curves in parts (a) and (b) of that figure.  These curves
are projections of the $A$- and $B$-manifolds in $\Phi_{4j}$, and
their intersections, the points labelled $P$ and $Q$ in the figure,
are the projections of intersections manifolds $I_1$ and $I_2$ as
illustrated in Fig.~\ref{ABintsect}, respectively.  This follows
because we defined $I_1$ as the intersection manifold upon which
$V<0$, and $V<0$ at point $P$ since it lies in the region $y<0$. Also,
the path (with direction indicated) surrounding the shaded area in
part (f) of Fig.~\ref{rps} is the projection of the loop (with
direction indicated) in $\Phi_{4j}$ illustrated in
Fig.~\ref{ABintsect1}.  We assign arrows to the path in $\Gamma$ by
following the flows generated by $J_{12}^2$ and $J_{23}^2$ in the
direction of increasing $\phi_{12}$ and $\phi_{23}$.  Thus, the path
The path $P\to Q\to P$ in part (f) of Fig.~\ref{rps} corresponds to
the sequence of rotations carried out on the tetrahedron in
Fig.~\ref{cycle}.  On reaching point $Q$, the path turns down, because
the rate of change of $J_{12}$ along the $J_{23}$-flow is the Poisson
bracket $\{J_{12},J_{23}\}$, which is negative when $V>0$, as follows
from (\ref{6jPB}).  One can also see this geometrically in
Fig.~\ref{cycle}; parts (b) and (c) of that figure correspond to point
$Q$, and it is clear that $J_{12}$ at first decreases, then increases
again, on carrying out the rotation about $\Jvec_{23}$, that is, on
passing from part (c) to part (d) of Fig.~\ref{cycle}.  

The symplectic area of the shaded region in part (f) of Fig.~\ref{rps}
is the same as the Ponzano-Regge phase, to within a constant that
depends on $j_r$, $r=1,\ldots,4$.  On quantized manifolds it will not
matter what region on $S^2$ is taken as the interior of the closed
loop on the sphere (the choices differ by multiples of the total area
of the sphere), but to be precise one must worry about this.  The
Ponzano-Regge phase is also $-2$ times the $F_4$-type generating
function (Goldstein 1980, Miller 1974).  Let us write
$(q,p)=(\phi_{12},J_{12})$, $(Q,P)=(\phi_{23},J_{23})$ and $F_4(p,P) =
-(1/2) S$, where $S$ is given by (\ref{thephase}).  Then we have
	\begin{equation}
	q = -\frac{\partial F_4}{\partial P}
	  =  \frac{\partial}{\partial J_5}
	\sum_{r=1}^6 J_r \phi_r,
	\label{F4q}
	\end{equation}
where we use index 5 instead of 12.  Here the angles $\phi_r$ must be
interpreted as functions of the $J$'s. 

A more geometrical way of stating the same thing is to express the
canonical transformation $(\phi_{12},J_{12}) \to (\phi_{23},J_{23})$
as a map between two spheres with symplectic forms (\ref{RPSomegas}).
Then the graph of the canonical transformation is a Lagrangian
manifold in $S^2\times S^2$ with symplectic form $dJ_{23}\wedge
d\phi_{23} - dJ_{12}\wedge d\phi_{12}$.  This Lagrangian manifold is
itself a sphere, whose projection onto $J_{12}$-$J_{23}$ space is
precisely the interior plus boundary of the oval curve in
Fig.~\ref{square}. The Lagrangian manifold has a double-valued
projection onto the interior of the oval region, and a single-valued
projection at the boundary (the caustics).  The two disks fit together
to form a sphere.

\subsection{The Ponzano-Regge Formula}
\label{PRformula}

If we factor out the phase
	\begin{equation}	
	\left(\frac{S_1+S_2}{2}\right)
	 - \left(\frac{\mu_1+\mu_2}{2}\right)\frac{\pi}{2}
	\label{splitphase}
	\end{equation}
from the quantity in the square bracket in (\ref{PRv1}), we obtain a
quantity proportional to
	\begin{equation}
	\cos\left[\left(\frac{S_1-S_2}{2}\right)
	 - (\mu_1-\mu_2)\frac{\pi}{4}
	\right].
	\label{PRphasev1}
	\end{equation}
However, in view of (\ref{S1S2defs}), (\ref{relphase}),
(\ref{thephase}) and (\ref{Psidef}), this can be written as a sign times
	\begin{equation}
	\cos\left(\Psi+\Delta\mu\frac{\pi}{4}\right),
	\label{PRcosV2}
	\end{equation}
where the sign is $(-1)^{j_2+j_4+j_{12}+j_{23}}$ and where
$\Delta\mu=\mu_2-\mu_1$.  Thus to obtain the argument of the cosine we
must compute the relative Maslov index between the two branches, just
as we have already computed the relative action $S=S_2-S_1$.  

This is the easiest of the Maslov index computations in deriving the
Ponzano-Regged formula.  We will omit details and simply remark that
the calculation can be carried out entirely on the reduced phase space
$\Gamma$, where it is not difficult to show that $\Delta\mu=1$.

There remains the overall phase, which, in view of the reality of
$6j$-symbol, must be a sign $\pm1$.  This phase cannot be computed
without taking into account the phase conventions for the two states
$\ket{A}$ and $\ket{B}$, as well as the absolute Maslov indices on the
intersection manifolds $I_1$ and $I_2$.  The analogous phase for the
$9j$-symbol was the most difficult part of the derivation of the
results presented in Haggard and Littlejohn (2010).  Since (in the
case of the $6j$-symbol) the answer is known, we can see that this
overall phase (combined with the $e^{i\pi/4}$ appearing in
(\ref{PRv1})) must be 1.  We had no such luxury in the case of the
$9j$-symbol, nor did we find it possible to guess the answer.  In any
case, the final result is the Ponzano-Regge formula,
	\begin{equation} 
	\left\{\begin{array}{ccc} j_1 & j_2 &
	j_{12} \\ j_3 & j_4 & j_{23} \end{array}\right\}
	=\frac{1}{\sqrt{12\pi|V|}} 
	\cos\left(\Psi+\frac{\pi}{4}\right).
	\label{PRv2}
	\end{equation}

For reasons of space we will omit the details in the calculation of
these final phases, also because we have made the main points we
wanted to make about the $6j$-symbol.  These are the importance of the
reduced phase space $\Gamma$, the geometry of symplectic reduction
connecting it with higher dimensional spaces, and the manner in which
elements of the semiclassical calculation (phases, amplitude
determinants, etc) can be mapped from one space to another.

\section{Conclusions}

A glance at the calculation of Roberts (1999) shows that it is much
easier and more elegant to compute the relative phase $S=S_2-S_1$ of
the $6j$-symbol in a symmetrical or $12j$-model than in the $4j$-model
presented in this paper.  The sum over edges times dihedral angles
appears almost immediately.  On the other hand, we have given a much
easier way of computing the amplitude determinant, reducing it to a
single Poisson bracket, whereas Roberts had to evaluate a large
determinant.  Obviously what is needed is a way of connecting the
various models at a semiclassical level, so that actions, amplitude
determinants, Maslov indices, etc, can be mapped from one model to
another and computed wherever most convenient.  We have investigated
this question and will report on our results in the future.  For now
we merely make a few comments.

Already in this paper we have mapped amplitude determinants between
various spaces, such as the Schwinger phase space $\Phi_{4j}$ and the
reduced phase space $\Gamma$, which are connected by symplecic
reduction.  One might guess, therefore, that the $12j$- or symmetrical
model of the $6j$-symbol and the $4j$-model are related by some kind
of symplectic reduction.  This is not the case, however, at least not
in the way that $\Phi_{4j}$ and $\Gamma$ are related.  One piece of
picture, however, is the following.

It is well known that a unitary map on a Hilbert space $U:\HS\to\HS$
corresponds semiclassically to a symplectic map or canonical
transformation $T:\Phi\to\Phi$, where $\Phi$ is the symplectic
manifold corresponding to $\HS$.  Also, the symplectic map $T$ is
conveniently viewed via its graph in $\Phi\times\Phi$, regarded as a
symplectic manifold in its own right with symplectic form
$\omega_1-\omega_2$, where the subscripts refer to the first and
second factors of $\Phi\times\Phi$ (Abraham and Marsden 1978).  With
this understanding, the graph of $T$ is a Lagrangian manifold, one
which supports semiclassically the operator $U$ in the same way as a
Lagrangian manifold in $\Phi$ supports a vector in $\HS$.  That is,
$U$, which begins as a map $:\HS\to\HS$, is reinterpreted as a
function $:\HS\otimes\HS^*\to\Complexes$, that is, a ``wave function''
on a doubled space.  The basic ideas inherent in this picture were
developed by H\"ormander (1971) and are also present in Miller (1974),
while an elementary explanation of the geometrical situation is given
by Littlejohn (1990).

It turns out that this picture generalizes to linear maps between
Hilbert spaces $M:\HS\to\HS'$, where the two Hilbert spaces need not
have the same dimensionality and where the map need not be unitary or
even invertbible.  The genralization involves again symplectic
reduction, but in a different manner than that in which it appears in
this paper.  This situation arises in the comparison of two models of
the $3j$-symbol, which is a simpler version of the comparison between
the $12j$- and $4j$-models of the $6j$-symbol.  The first is a $2j$-
or Clebsch-Gordan model, in which the Hilbert space is
$\CS_{j_1}\otimes\CS_{j_2}$, where the operator $\Jvec_3$ is a
function of $\Jvec_1$ and $\Jvec_2$, that is, $\Jvec_3 =\Jvec_1
+\Jvec_2$.  The second is a symmetrical or $3j$-model, in which the
Hilbert space is $\CS_{j_1}\otimes\CS_{j_2}\otimes\CS_{ j_3}$ and the
operator $\Jvec_3$ is independent, but in which we are interested only
in states satisfying the constraint $\Jvec_1+\Jvec_2+\Jvec_3=0$.  We
will report on these investigations in the future.
	
Roberts' (1999) derivation of the Ponzano-Regge formula effectively
proceeded by writing the $6j$-symbol as a scalar product
$\braket{B}{A}$, where $\ket{A}$ and $\ket{B}$ are given by
Fig.~\ref{robertsAB}, then writing the wave functions for $\ket{A}$
and $\ket{B}$ in the Bargmann (1962) or coherent state representation
to obtain an integral representation for the $6j$-symbol as an
integral over $\Complexes^{24}$.  In the coherent state representation
there is one copy of $\Complexes$ for each degree of freedom, while in
a $12j$-model there are two degrees of freedom for each $j$, hence 24
degrees of freedom total.  Roberts then used the stationary phase
approximation to evaluate the integral.  The Bargmann representation
has been used in a similar manner in several recent asymptotic studies
in the quantum gravity literature.

In I and in this paper, however, we have mostly worked in a
represenation-independent manner.  Our emphasis on Lagrangian
manifolds and other geometrical structures in phase space is part of
this approach.  From a certain point of view the coherent state
representation is just another representation, albeit a complex one,
so there is the question of whether it plays any privileged role or
offers any advantages.

For some purposes it certainly does, for example, the wave
functions in the coherent state representation are polynomials that
can be written down explicitly, and these in turn are useful for
deriving generating functions and other things, as shown by Schwinger
(1952) and Bargmann (1962).  Another point is that in the case of
$SU(2)$, the coherent state representation appears naturally in the
method of geometric quantization, that is, the holomorphic sections of
Hermitian line bundles over the symplectic manifold $S^2$ are
represented explicitly by Bargmann's entire analytic wave functions.

On the other hand, it is not obvious to us that the identification of
the stationary phase set or the other aspects of the semiclassical
calculation are easier or more transparent in the coherent state
representation than in the representation-independent approach of this
paper.  Moreover, if one focuses too narrowly on the stationary phase
evaluation of an integral, one misses the geometrical structures that
support the representation-independent approach. 

We had to assemble and extend several of the ideas presented in this
paper for our derivation of the asymptotic form of the $9j$-symbol,
and we have studied other extensions such as ``$g$-inserted'' spin
networks, in which a group element or $D$-matrix is inserted in the
edges of the spin network.  Such $g$-inserted spin networks are basic
amplitudes in loop quantum gravity (Rovelli 2004), taking one from
the basis of spin connections to the spin network basis.  The
asymptotics of these amplitudes leads to piecewise flat manifolds
similar to those introduced by Regge (1961). 

We will report on these and other developments in the future. 

\ack

The authors would like to acknowledge a number of stimulating and
helpful conversations with Annalisa Marzuoli and Mauro Carfora.

\section*{References}
\begin{harvard}

\item[] Abraham R and Marsden J E 1978 {\it Foundations of Mechanics}
(Reading, Massachusetts:  Benjamin/Cummings)

\item[] Alesci E, Bianchi E, Magliaro E and Perini C 2008 preprint
gr-qc 0809.3718

\item[] Ali\v saukas S 2000 {\it J Math Phys} {\bf 41} 7589

\item[] Anderson R W and Aquilanti V 2006 {\it J Chem Phys} {\bf
124} 214104

\item[] Aquilanti V, Cavalli S and De~Fazio D 1995 {\it J. Phys. Chem.}
{\bf 99} 15694

\item[] Aquilanti V, Cavalli S and Coletti C 2001 {\it Chem. Phys. Lett.}
{\bf 344} 587

\item[] Aquilanti V and Coletti C 2001 {\it Chem. Phys. Lett.} {\bf
344} 601 

\item[] Aquilanti V, Haggard H M, Littlejohn R G and Yu Liang 2007
{\it J Phys A} {\bf 40} 5637

\item[] Arnold V I 1989 {\it Mathematical Methods of Classical
Mechanics} (New York: Springer-Verlag)

\item[] Baez J C 1996 {\it Adv. Math.} {\bf 117} 253

\item[] Baez J C and Barrett J W 1999 {\it Adv. Theor. Math. Phys.}
{\bf 3} 815

\item[] Baez J C, Christensen J D and Egan G 2002 {\it Class. Quantum
Grav.} {\bf 19} 6489

\item[] Balazs N L and Jennings B K 1984 {\it Physics Reports} {\bf
104} 347

\item{} Balcar E and Lovesey S W 2009 {\it Introduction to the
Graphical Theory of Angular Momentum} (Berlin: Springer-Verlag)

\item[] Bargmann V 1962 {\it Rev. Mod. Phys.} {\bf 34} 829

\item[] Barrett J W 1998 {\it Adv. Theor. Math. Phys.} {\bf 2} 593

\item[] Barrett J W and Crane L 1998 {\it J. Math. Phys.} {\bf 39}
3296

\item[] Barrett J W, Dowdall R J, Fairbairn W J, Gomes H and Hellmann
F 2009 preprint gr-qc 0902.1170

\item[] Barrett J W and Steele C M 2003 {\it Class. Quantum Grav.}
{\bf 20} 1341

\item[] Barrett J W and Williams R M 1999 {\it Adv. Theor. Math. Phys.}
{\bf 3} 209

\item[] Berger M 1987 {\it Geometry I} (Berlin:  Springer-Verlag)

\item[] Berry M V 1977 {\it Phil Trans R Soc} {\bf 287} 237

\item[] Berry M V 1984 {\it Proc Roy Soc Lond A} {\bf 392}, 45

\item[] Bianchi E and Haggard H 2011 {\it Phys Rev Lett} {\bf 107}
011301

\item[] Biedenharn L C and Louck J D 1981 {\it The Racah-Wigner 
Algebra in Quantum Theory} (Reading, Massachusetts: Addison-Wesley)

\item[] Borodin K S, Kroshilin A E and Tolmachev V V (1978) {\it
Teoreticheskaya i Matematicheskaya Fizika} {\bf 34} 110, English
translation {\it Theoretical and Mathematical Physics} {\bf 34} 69

\item[] Brink D M and Satchler G R 1993 {\it Angular Momentum}
(Oxford: Oxford University Press)

\item[] Brunnemann J and Rideout D 2008 preprint gr-qc 0706.0469

\item[] \dash 2010 preprint gr-qc 1003.2348

\item[] Carbone G, Carfora M and Marzuoli A 2002 {\it Class. Quantum
Grav.} {\bf 19} 3761

\item[] Carlip S 1998 {\it Quantum Gravity in $2+1$ Dimensions} 
(Cambridge: Cambridge University Press)

\item[] Chakrabarti A 1964 {\it Ann. Inst. H Poincar\'e} {\bf 1} 301

\item[] Charles L 2008 preprint math 0806.1585

\item[] Conrady F and Freidel L 2008 {\it Phys. Rev. D} {\bf 78} 104023

\item[] Crippen G M and Havel T F 1988 {\it Distance Geometry and 
Molecular Confirmation} (Baldock, UK: Research Studies Press Ltd) 

\item[] Cushman R H and Bates L 1997 {\it Global Aspects of Classical 
Integrable Systems} (Basel: Birkh\"auser Verlag)

\item[] Cvitanovich P 2008 {\it Group Theory: Birdtracks, Lie's, and 
Exceptional Groups} (Princeton, Princeton University Press)

\item[] Danos M and Fano U 1998 {\it Phys. Rep.} {\bf 304} 155

\item[] De~Fazio D, Cavalli S and Aquilanti V 2003 {\it
Int. J. Quant. Chem.} {\bf 93} 91

\item[] Ding Y and Rovelli C 2010 preprint gr-qc 0911.0543

\item[] Edmonds A R 1960 {\it Angular Momentum in Quantum Mechanics}
(Princeton: Princeton University Press)

\item[] El Baz E and Castel B 1972 {\it Graphical Methods of Spin
Algebras} (New York: Marcel Dekker)

\item[] Freidel L and Louapre D 2003 {\it Class. Quantum Grav.}
{\bf 20} 1267

\item[] Goldstein H 1980 {\it Classical Mechanics} 2nd ed (Reading,
Massachusetts: Addison-Wesley)

\item[] Groenewold H J 1946 {\it Physica} {\bf 12} 405

\item[] Guillemin V and Sternberg S 1977 {\it Geometric Asymptotics}
(Providence, Rhode Island: American Mathematical Society)

\item[] \dash 1982 {\it Invent Math} {\bf 67} 515

\item[] Gurau R 2008 The Ponzano-Regge asymptotic of the $6j$-symbol:
an elementary proof {\it Preprint} math-ph 0808.3533

\item[] Gutzwiller M C 1967 {J Math Phys} {\bf 8} 1979

\item[] \dash 1969 {J Math Phys} {\bf 10} 1004

\item[] \dash 1970 {J Math Phys} {\bf 11} 1791

\item[] \dash 1971 {J Math Phys} {\bf 12} 343

\item[] Hackett J and Speziale S 2007 {\it Class. Quantum Grav.} {\bf 
24} 1525

\item[] Haggard H M and Littlejohn R G 2010 {\it Classical and Quantum
Gravity} {\bf 27} 135010

\item[] H\"ormander L 1971 {\it Acta Math} {\bf 127} 79

\item[] Kapovich M and Millson J J 1995 {\it J. Differential Geometry}
{\bf 42} 133

\item{} \dash 1996 {\it J. Differential Geometry} {\bf 44} 479

\item[] Kauffman L H and Lins 1994 {\em Temperley-Lieb Recoupling
Theory and Invariants of 3-Manifolds}, (Princeton, New Jersey:
Princeton University Press, New Jersey)

\item[] Kirillov A A 1976 {\it Elements of the Theory of
Representations} (New York: Springer-Verlag)

\item[] \dash 2004 {\it Lectures on the Orbit Method} (Providence, Rhode
Island:  American Mathematical Society)

\item[] Kurchan, Leboeuf P and Saraceno M 1989 {\it Phys. Rev. A} {\bf
40}, 6800

\item[] L\'evy-Leblond J-M and L\'evy-Nahas M 1965 {\it
J. Math. Phys.} {\bf 6} 1372

\item[] Lewandowski J 1996 preprint gr-qc 9602035

\item[] Lindgren I and Morrison J 1986 {\it Atomic Many-Body Theory}
(Springer-Verlag, New York)

\item[] Littlejohn R G 1986 {\it Phys. Reports} {\bf 138} 193

\item[] \dash 1990 {\it  J. Math.\ Phys.} {\bf 31} 2952

\item[] Littlejohn R G and Reinsch M 1995 {\it Phys.\ Rev.\ A} {\bf52}
2035

\item[] \dash 1997 {\it Rev Mod Phys} {\bf 69} 213

\item[] Littlejohn R G and Yu L 2009 {\it J. Phys. Chem. A} {\bf 113} 
14904

\item[] Major S A and Seifert M D 2001 preprint 0109056

\item[] Marsden J E and Ratiu T 1999 {\it Introduction to Mechanics and
  Symmetry} (New York: Springer-Verlag)

\item[] Martinez A 2002 {\it An Introduction to Semiclassical and
Microlocal Analysis} (New York: Springer-Verlag)

\item[] Marzuoli A and Rasetti M 2005 {\it Ann. Phys.} {\bf 318} 345

\item[] Maslov V P and Fedoriuk M V 1981 {\it Semi-Classical
Approximations in Quantum Mechanics} (Dordrecht: D. Reidel)

\item[] Messiah A 1966 {\it Quantum Mechanics} (New York: John Wiley)

\item[] Mishchenko A S, Shatalov V E and Sternin B Yu 1990 {\it Lagrangian
manifolds and the Maslov operator} (Berlin: Springer Verlag)

\item[] Miller W H 1974 {\it Adv. Chem. Phys.} {\bf 25} 69

\item[] Minkowski H 1897 {\it Nachr. Gess. Wiss.} 198

\item[] Moyal J E 1949 {\it Proc Camb Phil Soc} {\bf 45} 99

\item[] Neville D  1971 {\it J Math Phys} {\bf 12} 2438

\item[] \dash 2006 {\it Phys Rev D} {\bf 73} 124004

\item[] Nicolai H and Peeters K 2007, in {\it Approaches to Fundamental 
Problems}, Seiler E and Stamatescu I-O eds (Berlin: Springer) (Lecture
Notes in Physics v.~721) 151

\item[] Ooguri H 1992a {\it Nucl. Phys. B} {\bf 382} 276

\item[]\dash 1992b {\it Mod. Phys. Lett. A} {\bf 7} 2799

\item[] Ozorio de Almeida A M 1998 {\it Phys Rep} {\bf 295} 265

\item[] Penrose R 1971 in {\it Combinatorial Mathematics and its
Applications}, edited by D. Welsh (New York:  Academic Press)

\item[] Ponzano G and Regge T 1968 in {\it Spectroscopy and Group
Theoretical Methods in Physics} ed F Bloch \etal\ (Amsterdam:
North-Holland) p~1

\item[] Ragni M, Bitencourt A C P, Ferreira C da S, Aquilanti V, 
Anderson R W and Littlejohn R G 2010 {\it Int. J. Quantum Chem.} 
{\bf 110} 731

\item[] Regge T 1961 {\it Il Nuovo Cimento} {\bf 19}, 558

\item[] Regge T and Williams R M 2000 {\it J. Math. Phys.} {\bf 41} 3964

\item[] Roberts J 1999 {\it Geometry and Topology} {\bf 3} 21

\item[] Rovelli C 2004 {\it Quantum Gravity} (Cambridge University
Press)

\item[] Rovelli C and Smolin L 1995 {\it Phys. Rev. D} {\bf 52}, 5743

\item[] Rovelli C and Speziale S {\it Class. Quantum Grav.} {\bf 23} 
5861

\item[] Sastry K R S 2002 {\it Forum Geom} {\bf 2} 167

\item[] Schulten K and Gordon R G 1975a {\it J. Math. Phys.} {\bf 16} 1961

\item[] \dash 1975b {\it J. Math. Phys.} {\bf 16} 1971

\item[] Schwinger J 1952 {\it On Angular Momentum} U.S. Atomic Energy
Commission, NYO-3071, reprinted in Biedenharn L C and van Dam H 1965
{\it Quantum Theory of Angular Momentum} (New York: Academic Press)

\item[] Smolin L 2005 An Invitation to Loop Quantum Gravity {\it
Preprint} hep-th/0408048

\item[] Stedman G E 1990 {\it Diagram Techniques in Group Theory} 
(Cambridge: Cambridge University Press)

\item[] Taylor Y U and Woodward C T 2004 preprint math 0406.228

\item[] Thiemann T 2007 {\it Modern Canonical Quantum General
Relativity} (Cambridge:  Cambridge University Press)

\item[] Turaev and Viro 1992 {\it Topology} {\bf 31} 865

\item[] Varshalovich D A, Moskalev A N and Khersonskii V K 1981 {\it
Quantum Theory of Angular Momentum} (Singapore:  World Scientific)

\item[] Van der Veen R 2010 PhD Dissertation ``Asymptotics of quantum
spin networks''
http://www.science.uva.nl/research/math/Research/Dissertations/Veen2010.text.pdf

\item[] Voros A 1989 {\it Phys. Rev. A} {\bf 40}, 6814

\item[] Whittaker E T 1960 {\it A Treatise on the Analytical Dynamics
of Particles and Rigid Bodies} (Cambridge: Cambridge University Press)

\item[] Wormer P E S and Paldus J 2006 {\it Adv Quantum Chemistry}
{\bf 51} 59

\item[] Weissman Y 1982 {\it J. Chem. Phys.} {\bf 76}, 4067

\item[] Weyl H 1927 {\it Z Phys} {\bf 46} 1

\item[] Wigner E P 1932 {\it Phys Rev} {\bf 40} 749

\item[] \dash 1940 {\it On the matrices which reduce the Kronecker
products of representations of simply reducible groups} (unpublished),
reprinted in Biedenharn L C and van Dam H 1965 {\it Quantum Theory of
Angular Momentum} (New York: Academic Press)

\item[] \dash 1959 {\it Group Theory} (Academic Press, New York)

\item[] Woodhouse N M J 1991 {\it Geometric Quantization} (Oxford:
Oxford University Press)

\item[] Yutsis A P, Levinson I B and Vanagas V V 1962 {\it The Theory
of Angular Momentum} (S Monson, Jerusalem)

\end{harvard}

\end{document}